\documentclass[]{aastex701}
\usepackage{xspace}
\usepackage{amsmath}
\usepackage{csquotes}
\usepackage{graphicx}
\usepackage{float}

\newcommand{\Gaia}{\textit{Gaia}\xspace}

\newcommand{\kms}{\ensuremath{\mathrm{km\,s^{-1}}}}
\newcommand{\vsini}{\ensuremath{v\sin i}}

\usepackage[normalem]{ulem}
\newcommand\out{\bgroup\markoverwith
{\textcolor{red}{\rule[.5ex]{2pt}{1pt}}}\ULon}
\usepackage{xcolor}
\usepackage{caption}

\definecolor{CBgreen}{RGB}{0,158,115}

\shorttitle{Weighing mass via Rotational Broadening}
\shortauthors{Wang et al.}

\graphicspath{{./}{figures/}}
\begin{document}

\title{Weighing Hidden Companions of Compact Object Candidates via Rotational Broadening}

\author[0009-0005-4010-8213]{Rui Wang}
\affiliation{Department of Astronomy, Xiamen University, Xiamen, Fujian 361005, People’s Republic of China}
\email{rwang@stu.xmu.edu.cn}

\author[0000-0002-2419-6875]{Zhi-Xiang Zhang}
\affiliation{College of Physics and Information Engineering, Quanzhou Normal University, Quanzhou, Fujian 362000, People’s Republic of China}
\email[show]{zzx@qztc.edu.cn}

\author[0000-0003-3137-1851]{Wei-Min Gu}
\affiliation{Department of Astronomy, Xiamen University, Xiamen, Fujian 361005, People’s Republic of China}
\email[show]{guwm@xmu.edu.cn}

\author[0000-0002-2912-095X]{Hao-Bin Liu}
\affiliation{Department of Astronomy, Xiamen University, Xiamen, Fujian 361005, People’s Republic of China}
\email{liuhaobin@stu.xmu.edu.cn}

\author[0000-0002-5839-6744]{Tuan Yi}
\affiliation{Department of Astronomy, School of Physics, Peking University, Beijing 100871, People’s Republic of China}
\email{yituan@pku.edu.cn}

\author[]{Zhong-Rui Bai}
\affiliation{National Astronomical Observatories, Chinese Academy of Sciences, Beijing 100101, People’s Republic of China}
\email{zrbai@nao.cas.cn}

\begin{abstract}
The determination of unseen companion masses ($M_1$) is essential for identifying compact objects in binary systems, yet obtaining reliable orbital inclinations remains one of the most difficult challenges.
In this study, we focus on ten targets selected from a sample of 89 compact object candidates characterized by large mass functions, rapid rotation, and high-quality Large Sky Area Multi-object Fiber Spectroscopic Telescope (LAMOST) spectra.
We measure their projected rotational velocities ($v \sin i$) from the LAMOST medium-resolution spectra and, combined with stellar radii, derive orbital inclinations and the corresponding companion masses.
Our results show that five sources exhibit mass ratios $M_1 / M_2 > 2/3$, with no detectable spectral signatures of the unseen companions, providing strong evidence for their compact nature. 
Two particularly notable cases, J0341 and J0359, host companions with inferred masses of $1.39^{+0.09}_{-0.10}$ $M_\odot$ and $1.34^{+0.08}_{-0.09}$ $M_\odot$, respectively. These masses suggest that the invisible objects are either neutron stars or massive white dwarfs with masses close to the Chandrasekhar limit. If they are white dwarfs, these two targets are highly likely to be Type Ia supernova progenitors. This study highlights the potential of $v \sin i$ measurements as a systematic approach to unveiling compact objects in binaries.
\end{abstract}


\section{Introduction}\label{intro}

Searching for and confirming compact objects is fundamental for understanding the stellar evolution pathways, binary interactions, and the formation of exotic objects like neutron stars and black holes \citep{Tauris2017}. It is estimated that the Milky Way contains roughly \(10^7\) stellar-mass black holes in binaries \citep{2020A&A...638A..94O} and \(10^9\) neutron stars \citep{2010A&A...510A..23S}, yet only a tiny fraction of these objects have been identified through electromagnetic or gravitational wave observations \citep{corral2016,ligo2023}.

Traditional approaches to identifying compact objects have relied primarily on either pulsar surveys or accretion signatures. Large-scale radio surveys have discovered thousands of pulsars, providing a well-characterized population of isolated and binary neutron stars \citep{Manchester2005}. However, radio surveys are inherently limited to detecting only those neutron stars whose radio beams intersect the Earth, leaving a potentially large population of undetectable pulsars. In parallel, X-ray observations reveal accreting compact objects in binaries, including roughly 350 low-mass X-ray binaries \citep[LMXBs;][]{2023A&A...677A.134N} and over 169 high-mass X-ray binaries \citep[HMXBs;][]{2023A&A...675A.199A} in the Milky Way. The X-ray–based approach, meanwhile, is biased toward systems with strong accretion, potentially missing quiescent or weakly accreting objects. More recently, time-domain spectroscopic surveys and photometric variability studies have provided alternative pathways to identify compact companions in binaries through their dynamical effects (e.g., \citealp{2022NatAs...6.1203Y,2022MNRAS.510.4736K,2023MNRAS.521.4323E,2023AJ....165..187Q,2024ApJ...969..114L,2023MNRAS.518.2991S}).
%\textbf{\color{blue}\citealp{2023MNRAS.518.2991S}}).

The search for compact objects in quiescent binaries relies on the data accumulation of large time-domain spectroscopic surveys, which provide an efficient approach to identify compact object candidates from radial-velocity observations (\citealp{2019ApJ...872L..20G, 2022NatAs...6.1085S,2022A&A...664A.159M}). %;\textbf{\color{blue}\citealp{2022A&A...664A.159M}}).
In single-lined spectroscopic binaries (SB1s), the orbital period $P_{\mathrm{orb}}$ and radial velocity semi-amplitude $K_2$ of the visible star can be measured from multi-epoch spectra, and the mass function of the invisible companion can be derived as
\begin{equation}\label{eq1}
f(M_1) = \frac{(M_1 \sin i)^3}{(M_1 + M_2)^2} = \frac{K_2^3 P_{\mathrm{orb}}}{2\pi G},
\end{equation}
where $M_1$ and $M_2$ are the masses of the invisible and visible components, respectively, and $i$ is the orbital inclination. Given the measured mass function $f(M_1)$ and the mass of the visible star $M_2$, the mass of the invisible companion $M_1$ can be determined once the orbital inclination $i$ is known. In general, the orbital inclination is typically constrained using two methods:  modeling the ellipsoidal modulations in the photometric light curves \citep{2021MNRAS.501.2822G} or measuring the projected rotation broaden velocity $V_{\mathrm{rot}} \sin i$ (hereafter $v \sin i$) from high-resolution spectroscopy \citep{2020AJ....159...81M,zhang2024}. 

Light curve modeling is most widely used in the inclination determination. However, this approach is often affected by various factors, such as starspots, stellar pulsations, or uncertainties in the limb-darkening coefficients. These effects may lead to significant uncertainties in the derived inclination. As an alternative, $v \sin i$ measurements can provide an independent constraint on the inclination, assuming tidal synchronization and spin–orbit alignment \citep{2022A&A...664A.159M}. In tidally locked systems, the rotation velocity is given by $V_{\mathrm{rot}} = 2\pi R_{2}/P_{\mathrm{orb}}$, where $R_{2}$ is the radisee alsous of the visible star, so that $\sin i$ can be directly derived from the measured $v \sin i$, the visible star's radius, and the orbital period. Nevertheless, this method requires relatively high spectral resolution and has therefore been used only rarely in previous studies. In recent years, new spectroscopic surveys, such as Large Sky Area Multi-object Fiber Spectroscopic Telescope Medium-Resolution Spectroscopic Survey \citep[LAMOST-MRS;][]{Liuchao2020}, \textit{Gaia} Radial-Velocity Spectrometer \citep{gaia2016}, and APOGEE \citep{Majewski2017}, have provided large samples of medium- to high-resolution spectra, offering new opportunities to constrain orbital inclinations through $v \sin i$ measurements in tidally locked systems.

%In contrast, $v \sin i$ measurements provide a more robust approach. 
Despite these opportunities, most recent applications have not yet exploited  $v \sin i$ to its full potential. 
For instance, \citet{2024ApJ...969..114L} identified 89 compact object candidates from the LAMOST survey. For each system, the mass function was well constrained from its radial-velocity curve. However, only the mass function and minimum masses of the unseen companions were provided, without reasonable estimates of the compact object masses. To determine the masses of the invisible components, measurements of the orbital inclination are needed. Building on these developments, we utilize the LAMOST-MRS spectra ($R \sim 7500$) to measure $v \sin i$. This allows us to directly constrain the orbital inclinations of the systems and determine the masses of the compact companions. Given the scale and quality of the LAMOST MRS data, $v \sin i$ measurements hold significant promise for advancing future searches and characterization of compact objects in quiescent binaries.

In this work, we use $v \sin i$ measurements of the visible stars to estimate orbital inclinations and thereby constrain the masses of unseen companions in compact object candidates. Section~\ref{sample and data} introduces our sample selection and spectroscopic data. Section~\ref{result} describes $v \sin i$ measurements and mass estimates for the unseen companions. In Section~\ref{dis}, we discuss the fitting results and highlight two notable targets, and we present the summary in Section~\ref{conclusion}.

\section{Sample and Data}\label{sample and data}
\subsection{Sample Selection}\label{sample}
The targets of this paper are drawn from the compact object candidate catalog of \citet{2024ApJ...969..114L}, which is SB1s identified in the LAMOST survey. These objects exhibit significant radial velocity variability and relatively high mass functions, suggesting the presence of massive unseen companions. However, the analysis of \citet{2024ApJ...969..114L} was limited to deriving mass functions and minimum companion masses, without direct constraints on orbital inclinations. As a result, the masses of the compact companions remain uncertain. To constrain the nature of the unseen companions, we measure $v \sin i$ of the visible stars and use them together with the visible star masses and mass functions to determine the unseen companion masses. The visible stars in our sample are low- to intermediate-mass stars, with masses of $M_2 \simeq 0.5$–$2.5~M_\odot$ and effective temperatures ranging from $4100$ to $8500$ K, corresponding to spectral types from late K to early A (see Table~\ref{tab:parameter}).

%%%%%fig1
\begin{figure}
    \centering
    \includegraphics[width=\linewidth]{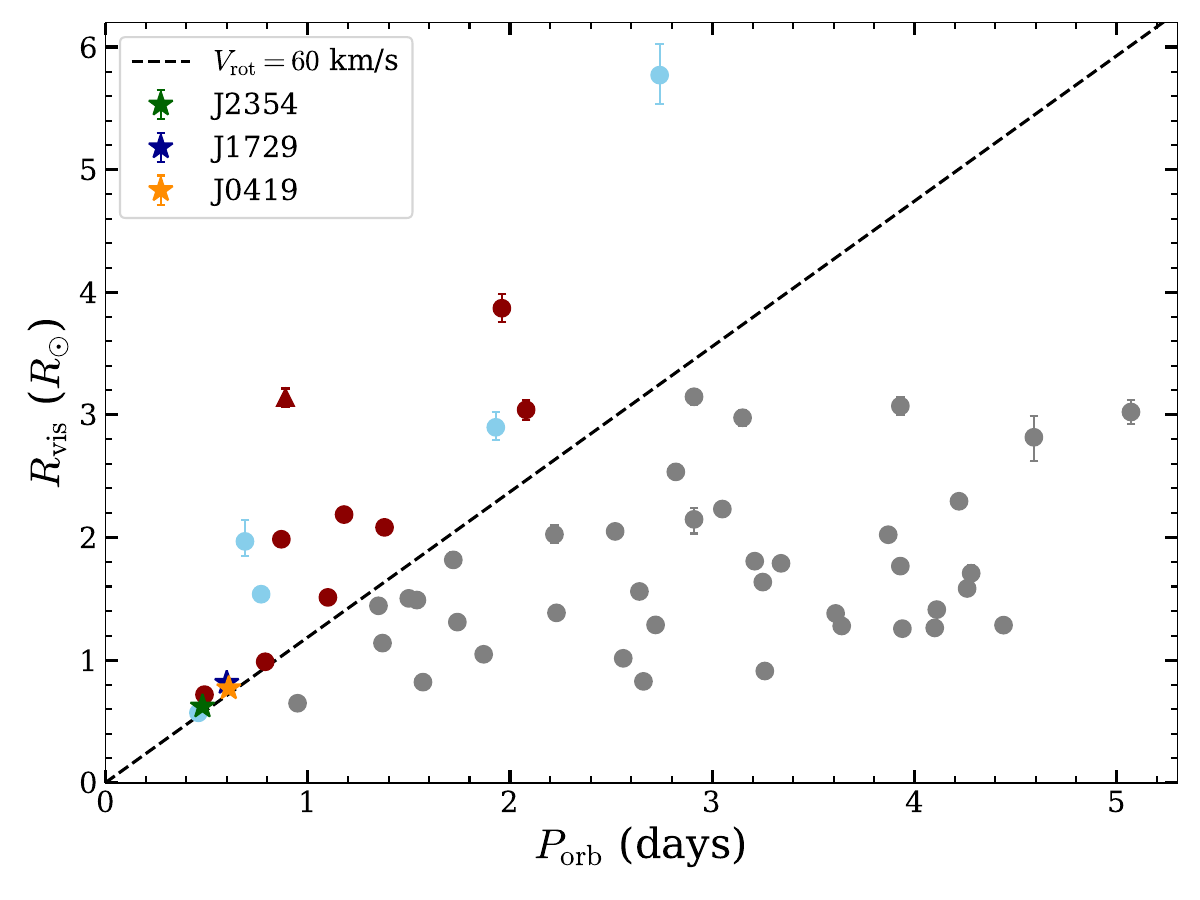}
    \caption{Orbital period vs visible star's radius for compact object candidates with $f(M_{1}) > 0.05~M_\odot$. The radii are obtained from \textit{Gaia} DR3. The dashed line corresponds to $V_{\mathrm{rot}} = 60~\mathrm{km\,s^{-1}}$. The sources above the line are shown as blue points for $\mathrm{SNR} < 30$, red points for $\mathrm{SNR} > 30$, and a red triangle for double-lined spectroscopic binaries. Three identified candidates, J2354 \citep{2023SCPMA..6629512Z}, J1729 \citep{2022ApJ...936...33Z}, and J0419 \citep{2022ApJ...933..193Z}, are highlighted with star markers.}
    \label{fig:Porb_vs_Rvis}
\end{figure}

To select suitable targets and ensure reliable $v \sin i$ measurements, we apply the following selection criteria to the sample from \citet{2024ApJ...969..114L}:
\begin{enumerate}
\item Mass function $f(M_{1}) > 0.05~M_{\odot}$;
\item $V_{\mathrm{rot}} = 2\pi R_{\mathrm{gaia}} / P_{\mathrm{orb}} > 60~\mathrm{km\,s^{-1}}$, where $R_{\mathrm{gaia}}$ is the radius of the visible star estimated from \textit{Gaia} Data Release 3 \citep[\textit{Gaia} DR3;][]{2023AA...674A...1G};
\item Each target is required to have spectra with SNR $> 30$ near either orbital phase 0.25 or 0.75.
\end{enumerate}
We adopt $f(M_1) > 0.05~M_\odot$ to avoid systems with extremely low mass ratios, in which even a normal main-sequence companion would contribute negligibly to the optical flux. In such cases, the absence of detectable secondary spectral features does not provide a meaningful constraint on whether the unseen companion is a compact object or a faint main-sequence star.
Among the 89 compact object candidates reported by \citet{2024ApJ...969..114L}, 11 candidates satisfy the above criteria. We inspect these 11 targets using the cross-correlation function (CCF) to check for potential double-lined binaries. Each observed spectrum is cross-correlated with a template constructed using the stellar parameters reported by the LAMOST catalog. One source (red triangle in Figure~\ref{fig:Porb_vs_Rvis}) shows a pronounced double-peaked CCF profile and is excluded from further analysis. The final sample thus contains 10 single-lined binaries, including two previously confirmed objects, J2354 and J1729, as shown in Figure~\ref{fig:Porb_vs_Rvis}. Basic properties of the selected systems are summarized in Table~\ref{tab:sample}.

\subsection{Spectroscopic Data}\label{data}

LAMOST is a 4-meter Schmidt telescope with 4000 fibers over a 5 degree field of view. It is capable of obtaining both low-resolution (LRS; R$\sim$1800) and medium-resolution (MRS; R$\sim$7500) spectra. The medium-resolution spectrograph comprises two arms \citep{Liuchao2020}: the blue arm (4950–5150~\text{\AA}) and the red arm (6300–6800~\text{\AA}). Each MRS observation consists of multiple consecutive exposures with 15–20 minute intervals. Both combined and individual-exposure spectra are provided for each observation. The resolution of LAMOST MRS spectra corresponds to an instrumental broadening of approximately $40~\mathrm{km\,s^{-1}}$. For all targets in our sample, the rotational broadening is comparable to or larger than the instrumental broadening. As a result, the line profiles are not dominated by the instrumental resolution, and the instrumental contribution is consistently accounted for by convolving the synthetic spectra with the instrumental line profile when measuring $v \sin i$.

In this work, we use MRS from LAMOST Data Release 11\footnote{\url{https://www.lamost.org/dr11/v1.1/medcas/search}} to measure $v \sin i$ of compact object candidates. For each target, the single-exposure spectra near phase 0.25 and 0.75 are selected. These spectra have minimal velocity shifts ($\sim 2~\mathrm{km\,s^{-1}}$) between two consecutive exposures, thus avoiding the smearing effect. An exception is J1729, which lacks a LAMOST spectrum near phase 0.25 or 0.75. The spectrum we used shows significant radial velocity variation, so smearing effects cannot be neglected. (see Section~\ref{results}). The Heliocentric Julian Dates (HJDs) of the selected spectra are listed in Table~\ref{tab:sample}.

\begin{deluxetable*}{lllrrrrrrrrr}
\tablecaption{Basic information of the 10 compact object candidates.\label{tab:sample}}
\tablewidth{1.0\textwidth}
% \tabletypesize{\scriptsize}
\tablehead{
\colhead{Objname} & \colhead{R.A.} & \colhead{Decl.} & \colhead{HJD} & \colhead{SNR} &  \colhead{SNR} & \colhead{$P_{\rm orb}$ } & \colhead{$K_{2}$} &\colhead{Distance} &\colhead{$R_{\mathrm{gaia}}$ }&\colhead{$f(M_{1})$}\\
\colhead{} & \colhead{} & \colhead{} & \colhead{}& \colhead{(blue)}  & \colhead{(red)}& \colhead{(days)} & \colhead{(km\,s$^{-1}$)}& \colhead{(pc)} &\colhead{($R_\odot$)} & \colhead{($M_\odot$)} 
}
\startdata
J0106  & 01:06:41.31 & +02:56:21.7 & 2458822.00950 & 90.0  & 134.2 & 0.87& 84.44 & 1228$\pm$32 & 1.99$\pm$0.03 & 0.059 \\
J0117  & 01:17:26.82 & +45:20:13.2 & 2459509.16928 & 132.2 & 153.5 &2.08 & 78.89 &1597$\pm$37  & 3.04$\pm$0.08 & 0.104 \\
J0341  & 03:41:21.29 & +57:33:15.0 & 2458863.08442 & 77.7  & 146.7 &1.10 & 65.06 &276$\pm$2    & 1.51$\pm$0.03 & 0.060 \\
J0359  & 03:59:12.12 & +57:02:28.8 & 2458478.12300 & 98.7  & 124.2 &1.38 & 77.48 & 1061$\pm$17 & 2.08$\pm$0.05 & 0.066 \\
J0635  & 06:35:52.55 & +18:42:33.7 & 2458415.36389 & 98.7  & 124.2 & 1.96& 74.73 & 1179$\pm$29 & 3.89$\pm$0.13 & 0.082 \\
J0853  & 08:53:04.59 & +13:20:32.3 & 2458134.17416 & 85.6  & 146.7 & 0.79 & 128.69&325$\pm$3    & 0.99$\pm$0.03 & 0.161 \\
J1429  & 14:29:53.75 & +46:03:53.8 & 2458567.26083 & 30.8  & 74.8  &0.49 & 100.55 & 118$\pm$1   & 0.72$\pm$0.02 & 0.052 \\
J1729  & 17:29:00.17 & +65:29:52.8 & 2458631.19465 & 96.4  & 152.5 &0.60  &93.87  & 229$\pm$2   & 0.81$\pm$0.02 & 0.123 \\
J2023  & 20:23:09.48 & +40:19:16.5 & 2458415.98047 &  124.7& 130.1 &1.18 & 94.04 &684$\pm$6    & 2.19$\pm$0.05 & 0.101 \\
J2354  & 23:54:56.76 & +33:56:25.7 & 2458831.95948 & 51.3  & 102.0 &0.48 & 210.50 & 127$\pm$1   & 0.62$\pm$0.02 & 0.525 \\
\enddata
\tablecomments{Columns are: (1) Object name; (2) Right ascension; (3) Declination; (4) Heliocentric Julian Date of observation; (5) Signal-to-noise ratio in the blue arm of the LAMOST MRS spectra; (6) Signal-to-noise ratio in the red arm; (7) Orbital period in days; (8) Radial velocity semi-amplitude of the visible star; (9) Distance in pc; (10) Radius of the visible star estimated from \textit{Gaia} DR3; (11) Mass function $f(M_1)$.}
\end{deluxetable*}

\section{Analysis and Results}\label{result}
\subsection{\texorpdfstring{$v \sin i$}{vsini} Measurement}\label{measure}
We measure $v \sin i$ by fitting synthetic spectra to the observed LAMOST MRS spectra. Synthetic spectra are interpolated from the high-resolution MARCS and BT-Cond model grids \citep{2001ApJ...556..357A,2008A&A...486..951G} using the \texttt{stellarSpecModel}\footnote{\url{https://github.com/zhang-zhixiang/stellarSpecModel}} package \citep{stellarSpecModel2025}. The models we used have a sampling resolution of $4~\mathrm{km\,s^{-1}}$. A macroturbulent velocity of $2~\mathrm{km\,s^{-1}}$ is also adopted. This value is typical for low- to intermediate-mass stars \citep{2019KPCB...35..129S}, which dominate our sample, in contrast to massive O/B stars, where macroturbulence can reach tens of $\mathrm{km\,s^{-1}}$ \citep{2017A&A...597A..22S}. For the former, spectral line broadening is dominated by rotation and instrumental broadening, while the contribution from macroturbulence being minor \citep{2005oasp.book.....G}. The grids are parameterized by effective temperature ($T_{\rm eff}$), surface gravity ($\log g$), and metallicity ([Fe/H]). The MARCS grid covers $2700~\mathrm{K} \leq T_{\rm eff} \leq 8000~\mathrm{K}$, 
$2.0 \leq \log g \leq 5.0$, and $-4.0 \leq \mathrm{[Fe/H]} \leq 1.0$, 
while the BT-Cond grid spans $3500~\mathrm{K} \leq T_{\rm eff} \leq 25000~\mathrm{K}$, 
$1.5 \leq \log g \leq 4.5$, and $-4.0 \leq \mathrm{[Fe/H]} \leq 0.5$. 
For stars with $T_{\rm eff} < 8000~\mathrm{K}$, MARCS models are adopted, 
whereas for hotter stars the BT-Cond models are used.

Spectral line broadening in stellar spectra primarily arises from macroturbulence, stellar rotation, and instrumental effects. In generating the synthetic templates, we adopt a fixed macroturbulent velocity of $2~\mathrm{km\,s^{-1}}$, which has a negligible impact on the fitting results. Rotational broadening is applied to the synthetic spectra by convolving with a standard rotational profile defined as:
\begin{equation}\label{eq2}
    G(x)=\left\{\begin{array}{ll}
\frac{2(1-\epsilon)\sqrt{1-x^{2}} + \frac{\pi \epsilon}{2}(1-x^{2})}{\pi \left(1-\frac{\epsilon}{3}\right)} & \text{for } x \leq 1 \\
0 & \text{for } x > 1
\end{array}\right.,
\end{equation}
where $x = V / v \sin i$, and $\epsilon$ is the limb-darkening coefficient \citep{2011AA...529A..75C, 2017AA...600A..30C}.

%%%%%%fig2
\begin{figure*}
    \centering
    \includegraphics[width=0.85\linewidth]{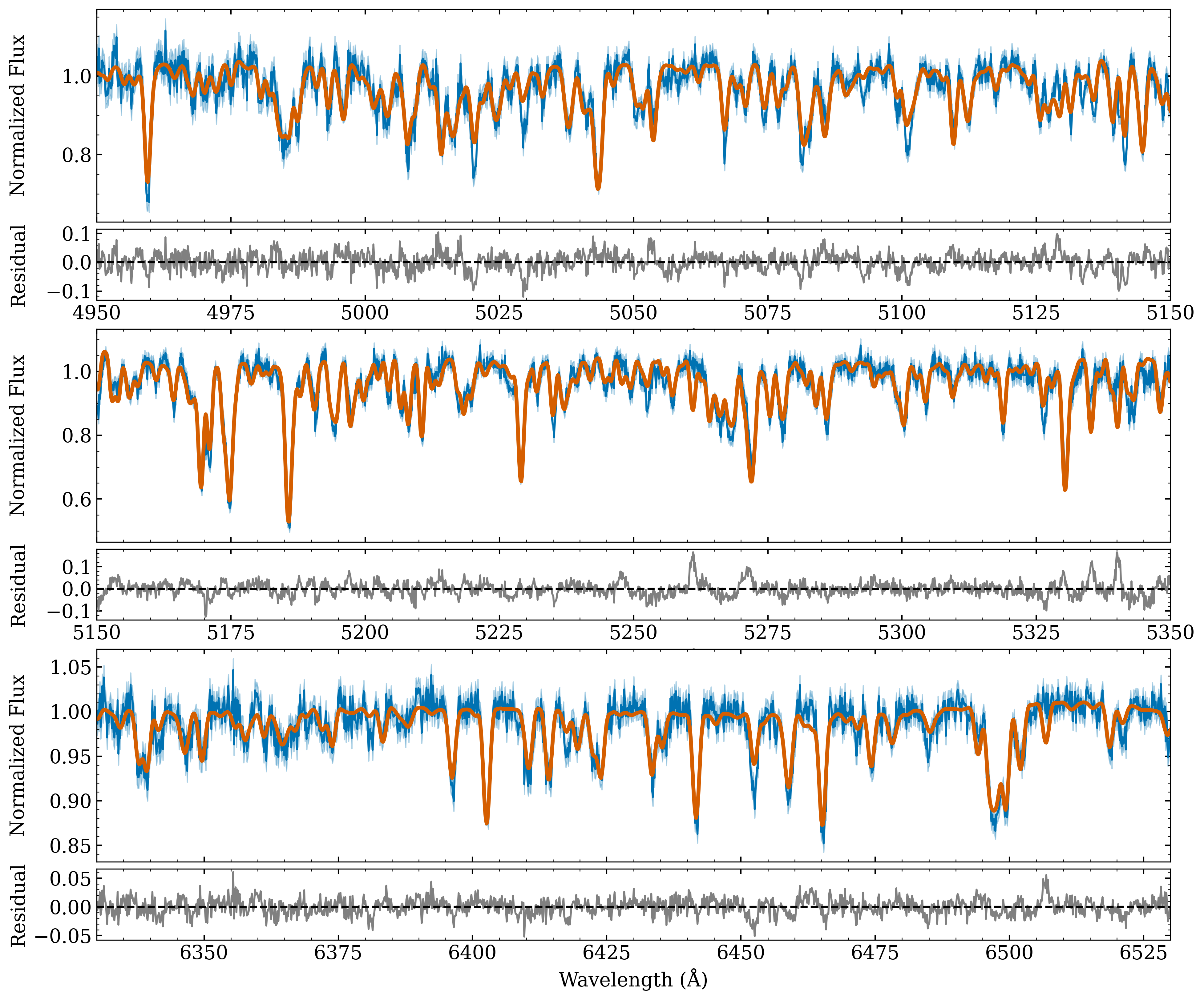}
    \caption{Spectral fits for J0341 in our sample. The observed LAMOST MRS spectra (blue) are shown with best-fit model spectra (orange) over the wavelength range $4950$–$6530~\text{\AA}$.}
    \label{fig:spec}
\end{figure*}

Given the resolution of the LAMOST MRS spectra ($R \sim 7500$), the instrumental broadening ( $\sim 40~\kms$) is non-negligible. We therefore model this effect by extracting the instrumental line spread function (LSF) from isolated emission lines in the LAMOST arc lamp spectra and adopting these profiles as convolution kernels. The final model spectrum, $F_{\rm model}(\lambda)$, is constructed by sequentially convolving the template spectrum, $F_0(\lambda)$, with the rotational and instrumental broadening kernels:
\begin{equation}\label{eq3}
    F_{\rm model}(\lambda) = \left[ F_0(\lambda) \otimes G(v \sin i) \right] \otimes P_{\rm instr}(\lambda),
\end{equation}
where $G(v \sin i)$ is the rotational broadening kernel defined in Equation~\ref{eq2}, $P_{\rm instr}(\lambda)$ is the instrumental LSF profile.

\begin{deluxetable*}{llrccccccc}
\tablecaption{Stellar Parameters Derived from Spectral and SED Fitting.\label{tab:parameter}}
\tablewidth{1.0\textwidth}
%\tabletypesize{\scriptsize}
\tablehead{
\colhead{Objname}  & \colhead{$T_{\text{spec}}$} & \colhead{$v \sin i$} & \colhead{Limb($\epsilon$)}  & \colhead{$i$}  &  \colhead{$T_{\text{SED}}$} & \colhead{$R_{2}$} & \colhead{$M_{2}$ }&\colhead{$M_{1}$}&\colhead{$f$}\\
\colhead{}  & \colhead{(K)}& \colhead{(km\,s$^{-1}$)} &\colhead{} & \colhead{(degree)} & \colhead{(K)} & \colhead{($R_\odot$)}& \colhead{($M_\odot$)} & \colhead{($M_\odot$)}&\colhead{} 
}

\startdata
J0106 & $5901^{+39}_{-41}$ & $107.9^{+2.2}_{-1.8}$ & $0.14^{+0.18}_{-0.1}$ & $67.9^{+4.9}_{-6.2}$ & $5396^{+51}_{-50}$ & $2.00^{+0.08}_{-0.06}$ & $1.01^{+0.12}_{-0.14}$ & $0.57^{+0.05}_{-0.05}$ & $1.05^{+0.06}_{-0.08}$ \\
J0117 & $6180^{+19}_{-22}$ & $56.6^{+1.4}_{-1.5}$ & $0.24^{+0.17}_{-0.15}$ & $50.6^{+2.4}_{-2.7}$ & $6179^{+11}_{-11}$ & $3.01^{+0.08}_{-0.07}$ & $1.29^{+0.04}_{-0.04}$ & $1.08^{+0.06}_{-0.06}$ & $0.84^{+0.02}_{-0.02}$ \\
J0341 & $5993^{+6}_{-12}$ & $36.2^{+1.4}_{-1.3}$ & $0.26^{+0.15}_{-0.17}$ & $31.9^{+1.5}_{-1.6}$ & $5991^{+12}_{-12}$ & $1.49^{+0.03}_{-0.03}$ & $1.20^{+0.01}_{-0.02}$ & $1.39^{+0.09}_{-0.10}$ & $0.66^{+0.01}_{-0.01}$ \\
J0359 & $8528^{+12}_{-14}$ & $53.1^{+1.7}_{-1.4}$ & $0.23^{+0.18}_{-0.16}$ & $42.1^{+2.2}_{-2.3}$ & $8529^{+13}_{-13}$ & $2.16^{+0.06}_{-0.08}$ & $1.98^{+0.03}_{-0.07}$ & $1.34^{+0.08}_{-0.09}$ & $0.67^{+0.03}_{-0.02}$ \\
J0635 & $8052^{+14}_{-15}$ & $96.5^{+0.3}_{-0.6}$ & $0.02^{+0.03}_{-0.02}$ & $73.4^{+4.3}_{-5.5}$ & $8074^{+11}_{-11}$ & $3.89^{+0.10}_{-0.11}$ & $2.44^{+0.04}_{-0.05}$ & $1.04^{+0.03}_{-0.04}$ & $0.86^{+0.02}_{-0.02}$ \\
J0853 & $5572^{+49}_{-84}$ & $56.4^{+1.6}_{-1.5}$ & $0.23^{+0.18}_{-0.16}$ & $62.8^{+4.0}_{-4.6}$ & $5214^{+82}_{-57}$ & $0.99^{+0.02}_{-0.03}$ & $0.86^{+0.02}_{-0.01}$ & $0.89^{+0.05}_{-0.05}$ & $0.61^{+0.02}_{-0.01}$ \\
J1429 & $4114^{+34}_{-34}$ & $62.1^{+2.1}_{-2.0}$ & $0.29^{+0.14}_{-0.19}$ & $49.6^{+2.7}_{-2.9}$ & $4070^{+16}_{-18}$ & $0.79^{+0.02}_{-0.02}$ & $0.53^{+0.03}_{-0.03}$ & $0.50^{+0.04}_{-0.03}$ & $0.78^{+0.03}_{-0.03}$ \\
J1729  & $4957^{+10}_{-10}$ & $70.3^{+1.4}_{-1.1}$ & $0.21^{+0.16}_{-0.14}$ & $--$ & $4896^{+25}_{-25}$ & $0.77^{+0.01}_{-0.01}$ & $0.81^{+0.07}_{-0.07}$ & $--$ & $--$\\
J2023 & $7551^{+37}_{-24}$ & $93.4^{+1.6}_{-1.8}$ & $0.40^{+0.07}_{-0.14}$ & $72.7^{+3.8}_{-4.9}$ & $7525^{+16}_{-16}$ & $2.28^{+0.03}_{-0.04}$ & $1.76^{+0.02}_{-0.03}$ & $0.95^{+0.03}_{-0.03}$ & $0.80^{+0.01}_{-0.01}$ \\
J2354 & $4206^{+13}_{-13}$ & $69.7^{+1.5}_{-1.1}$ & $0.17^{+0.18}_{-0.12}$ & $78.6^{+5.3}_{-5.1}$ & $4070^{+40}_{-40}$ & $0.66^{+0.02}_{-0.01}$ & $0.73^{+0.06}_{-0.05}$ & $1.34^{+0.05}_{-0.07}$ & $0.61^{+0.02}_{-0.02}$ \\
\enddata
\tablecomments{The spectroscopic parameters ($T_{\rm spec}$, $v \sin i$, Limb($\epsilon$)) are derived from MCMC fitting of LAMOST MRS spectra. The orbital inclinations $i$ are derived from the measured $v \sin i$ and the visible star radii $R_2$. The effective temperature ($T_{\text{SED}}$) and stellar radius ($R_2$) of the visible star are constrained via SED fitting. The mass of the visible star ($M_2$) is inferred from stellar evolutionary models \citep{2015ascl.soft03010M, 2016ApJS..222....8D} given the derived parameters, and the compact object mass ($M_1$) is calculated using the binary mass function. The $f$ denotes the Roche-lobe filling factor of the visible star.}
\end{deluxetable*}

For each target, we select three wavelength intervals (4950–5150~\text{\AA}, 5150–5350~\text{\AA}, and 6330–6530~\text{\AA}) to perform spectral fitting. These regions contain prominent absorption lines suitable for constraining $v \sin i$. For each segment, an LSF profile extracted from the arc lamp spectrum in the corresponding wavelength range is used to convolve with the synthetic templates, ensuring realistic modeling of the instrumental broadening. As an illustrative target, for J0117, the emission lines of 4965.08~\text{\AA} (Ar\,\textsc{II}), 5231.16~\text{\AA} (Th\,\textsc{I}), and 6457.28~\text{\AA} (Th\,\textsc{I}) in the arc lamp spectrum are used as the corresponding instrumental broadening profiles.

Previous studies demonstrated that the limb-darkening coefficients of individual spectral lines are significantly smaller than those derived from broadband photometry \citep[e.g.,][]{Collins1995,Shahbaz2003}. Adopting broadband coefficients may introduce systematic biases when modeling absorption-line profiles. For this reason, both rotational broadening and $\epsilon$ are treated as free parameters in our fitting procedure. Consequently, we simultaneously fit for $T_{\mathrm{eff}}$, $\log g$, [Fe/H], $v \sin i$, and $\epsilon$.

We perform spectral fitting using a Markov Chain Monte Carlo (MCMC) approach to sample the posterior probability distributions of the model parameters. The fitting is carried out using the \texttt{emcee} package \citep{2013PASP..125..306F}, which implements the affine-invariant ensemble sampler. The posterior distribution is evaluated by combining priors with a likelihood function, where the likelihood is given by
\begin{equation}\label{eq4}
    \ln \mathcal{L}(\theta) = -\frac{1}{2} \sum_{i} \left[
    \frac{\left(f_{\mathrm{obs},i} - f_{\mathrm{model},i}(\theta)\right)^2}
    {\sigma_i^2 + \sigma_{\mathrm{sys}}^2}
    + \ln\left(2\pi\,(\sigma_i^2 + \sigma_{\mathrm{sys}}^2)\right)
    \right],
\end{equation}
where $f_{\mathrm{obs},i}$ and $f_{\mathrm{model},i}(\theta)$ are the observed and model fluxes, respectively, and $\theta$ represents model parameters. For the limb-darkening coefficient $\epsilon$, a uniform prior is set between 0 and the tabulated values provided by \citet{2017AA...600A..30C}. The prior on $\log g$ is informed by stellar evolutionary models \citep[isochrones;][]{2015ascl.soft03010M}. The sampler is then run for 10000 steps to generate the posterior distributions. Before fitting, the template spectra are shifted in radial velocity to achieve consistency with the observed spectra. The radial velocities are determined via CCF method using templates generated from the stellar parameters provided by the LAMOST catalog.

The best-fit stellar parameters are determined from the posterior distributions obtained through MCMC sampling. The fitting results for all targets are summarized in Table~\ref{tab:parameter}. Uncertainties correspond to the 16th and 84th percentiles of the posterior distributions. Figure~\ref{fig:spec} presents the best-fit synthetic spectra (orange) compared with the observed spectrum (blue) for one representative target (J0341) in three wavelength segments. The residuals are shown below each panel. The model reproduces the observed spectra well across all segments, and the residuals show no significant systematic patterns. All spectral fitting results are presented in Appendix~\ref{specfit}, and the corresponding posterior distributions are shown in Appendix~\ref{corner_spec}.

\subsection{Spectral Energy Distributions}\label{sec:sed}

\begin{figure*}
    \centering
    \includegraphics[width=0.85\linewidth]{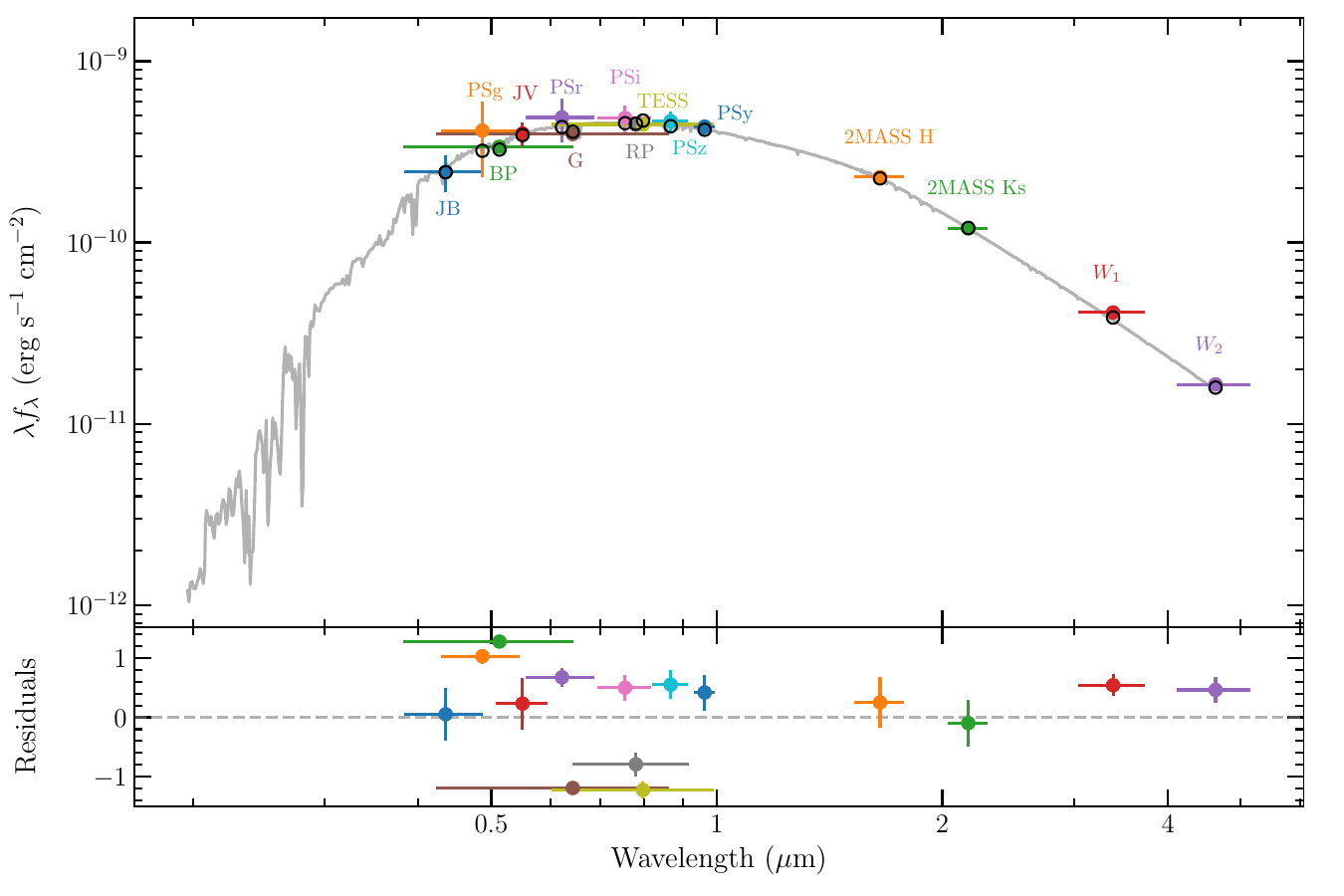}
    \caption{SED fitting result for J0341. Top panel: Colored circles with error bars indicate observed photometric measurements, while the gray curve represents the best-fit model spectrum derived from the SED fitting. Filter labels are placed adjacent to their corresponding data points. Bottom panel: The residuals of the fit, normalized by the photometric uncertainties.}
    \label{fig:SED}
\end{figure*}

\begin{figure*}
    \centering
    \includegraphics[width=0.80\linewidth]{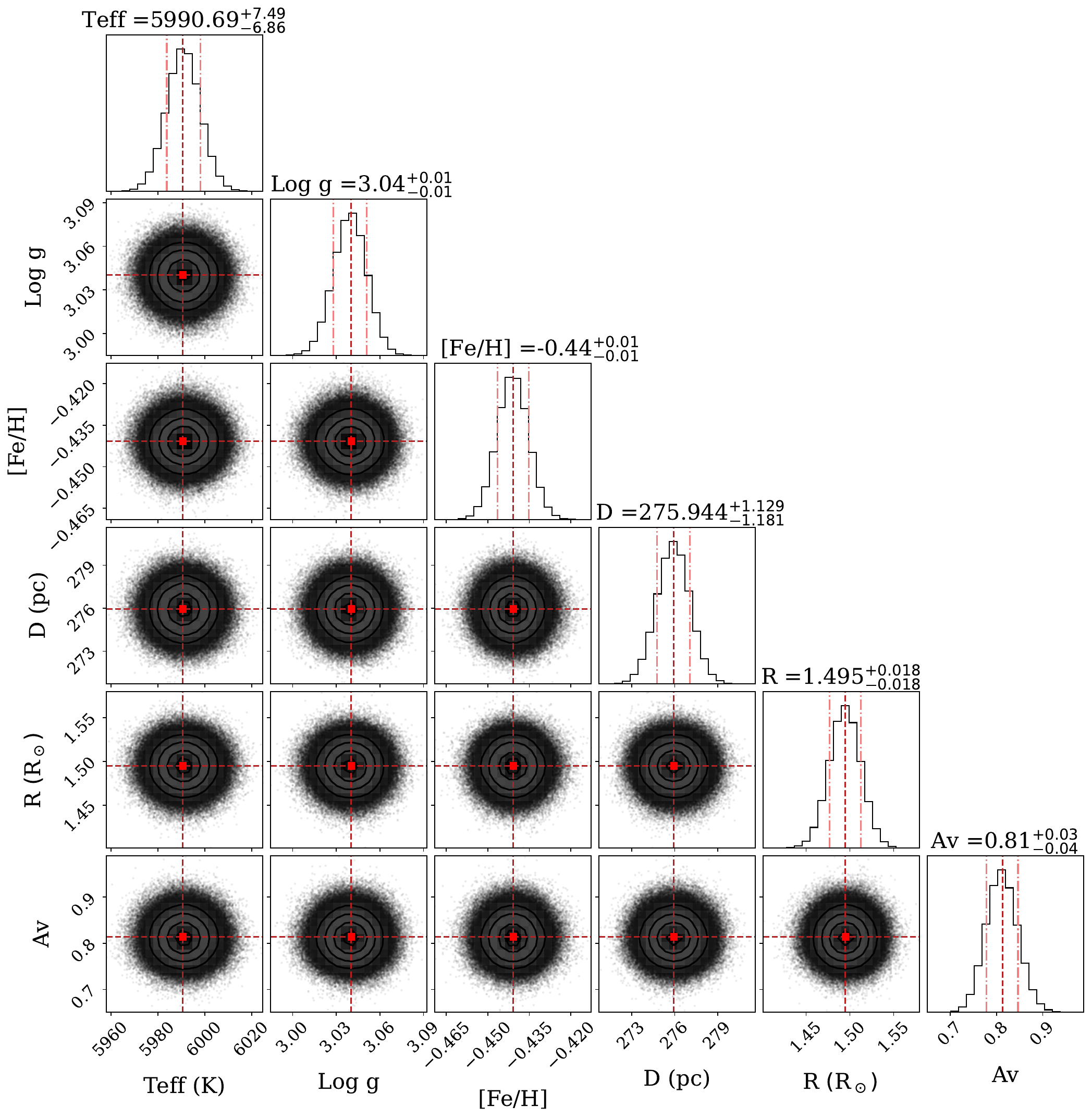}
    \caption{Posterior distributions of stellar parameters for target J0341 from SED fitting using \texttt{AstroARIADNE}.}
    \label{fig:corner}
\end{figure*}

To constrain the radii of the visible stars, we perform spectral energy distribution (SED) fitting using multi-band photometric data. The fitting is carried out with the \texttt{astroARIADNE}\footnote{https://github.com/jvines/astroARIADNE} package \citep{2022MNRAS.513.2719V}, which utilizes a Bayesian framework to compare observed photometry with synthetic fluxes generated from various stellar atmosphere models, including Phoenix v2 \citep{2013A&A...553A...6H}, BT-Models \citep{2012RSPTA.370.2765A}, Castelli \& Kurucz \citep{2003IAUS..210P.A20C}, and Kurucz 1993 \citep{1993KurCD..13.....K}. The best-fitting parameters are determined through nested sampling. In our analysis, the effective temperature derived from spectroscopic fitting and the distance inferred from \textit{Gaia} DR3 parallaxes \citep{2023AA...674A...1G} are used as priors. Figures~\ref{fig:SED} and \ref{fig:corner} show the SED fit and the corresponding posterior distributions for one representative target (J0341). The best-fit SEDs for all sources are presented in Appendix~\ref{SEDfit}. The best-fit effective temperatures ($T_{\text{SED}}$) and radii ($R_{2}$) for all visible stars are summarized in Table~\ref{tab:parameter}. To estimate the masses of the visible components ($M_{2}$), 
we use the \texttt{isochrones} Python package \citep{2015ascl.soft03010M} to interpolate grids of stellar evolutionary models, adopting the MIST models \citep{2016ApJS..222....8D} to derive their masses. The best-fit parameters from the SED served as inputs for these determinations. The masses of the visible components for all targets are listed in Table~\ref{tab:parameter}.

%%%%%%fig3

To further characterize the geometric configuration of the visible stars, we estimated their Roche-lobe radii ($R_{\mathrm{L2}}$) using the analytic approximation given by \citet{1983ApJ...268..368E},
\begin{equation}
\frac{R_{\mathrm{L2}}}{a}=\frac{0.49 q^{2/3}}{0.6 q^{2/3} + \ln(1+q^{1/3})},
\end{equation}
where $q = M_2/M_1$ is the mass ratio of the visible star to its companion and $a$ is the orbital separation derived from Kepler's third law. The Roche-lobe filling factor was then computed as $f = R_2/R_{\mathrm{L2}}$. Uncertainties in the filling factors were estimated by Monte Carlo sampling. The resulting filling factors are summarized in Table~\ref{tab:parameter}.

These filling factors provide a geometric measure of how close the visible stars are to their Roche lobes and serve as a basis for comparison with the observed photometric variability. Comparison with the photometric light curves (Figure~\ref{fig:lc_all}) shows that the systems exhibiting ellipsoidal modulation generally have filling factors $f \gtrsim 0.8$ and peak-to-peak amplitudes of $\sim10\%$. In contrast, the systems lacking clear ellipsoidal signatures typically have more moderate filling factors ($f \approx 0.6$) and display weaker variability of $\sim5\%$.

One notable case is J0106, for which the inferred filling factor slightly exceeds unity ($f = 1.05^{+0.06}_{-0.08}$). This may be attributed to the measurement uncertainties arising from the visible star being close to filling its Roche lobe. Correspondingly, this source shows the largest photometric amplitude in the sample, consistent with its near Roche-lobe–filling configuration.

%%%%%fig4

\subsection{Mass Constraints of the Unseen Companions}
\label{mass}
We constrain the masses of the unseen components ($M_1$) using the mass function (Equation~\ref{eq1}). For our targets, the systems are assumed to be tidally synchronized and spin–orbit aligned. In this case, the equatorial rotational velocity is given by $V_{\mathrm{rot}} = 2\pi R_{2}/P_{\mathrm{orb}}$, which allows the orbital inclination to be derived from the measured $v \sin i$ and the visible star's radius $R_2$. The resulting inclination values for all systems are listed in Table~\ref{tab:parameter}. The mass of the visible star $M_2$ is independently determined from stellar evolution models (see Section~\ref{sec:sed}). With $f(M_1)$, $\sin i$, and $M_2$ known, the mass function is numerically solved to obtain $M_1$ for each system. The unseen companion masses for all sources are listed in Table~\ref{tab:parameter}. Figure~\ref{fig:M2_vs_M1} shows the derived $M_1$ and $M_2$ values for our sample, along with reference lines for $M_1 = M_2$, $M_1 = 2/3M_2$, and $M_1 = 1/2M_2$. 

%%%%%fig4
\begin{figure*}
    \centering
    \includegraphics[width=0.9\linewidth]{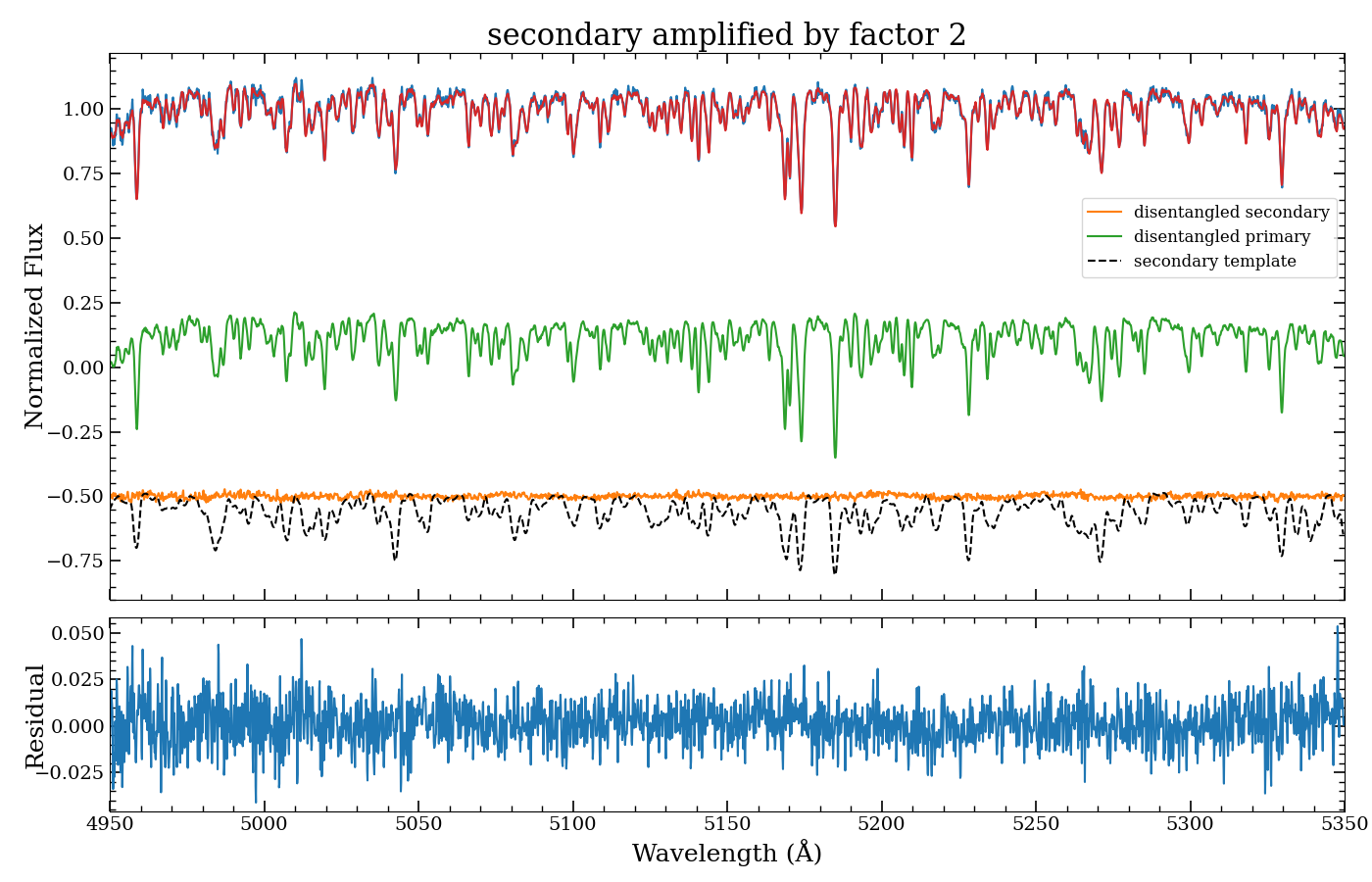}
    \caption{Example of spectral disentangling for the target J0341. Top panel: the original observed spectrum (blue), the reconstructed spectrum from the disentangling procedure (red), and the individual spectra of the primary (green) and secondary (orange) components. Secondary spectrum rescaled by a factor of 2. Stellar template (black dashed line) corresponds to 1.39~$M_\odot$ ($T_{\rm eff}$ = 7017~K, $\log g$ = 4.0, $v \sin i = 80~\mathrm{km\,s^{-1}}$). Bottom panel: the residuals between the reconstructed and observed spectra, showing that the disentangling reproduces the observed spectrum within the noise level.}
    \label{distangle}
\end{figure*}

Among all targets, six objects have mass ratios $M_{1}/M_{2} > 2/3$. For the remain three objects, the mass ratios lie between 0.43 and 0.56. Based on our experience, when the flux contribution of the secondary exceeds $\sim$10\%, significant double-peaked structures should appear in the CCF profile if the radial-velocity separation between the two components is large. However, for all the targets analyzed in this work, we only detect single-peaked CCFs, and the synthetic templates provide excellent matches to the observed spectra (see Figure \ref{fig:spec} and Appendix \ref{specfit}). To further test whether a luminous secondary component could be hidden in the spectra, we performed spectral disentangling for all systems (Section~\ref{specdisentangle}). For all targets, the extracted secondary spectra are essentially flat and show no identifiable stellar absorption features.

In our sample, two targets, J2354 and J1729, have already been investigated individually in previous studies \citep{2022ApJ...936...33Z,2023SCPMA..6629512Z}, and we will discuss them in detail in Section~\ref{results}. Among the remaining objects, the unseen companions of J0341 and J0359 have inferred masses of $1.39^{+0.09}_{-0.10}$~$M_\odot$ and $1.34^{+0.08}_{-0.09}$~$M_\odot$, respectively, strongly suggesting that they are candidates for neutron stars or massive white dwarfs. We will discuss them more detail in Section \ref{twotargets}.

\subsection{Spectral Disentangling}\label{specdisentangle}
Based on the masses of the unseen companions derived in Section~\ref{mass}, the targets are most likely compact objects. To further test this interpretation and to assess whether a luminous secondary component could be present in the spectra, we performed spectral disentangling for each object.

We applied a shift-and-add procedure \citep{1998PASP..110.1416M,2020AN....341..628Q,2020A&A...639L...6S,2022NatAs...6.1085S,2022A&A...665A.148S} to iteratively reconstruct the primary (the visible star) and secondary (the companion) spectra.
To ensure consistency for the subsequent analysis, the spectra were carefully normalized.
Assuming the radial velocity (RV) semi-amplitudes of the primary and secondary components are $K_\text{pri}$ and $K_\text{sec}$, respectively, the RV of each component at a given orbital phase $\phi$ is denoted as $RV_\text{pri}(\phi)$ and $RV_\text{sec}(\phi)$. The disentangling process was performed as follows:
\begin{enumerate}
    \item All the normalized observed spectra were first shifted to the rest frame of Component 1 based on $RV_\text{pri}(\phi)$ and co-added to create an initial mean template for the primary star ($S_\text{pri}$).
    \item This template $S_\text{pri}$ was then shifted back to the observer's frame and subtracted from each original observed spectrum to isolate the contribution of the secondary component. 
    \item The resulting residuals were subsequently shifted to the rest frame of Component 2 ($RV_\text{sec} = 0$) and merged to generate the secondary template ($S_\text{sec}$).
    \item Using the newly derived $S_\text{sec}$, the process was reversed to update and refine $S_\text{pri}$.
\end{enumerate}
This procedure was carried out iteratively until the profiles of both $S_\text{pri}$ and $S_\text{sec}$ reached a stable state and the residuals between the reconstructed composite spectra and the observations no longer showed significant improvement. To facilitate the assessment of potential companion features, the disentangled secondary spectra were subsequently rescaled according to the expected flux ratios derived from the estimated secondary masses, assuming the mass–luminosity relation for main sequence stars. Representative stellar templates corresponding to the inferred secondary masses were also plotted for comparison, illustrating the expected spectral appearance of a putative main-sequence companion (see Figure~\ref{distangle} and Appendix~\ref{disen}). The stellar template spectra were generated using the \texttt{StellarSpecModel} package, and each spectrum was convolved assuming $v \sin i = 80~\mathrm{km\,s^{-1}}$.

Since the secondary semi-amplitude $K_\text{sec}$ is unconstrained, we performed a grid search over plausible values of $K_\text{pri}$ and $K_\text{sec}$. For each pair, the recombined model spectra were compared to the observations using a chi-squared statistic:
\begin{equation}
    \chi^2(K_\text{pri}, K_\text{sec}) = \sum_{i=1}^{N_{\rm epochs}} \sum_{k=1}^{N_\lambda}
    \frac{[O_i(\lambda_k) - \mathrm{Model}_{i}(\lambda_k)]^2}{\sigma_{i,k}^2},
\end{equation}
where $N_{\rm epochs}$ is the number of observed spectra, $N_\lambda$ is the number of wavelength bins, $\mathrm{Model}_{i}$ is the expected model spectrum of observation phase $\phi$, and $\sigma_{i,k}$ is the flux uncertainty. The semi-amplitudes $K_\text{pri}$ and $K_\text{sec}$ that minimize $\chi^2$ were adopted as the best-fit values, and the corresponding reconstructed spectra were taken as the final estimates of the primary and secondary components.

We performed spectral disentangling for all unclassified sources in our sample, and the full set of results is presented in Appendix~\ref{disen}. To enhance the sensitivity to secondary spectral features, we selected LAMOST spectra with signal-to-noise ratios greater than 30 for each target. For each target, at least 20 individual-exposure spectra were selected, and they basically cover the entire phase on average. We performed spectral disentangling using the co-added spectra to improve the signal-to-noise ratio.
The results show that, after rescaling the secondary spectra and plotting representative main sequence templates, the extracted secondary components for most targets remain virtually featureless and show no identifiable stellar absorption features (e.g., J0341, Figure~\ref{distangle}). A clear exception is J0117, whose rescaled secondary spectrum closely matches the corresponding stellar template, with consistent absorption features across multiple wavelength regions. This agreement strongly supports the presence of a luminous stellar companion in this system (Figure~\ref{fig_spec_1}).

The situation is different for J0106. Although the rescaled secondary spectrum appears to display absorption features broadly resembling those of the template (see Appendix~\ref{disen}), this behavior is not expected for a genuine stellar companion, but is a known artifact when the assumption of phase-invariant component spectra is violated. J0106 is the only system in our sample with a Roche-lobe filling factor close to unity ($f = 1.05^{+0.06}_{-0.08}$), implying strong tidal distortion of the visible star. These effects may introduce phase-dependent variations in the line profiles, which are then mapped by the disentangling algorithm into a spurious secondary component. Moreover, given the large amplification factor (50) applied during rescaling, even very small residuals from the primary may resemble stellar-like absorption features. We therefore interpret the weak features in the extracted secondary spectrum of J0106 as arising from phase-dependent distortions of the primary spectrum rather than from a physically distinct stellar companion.

For J0359, the rescaled secondary spectrum exhibits a weak absorption feature around 4980–4990~\text{\AA} that roughly coincides with a predicted template line. However, no comparable correspondence is found in other wavelength regions, where prominent template features are not recovered. Given the current SNR, the evidence is insufficient to robustly confirm the presence of a visible stellar companion. Higher SNR observations will be required to further clarify the nature of the secondary. In the following discussion, we therefore treat J0359 as a compact object candidate for the purpose of evaluating its dynamical properties, while noting that a double-lined visible binary configuration cannot be ruled out based on the present data quality.

%%%%%%fig5
\begin{figure}
    \centering
    \includegraphics[width=\linewidth]{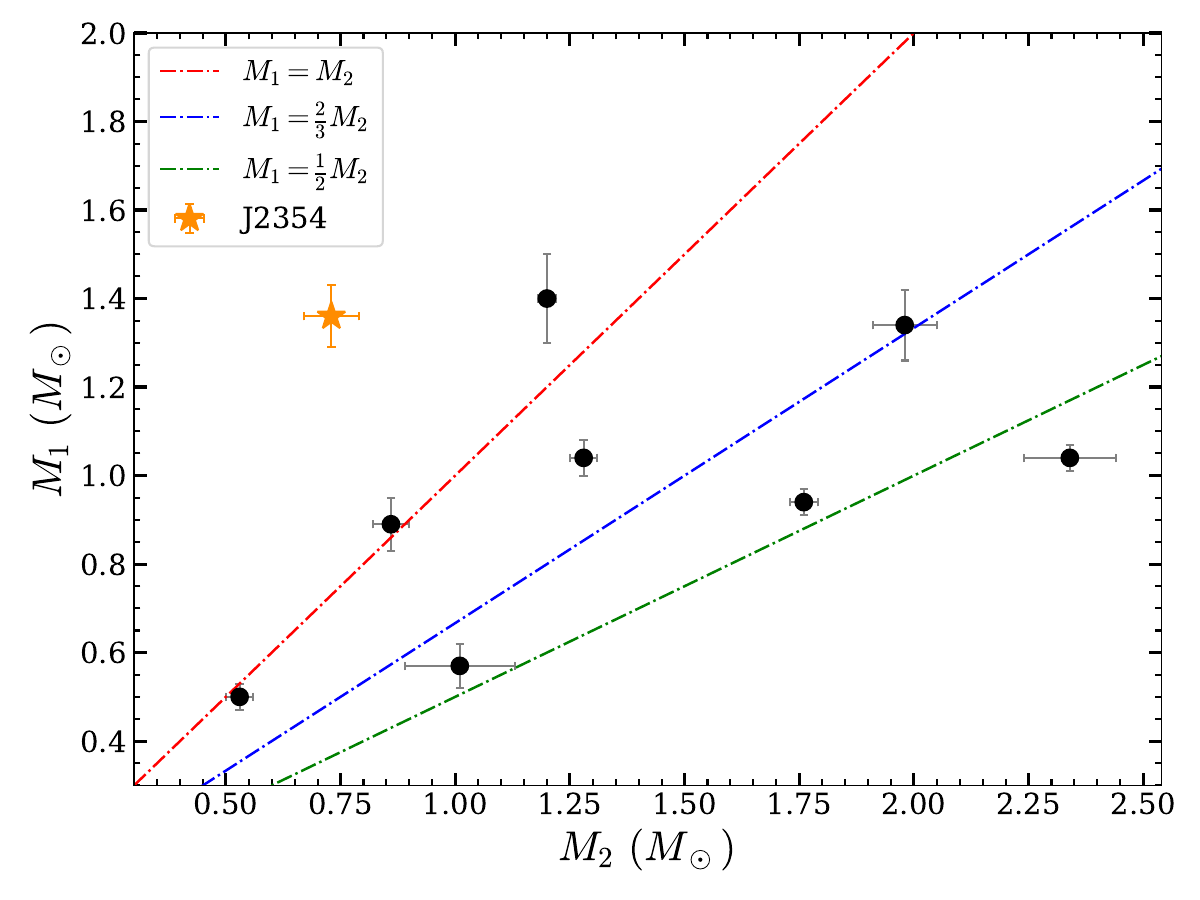}
    \caption{The mass of the visible star ($M_2$) versus that of the unseen companions ($M_1$). The red, blue, and green dashed–dotted lines indicate $M_1 = M_2$, $M_1 = 2/3M_2$, and $M_1 = 1/2M_2$, respectively. The confirmed compact object system J2354 is marked with a orange star, while other targets are shown as black circles with errors.}
    \label{fig:M2_vs_M1}
\end{figure}

\section{Discussion}\label{dis}
\subsection{Assumptions on Tidal Synchronization and Spin–Orbit Alignment}\label{assum}

Our mass constraints are based on the assumptions of tidal synchronization and spin–orbit alignment of the visible star. We briefly discuss their physical validity and applicability to our systems.

The assumptions of synchronization and alignment are generally expected to hold in close binaries, where tidal forces efficiently transfer angular momentum between the stellar rotation and orbital motion \citep[e.g.,][]{1977A&A....57..383Z, 1981A&A....99..126H, 2019A&A...628A..29C}. In such systems, the stellar spin is usually aligned with the orbital axis and the rotation period equals the orbital period. However, these conditions may break down in systems with wider separations, eccentric orbits, or young ages, where tidal interactions are weaker and the timescale for synchronization or alignment exceeds the system’s age. In those cases, inclination estimates derived from $v \sin i$ may be biased.

Given the large Roche-lobe filling factors of our systems ($f \gtrsim 0.6$), the low- to intermediate-mass nature of the visible stars, and the short orbital periods ($\lesssim 2$~days), tidal interactions are expected to be strong. Under these conditions, the alignment timescale is expected to be comparable to the synchronization timescale \citep{1981A&A....99..126H}, and both are expected to be much shorter than the evolutionary lifetimes of the stars. Using Equation~(27) in \citet{Hurley2002}, we estimate a tidal synchronization timescale of order $\sim 10^6$~yr for our systems, which is significantly shorter than the characteristic evolutionary timescale of $\sim 10^9$~yr. While non-synchronous rotation or spin–orbit misalignment cannot be strictly excluded, these estimates indicate that tidal synchronization and alignment are physically plausible for our targets. Accordingly, our analysis of the unseen companion masses is carried out under the assumption that the systems are tidally synchronized and spin–orbit aligned.

\subsection{Analysis and Validation of the Fitting Results}
\label{results}
Two targets, J2354 and J1729, in our sample have been previously confirmed. In which J2354 has been identified as a neutron star candidate with the compact object mass estimated to be $1.40^{+0.09}_{-0.08}~M_\odot$ in \citet{2023SCPMA..6629512Z}. The $v \sin i$ was previously measured using CFHT (Canada France Hawaii Telescope) spectra, which have a high spectral resolution of $R \sim 68000$ and cover a wavelength range of $3690$--$10480~\text{\AA}$. The observations were taken close to orbital phase 0.75 of the visible star. A value of $v \sin i = 68.37\pm0.09~\mathrm{km\,s^{-1}}$ was finally measured in \citet{2023SCPMA..6629512Z}. To test the consistency of our method, we measure $v \sin i$ from the LAMOST spectrum of J2354 at the same phase and obtain $v \sin i = 69.7^{+1.5}_{-1.1}$~$\mathrm{km\,s^{-1}}$. The close agreement between these two independent measurements demonstrates that our spectral fitting produces reliable $v \sin i$ values. The spectral fitting results for J2354 are presented in Appendix~\ref{specfit}.

J1729 has been identified as a white dwarf binary \citep{2022ApJ...936...33Z}, with the mass of its invisible companion estimated to be $\gtrsim 0.63\,M_\odot$. In our analysis, the measured $v \sin i$ and the estimated radius of the visible star yield a value of $\sin i = 1.03$, which is unphysical. Therefore, a reliable mass determination for the unseen companion cannot be obtained for this object. We identify several factors that may account for this outcome. First, the light curve of J1729 exhibits strong temporal evolution (see Figure 5 of \citet{2022ApJ...936...33Z}), indicating significant starspot activity on the visible component. In such cases, the stellar radius estimate may suffer from substantial systematic uncertainties. Second, the spectra used for the $v \sin i$ measurement were obtained near orbital phase 0.54, rather than close to quadrature (phases 0.25 or 0.75), which likely introduced velocity smearing and led to an overestimated $v \sin i$. These effects, together with observational uncertainties, naturally explain the slightly unphysical value of $\sin i$. Although we cannot provide a robust inclination estimate, the results suggest that the system is nearly edge-on to the observer. The spectral fitting results for J1729 are presented in Appendix~\ref{specfit}.

For the other eight targets, we checked the \textit{Gaia} archive and found reported $v \sin i$ values \citep{2023A&A...674A...8F} for only two: J0635 ($92.7 \pm 20.4~\mathrm{km\,s^{-1}}$) and J2023 ($80.2 \pm 29.8~\mathrm{km\,s^{-1}}$). Both are broadly consistent with our measurements (see Table~\ref{tab:parameter}), providing additional validation of our results. 

As detailed in Appendix~\ref{specfit}, the spectral fitting for each target was performed in three segments, with residuals that are small and randomly distributed. The effective temperatures derived from spectral fitting are also in good agreement with those obtained from SED fitting, except for two objects, J0106 and J0853, for which differences of about 400–500 K are found. In these two sources, the light curves (Figure~\ref{fig:lc}) exhibit variability that departs from standard ellipsoidal modulation, with amplitudes and morphologies varying across different sectors. Such variability is most likely caused by stellar surface activity, including the presence and evolution of starspots, which produces non-uniform temperature distributions across the stellar surface. This effect can bias the SED-based estimates and thereby account for the observed temperature differences.

\subsection{Inclination Constraints from Ellipsoidal Light Curve Modeling}\label{lkmodel}

\begin{figure*}
    \centering
    \includegraphics[width=0.80\linewidth]{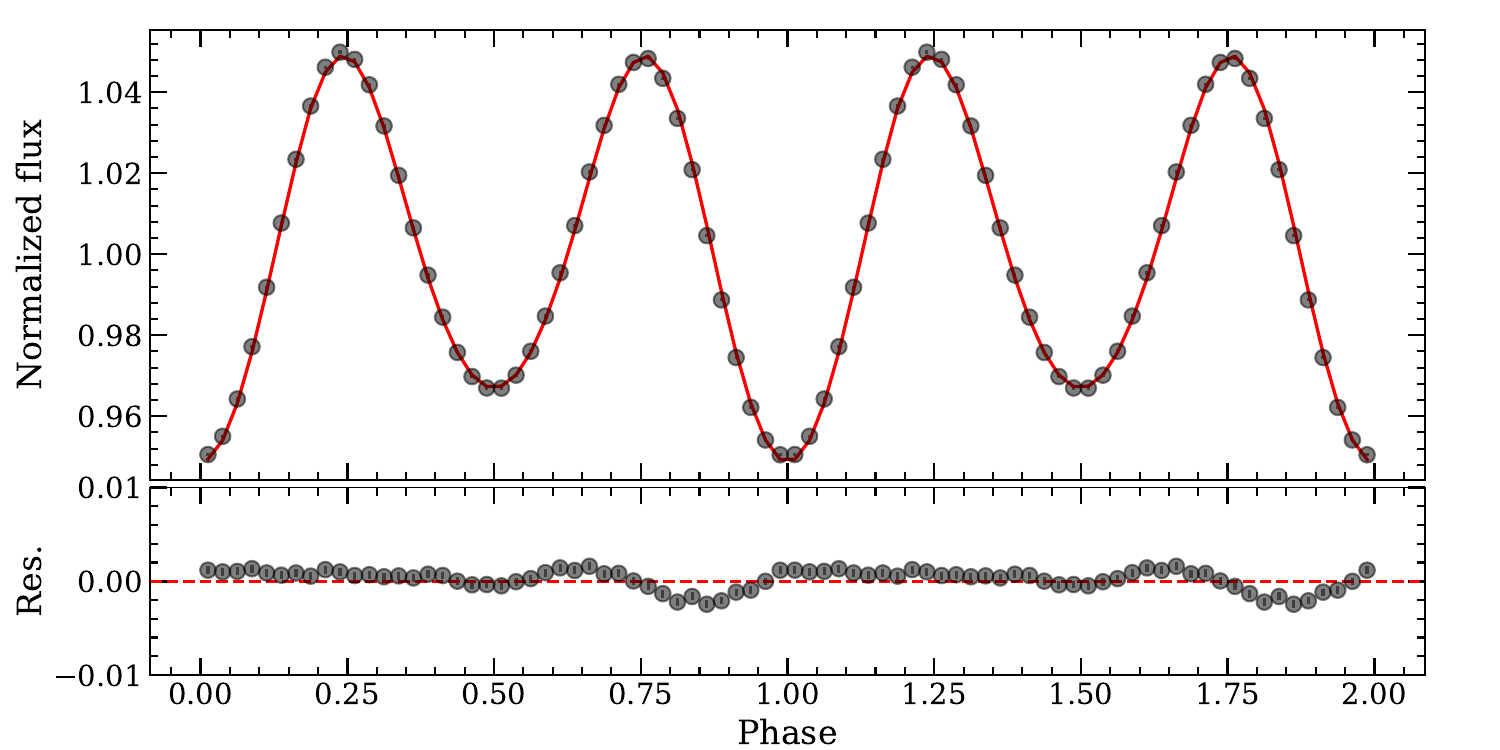}
    \caption{\texttt{PHOEBE}'s best-fitting results. In upper panel, the black dots show the rebinned TESS light curve, and the red curve is the best fitting model. The residuals between the observed data and the model is shown in bottom panel.}
    \label{fig:J0635lc}
\end{figure*}

Light curve modeling is a frequently-used independent method to constrain the binary inclination.
To compare the fitting results of light curve modeling with \vsini\ measurement,
we performed light curve modeling for systems exhibiting clean, large-amplitude ellipsoidal like variability. 
Two targets in our sample, J0635 and J2023, were selected for detailed modeling because they met the above conditions. Other sources show asymmetric or complex variability features, likely caused by starspots or pulsations (see Figure~\ref{fig:lc_all}), and are therefore excluded from photometric modeling. 

We performed the light curve modeling using the \texttt{PHOEBE} software package \citep{2005ApJ...628..426P,2016ApJS..227...29P,2020ApJS..250...34C} to fit the TESS photometric data. In our model, the unseen companion was treated as a dark compact object (\texttt{distortion\_method = none}), contributing no flux or eclipse effects. Instead of setting the companion mass $M_1$ as a free parameter, we dynamically calculated it using the binary mass function $f(M_1)$, given the visible star's mass $M_2$, and the orbital inclination $i$. The free parameters in our fitting process include the orbital inclination $i$, the mass $M_2$, radius $R_2$ of the visible star, and the gravity darkening coefficient $\beta_\mathrm{grav}$. To further constrain the solution, Gaussian priors on $M_2$ and $R_2$ were imposed based on the results of from the SED fitting (see Table~\ref{tab:parameter}). All other system parameters were fixed to values inferred from spectroscopy or previous analyses. Linear limb darkening was adopted for the visible star, with the limb-darkening coefficient interpolated from the tables of \citet{2017AA...600A..30C} according to the stellar atmospheric parameters. The light-curve variations are assumed to be dominated by ellipsoidal modulation induced by tidal distortion of the visible star.

\begin{figure*}
    \centering
    \includegraphics[width=0.80\linewidth]{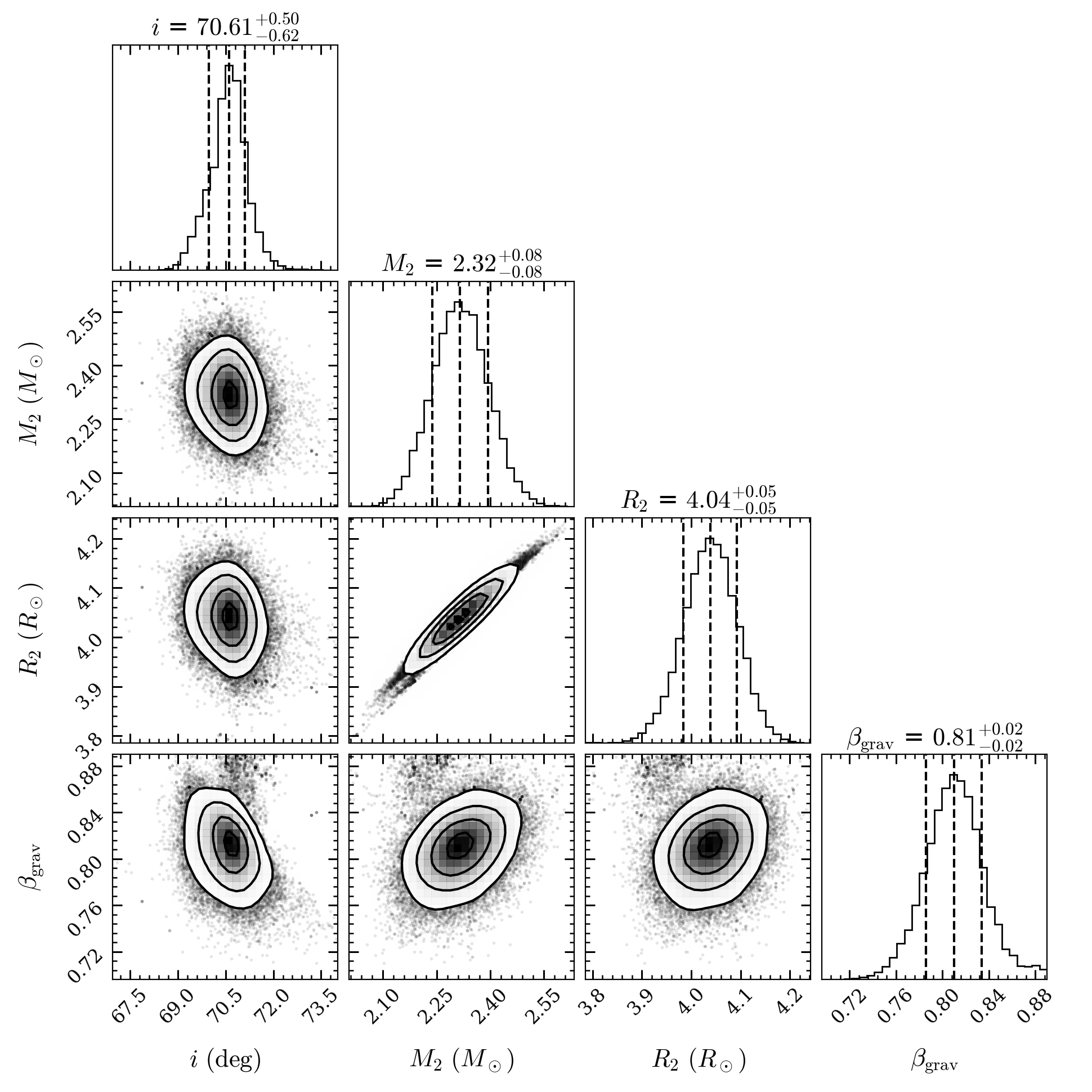}
    \caption{Parameter distributions from the \texttt{PHOEBE} light curve fitting of J0635.}
    \label{fig:J0635corner}
\end{figure*}

The posterior distributions were sampled using the \texttt{emcee} Markov Chain Monte Carlo ensemble sampler. Multiple parallel chains of 10,000 steps were run. The best-fit model and posterior distributions for J0635 are shown in Figures~\ref{fig:J0635lc} and~\protect\ref{fig:J0635corner}. The \texttt{PHOEBE} model reproduces the observed TESS light curve of J0635 well, with best-fit parameters $M_2 = 2.32^{+0.08}_{-0.08}~M_\odot$, $R_2 = 4.04^{+0.05}_{-0.05}~R_\odot$, and $i = 70.61^{+0.50}_{-0.62}$ degree (see Figure~\protect\ref{fig:J0635corner}), consistent with the estimation based on \vsini\ measurement (see Table~\ref{tab:parameter}). J2023 shows a similar level of agreement (Appendix~\ref{phoebe}). 
The consistency between the two independent methods highlights the complementarity of spectroscopic and photometric constraints on the binaries inclination.

\subsection{The Two Notable Massive Compact Object Candidates: J0341 and J0359}
\label{twotargets}

Mass estimates of $1.39^{+0.09}_{-0.10}$~$M_\odot$ and $1.34^{+0.08}_{-0.09}$~$M_\odot$ for the unseen companions of J0341 and J0359 point to the possibility of neutron stars or massive white dwarfs. These results represent a significant outcome enabled by $v \sin i$ measurements from LAMOST medium-resolution spectra. To assess the reliability of these findings, we conducted multiple independent analyses. The unseen component masses are generally derived from the mass function, which depends on $K_{2}$, $P_{\mathrm{orb}}$, inclination, and $M_{2}$ (see Equation~\ref{eq1}). For J0341 and J0359, radial velocity data reported by \citet{2024ApJ...969..114L} cover most orbital phases with well-fitted curves, indicating that $K_{2}$ and $P_{\mathrm{orb}}$ are highly reliable. We thus focus on evaluating the orbital inclinations and visible-star masses.

The $v \sin i$ measurements for J0341 and J0359 are $36.2^{+1.4}_{-1.3}$~$\mathrm{km\,s^{-1}}$ and $53.1^{+1.7}_{-1.4}$~$\mathrm{km\,s^{-1}}$, respectively. We display their spectral fitting results in Figure~\ref{fig:fit_page2}.  
It can be easily seen that their line widths are markedly narrower than those of the other objects, consistent with their relatively low $v \sin i$. Compared with minor line mismatches seen in other sources, the model and observed spectra of these two targets exhibit an almost perfect match, highlighting the reliability of the $v \sin i$ measurements. Moreover, we examined the arc lamp spectra obtained during the observations of these two objects. The corresponding instrumental broadenings are $41~\mathrm{km\,s^{-1}}$ and $40~\mathrm{km\,s^{-1}}$, respectively. These values are comparable to, but do not dominate over, the rotational broadening. In this regime, the instrumental broadening contribution can be properly accounted for by convolving the template with the instrumental line profile, thereby ensuring the robustness of the $v \sin i$ determination. Nevertheless, to completely exclude potential biases introduced by instrumental resolution, we plan to apply for high-resolution follow-up spectroscopy, which will allow us to further verify the accuracy of the $v \sin i$ measurements.

Reliable inclination estimates also require accurate radii of the visible stars, which in turn depend on well-determined effective temperatures and distances. For J0341 and J0359, the temperatures derived from spectral fitting are
consistent with those inferred from SED fitting (see Table \ref{tab:parameter}). In the case of J0341, the \Gaia DR3 effective temperature (5997 K) is essentially identical to the spectroscopic estimate (5993 K), reinforcing the robustness of the temperature
determination. For J0359, the Gaia estimate ($8997$ K) is somewhat higher than the spectroscopic value ($8528$ K). This kind of discrepancy is known to arise in \Gaia GSP-Phot results due to the temperature–extinction degeneracy, especially for stars with uncertain line-of-sight extinction and moderate to high A0 type (see \citet{Andrae2023,Avdeeva2024}). We therefore regard the spectroscopic temperature estimates as more reliable for this object.

The distances of these two sources are directly calculated by using the parallax measurements in \Gaia DR3, yielding $276 \pm 2$~pc for J0341 and $1061 \pm 17$~pc for J0359. Both lie within the regime where \Gaia provides highly reliable parallax measurements. To further assess the robustness of the astrometric solutions, we inspected the \Gaia astrometric quality indicators. The \texttt{RUWE} values are 1.3 for J0341 and 0.96 for J0359, both of which are below the commonly adopted threshold of 1.4, which supports the reliability of the parallaxes \citep{Lindegren2018,Lindegren2021}. 

We further checked the \Gaia Non-Single-Star (NSS) catalog \citep{2023AA...674A...1G}, and found that J0341 is listed in the \texttt{nsstwobody\_orbit} table, corresponding to orbital solutions compatible with a single-lined spectroscopic binary (SB1). The fitted parameters in that catalog are broadly consistent with our independent estimates. However, no constraint on the orbital inclination is provided. A joint fit of the radial-velocity data and the astrometric solution, which may become available in future \Gaia releases, would yield another independent determination of the orbital inclination. 

Estimating the masses of the visible stars is more complex. For the two targets, the visible components have effective temperatures of 5991~K and 8529~K, and radii of 1.49\,$R_\odot$ and 2.16\,$R_\odot$, respectively, suggesting that they are beginning to leave the main sequence. Based on the well-constrained temperatures, radii, and metallicities, we employed MIST single-star evolutionary models \citep[\texttt{isochrones};][]{2015ascl.soft03010M} to derive approximate masses of $1.20^{+0.01}_{-0.02}~M_\odot$ and $1.98^{+0.03}_{-0.07}~M_\odot$ for J0341 and J0359. Although these mass estimates are not rigorous, given that both binaries must have experienced mass transfer during their evolutionary histories, they still provide a reasonable first-order approximation.
In principle, a more realistic determination requires full binary evolution modeling \citep[e.g.,][]{Hurley2002,Paxton2015}, but such detailed work is beyond the scope of this study.

Currently, it remains unclear whether the unseen companions in these two systems are massive white dwarfs (WD) or neutron stars. If the compact objects are WDs, explaining the origin of their unusually high mass would be challenging, 
making them particularly intriguing cases \citep{Althaus2022}. Since the visible stars have already evolved beyond the main sequence, they will inevitably undergo further expansion, leading to mass transfer in the near future. In the case of massive WDs, as the inferred companion masses are already close to the Chandrasekhar limit, such mass transfer could eventually trigger Type Ia supernovae, making these binaries strong candidate progenitor systems \citep{2019ApJ...878L..38T,2023Natur.615..605G}. Alternatively, if the companions are neutron stars, the onset of mass transfer would turn these systems into X-ray binaries \citep{2002ApJ...565.1107P, 2024Ap&SS.369...16L}. In either scenario, these two binaries represent highly intriguing targets that warrant detailed follow-up investigations.

\section{Summary}\label{conclusion}

In this work, we have measured $v \sin i$ for ten compact object candidates from \citet{2024ApJ...969..114L}, selected based on their high mass functions, rapid rotation, and high-quality LAMOST spectra. CCF analysis confirmed that all selected systems are single-lined spectroscopic binaries. The consistency between our $v \sin i$ measurement ($69.7^{+1.5}_{-1.1}$~$\mathrm{km\,s^{-1}}$) and the high-resolution CFHT result ($68.37\pm0.09~\mathrm{km\,s^{-1}}$; see \citet{2023SCPMA..6629512Z}) for a previously confirmed neutron star candidate, J2354, together with the broad agreement between our measurements and the available \textit{Gaia} $v \sin i$ values for the other two targets in the sample, supports the reliability of our $v \sin i$ determinations.

For each system, we performed SED fitting using spectroscopic temperatures and \textit{Gaia}-based distances as priors to derive the radii of the visible stars. By combining these radii with orbital inclinations inferred from $v \sin i$, the masses of the unseen companions were estimated. In our sample, six targets exhibit mass ratios of the unseen companions to the visible stars of $M_1 / M_2 > 2/3$. However, spectral disentangling analysis shows that J0117 hosts a luminous stellar companion rather than a compact object. Excluding this system, the remaining five can therefore be regarded as compact object candidates. Among them, J2354 is a previously confirmed neutron star candidate. For the other four targets, the masses of the unseen companions range from 0.50 to 1.39~$M_\odot$. Most are likely white dwarfs, except for J0341 and J0359, whose unseen companion masses are $1.39^{+0.09}_{-0.10}$~$M_\odot$ and $1.34^{+0.08}_{-0.09}$~$M_\odot$, making them strong candidates for neutron stars or massive white dwarfs. We performed detailed analyses of the $v \sin i$ and stellar parameters for these two sources (see Section~\ref{twotargets}), confirming the reliability of the results and highlighting them as promising targets for follow-up high-resolution spectroscopic studies. 
Overall, our findings highlight the potential of medium-resolution surveys such as LAMOST to systematically uncover hidden compact companions in binary systems.

\begin{acknowledgements}

We thank Shilong Liao for helpful discussion on \Gaia issues and the referee for constructive suggestions that improved the paper. This work was supported by the National Key R\&D Program of China under grants 2021YFA1600401 and 2023YFA1607901, the National Natural Science Foundation of China under grants 12433007, 12221003, and 12263003, the Research Start-up Foundation of Quanzhou Normal University H25052, and the fellowship of China National Postdoctoral Program for Innovation Talents under grant BX20230020. We acknowledge the science research grants from the China Manned Space Project with No.CMS-CSST-2025-A13. This work utilizes data from the LAMOST survey. The Large Sky Area Multi-Object Fiber Spectroscopic Telescope (LAMOST), also known as the Guoshoujing Telescope, is a major national scientific facility constructed by the Chinese Academy of Sciences. This work makes use of data from the TESS mission, which were obtained from the Mikulski Archive for Space Telescopes (MAST) at the Space Telescope Science Institute.
\end{acknowledgements}

\software{
    astroARIADNE \citep{2022MNRAS.513.2719V},
    stellarSpecModel \citep{stellarSpecModel2025},
    spectool \citep{spectool2025},
    emcee \citep{2013PASP..125..306F},
    Astropy \citep{2013A&A...558A..33A,2018AJ....156..123A},
    SciPy \citep{2020NatMe..17..261V},
    Isochrones \citep{2015ascl.soft03010M}
}

\bibliographystyle{aasjournalv7}
\bibliography{refs}

@ARTICLE{ligo2023,
       author = {{Abbott}, R. and {Abbott}, T.~D. and {Acernese}, F. and {Ackley}, K. and {Adams}, C. and {Adhikari}, N. and {Adhikari}, R.~X. and {Adya}, V.~B. and {Affeldt}, C. and {Agarwal}, D. and {Agathos}, M. and {Agatsuma}, K. and {Aggarwal}, N. and {Aguiar}, O.~D. and {Aiello}, L. and {Ain}, A. and {Ajith}, P. and {Akutsu}, T. and {de Alarc{\'o}n}, P.~F. and {Akcay}, S. and {Albanesi}, S. and {Allocca}, A. and {Altin}, P.~A. and {Amato}, A. and {Anand}, C. and {Anand}, S. and {Ananyeva}, A. and {Anderson}, S.~B. and {Anderson}, W.~G. and {Ando}, M. and {Andrade}, T. and {Andres}, N. and {Andri{\'c}}, T. and {Angelova}, S.~V. and {Ansoldi}, S. and {Antelis}, J.~M. and {Antier}, S. and {Antonini}, F. and {Appert}, S. and {Arai}, Koji and {Arai}, Koya and {Arai}, Y. and {Araki}, S. and {Araya}, A. and {Araya}, M.~C. and {Areeda}, J.~S. and {Ar{\`e}ne}, M. and {Aritomi}, N. and {Arnaud}, N. and {Arogeti}, M. and {Aronson}, S.~M. and {Arun}, K.~G. and {Asada}, H. and {Asali}, Y. and {Ashton}, G. and {Aso}, Y. and {Assiduo}, M. and {Aston}, S.~M. and {Astone}, P. and {Aubin}, F. and {Austin}, C. and {Babak}, S. and {Badaracco}, F. and {Bader}, M.~K.~M. and {Badger}, C. and {Bae}, S. and {Bae}, Y. and {Baer}, A.~M. and {Bagnasco}, S. and {Bai}, Y. and {Baiotti}, L. and {Baird}, J. and {Bajpai}, R. and {Ball}, M. and {Ballardin}, G. and {Ballmer}, S.~W. and {Balsamo}, A. and {Baltus}, G. and {Banagiri}, S. and {Bankar}, D. and {Barayoga}, J.~C. and {Barbieri}, C. and {Barish}, B.~C. and {Barker}, D. and {Barneo}, P. and {Barone}, F. and {Barr}, B. and {Barsotti}, L. and {Barsuglia}, M. and {Barta}, D. and {Bartlett}, J. and {Barton}, M.~A. and {Bartos}, I. and {Bassiri}, R. and {Basti}, A. and {Bawaj}, M. and {Bayley}, J.~C. and {Baylor}, A.~C. and {Bazzan}, M. and {B{\'e}csy}, B. and {Bedakihale}, V.~M. and {Bejger}, M. and {Belahcene}, I. and {Benedetto}, V. and {Beniwal}, D. and {Bennett}, T.~F. and {Bentley}, J.~D. and {Benyaala}, M. and {Bergamin}, F. and {Berger}, B.~K. and {Bernuzzi}, S. and {Berry}, C.~P.~L. and {Bersanetti}, D. and {Bertolini}, A. and {Betzwieser}, J. and {Beveridge}, D. and {Bhandare}, R. and {Bhardwaj}, U. and {Bhattacharjee}, D. and {Bhaumik}, S. and {Bilenko}, I.~A. and {Billingsley}, G. and {Bini}, S. and {Birney}, R. and {Birnholtz}, O. and {Biscans}, S. and {Bischi}, M. and {Biscoveanu}, S. and {Bisht}, A. and {Biswas}, B. and {Bitossi}, M. and {Bizouard}, M. -A. and {Blackburn}, J.~K. and {Blair}, C.~D. and {Blair}, D.~G. and {Blair}, R.~M. and {Bobba}, F. and {Bode}, N. and {Boer}, M. and {Bogaert}, G. and {Boldrini}, M. and {Bonavena}, L.~D. and {Bondu}, F. and {Bonilla}, E. and {Bonnand}, R. and {Booker}, P. and {Boom}, B.~A. and {Bork}, R. and {Boschi}, V. and {Bose}, N. and {Bose}, S. and {Bossilkov}, V. and {Boudart}, V. and {Bouffanais}, Y. and {Bozzi}, A. and {Bradaschia}, C. and {Brady}, P.~R. and {Bramley}, A. and {Branch}, A. and {Branchesi}, M. and {Brandt}, J. and {Brau}, J.~E. and {Breschi}, M. and {Briant}, T. and {Briggs}, J.~H. and {Brillet}, A. and {Brinkmann}, M. and {Brockill}, P. and {Brooks}, A.~F. and {Brooks}, J. and {Brown}, D.~D. and {Brunett}, S. and {Bruno}, G. and {Bruntz}, R. and {Bryant}, J. and {Bulik}, T. and {Bulten}, H.~J. and {Buonanno}, A. and {Buscicchio}, R. and {Buskulic}, D. and {Buy}, C. and {Byer}, R.~L. and {Cadonati}, L. and {Cagnoli}, G. and {Cahillane}, C. and {Bustillo}, J. Calder{\'o}n and {Callaghan}, J.~D. and {Callister}, T.~A. and {Calloni}, E. and {Cameron}, J. and {Camp}, J.~B. and {Canepa}, M. and {Canevarolo}, S. and {Cannavacciuolo}, M. and {Cannon}, K.~C. and {Cao}, H. and {Cao}, Z. and {Capocasa}, E. and {Capote}, E. and {Carapella}, G.},
        title = "{Population of Merging Compact Binaries Inferred Using Gravitational Waves through GWTC-3}",
      journal = {Physical Review X},
         year = 2023,
        month = jan,
       volume = {13},
       number = {1},
          eid = {011048},
        pages = {011048},
          doi = {10.1103/PhysRevX.13.011048},
archivePrefix = {arXiv},
       eprint = {2111.03634},
 primaryClass = {astro-ph.HE},
       adsurl = {https://ui.adsabs.harvard.edu/abs/2023PhRvX..13a1048A}
}

@ARTICLE{2017A&A...597A..22S,
       author = {{Sim{\'o}n-D{\'\i}az}, S. and {Godart}, M. and {Castro}, N. and {Herrero}, A. and {Aerts}, C. and {Puls}, J. and {Telting}, J. and {Grassitelli}, L.},
        title = "{The IACOB project . III. New observational clues to understand macroturbulent broadening in massive O- and B-type stars}",
      journal = {\aap},
         year = 2017,
        month = jan,
       volume = {597},
          eid = {A22},
        pages = {A22},
          doi = {10.1051/0004-6361/201628541},
archivePrefix = {arXiv},
       eprint = {1608.05508},
 primaryClass = {astro-ph.SR},
       adsurl = {https://ui.adsabs.harvard.edu/abs/2017A&A...597A..22S}
}

@ARTICLE{2022A&A...665A.148S,
       author = {{Shenar}, T. and {Sana}, H. and {Mahy}, L. and {Ma{\'\i}z Apell{\'a}niz}, J. and {Crowther}, Paul A. and {Gromadzki}, M. and {Herrero}, A. and {Langer}, N. and {Marchant}, P. and {Schneider}, F.~R.~N. and {Sen}, K. and {Soszy{\'n}ski}, I. and {Toonen}, S.},
        title = "{The Tarantula Massive Binary Monitoring. VI. Characterisation of hidden companions in 51 single-lined O-type binaries: A flat mass-ratio distribution and black-hole binary candidates}",
      journal = {\aap},
         year = 2022,
        month = sep,
       volume = {665},
          eid = {A148},
        pages = {A148},
          doi = {10.1051/0004-6361/202244245},
archivePrefix = {arXiv},
       eprint = {2207.07674},
 primaryClass = {astro-ph.SR},
       adsurl = {https://ui.adsabs.harvard.edu/abs/2022A&A...665A.148S}
}

@ARTICLE{2020A&A...639L...6S,
       author = {{Shenar}, T. and {Bodensteiner}, J. and {Abdul-Masih}, M. and {Fabry}, M. and {Mahy}, L. and {Marchant}, P. and {Banyard}, G. and {Bowman}, D.~M. and {Dsilva}, K. and {Hawcroft}, C. and {Reggiani}, M. and {Sana}, H.},
        title = "{The ``hidden'' companion in LB-1 unveiled by spectral disentangling}",
      journal = {\aap},
         year = 2020,
        month = jul,
       volume = {639},
          eid = {L6},
        pages = {L6},
          doi = {10.1051/0004-6361/202038275},
archivePrefix = {arXiv},
       eprint = {2004.12882},
 primaryClass = {astro-ph.SR},
       adsurl = {https://ui.adsabs.harvard.edu/abs/2020A&A...639L...6S}
}

@ARTICLE{2020AN....341..628Q,
       author = {{Quintero}, Edwin A. and {Eenens}, Philippe and {Rauw}, Gregor},
        title = "{The massive binary system 9 Sgr revisited: New insights into disentangling methods}",
      journal = {Astronomische Nachrichten},
         year = 2020,
        month = jul,
       volume = {341},
       number = {628},
        pages = {628-637},
          doi = {10.1002/asna.202013696},
archivePrefix = {arXiv},
       eprint = {2010.01181},
 primaryClass = {astro-ph.SR},
       adsurl = {https://ui.adsabs.harvard.edu/abs/2020AN....341..628Q}
}

@ARTICLE{2019KPCB...35..129S,
       author = {{Sheminova}, V.~A.},
        title = "{Turbulence and Rotation in Solar-Type Stars}",
      journal = {Kinematics and Physics of Celestial Bodies},
         year = 2019,
        month = may,
       volume = {35},
       number = {3},
        pages = {129-142},
          doi = {10.3103/S088459131903005X},
archivePrefix = {arXiv},
       eprint = {1907.12241},
 primaryClass = {astro-ph.SR},
       adsurl = {https://ui.adsabs.harvard.edu/abs/2019KPCB...35..129S}
}

@BOOK{2005oasp.book.....G,
       author = {{Gray}, David F.},
        title = "{The Observation and Analysis of Stellar Photospheres}",
         year = 2005,
          doi = {10.1017/CBO9781316036570},
       adsurl = {https://ui.adsabs.harvard.edu/abs/2005oasp.book.....G}
}

@ARTICLE{1983ApJ...268..368E,
       author = {{Eggleton}, P.~P.},
        title = "{Aproximations to the radii of Roche lobes.}",
      journal = {\apj},
         year = 1983,
        month = may,
       volume = {268},
        pages = {368-369},
          doi = {10.1086/160960},
       adsurl = {https://ui.adsabs.harvard.edu/abs/1983ApJ...268..368E}
}

@ARTICLE{2005ApJ...628..426P,
       author = {{Pr{\v{s}}a}, A. and {Zwitter}, T.},
        title = "{A Computational Guide to Physics of Eclipsing Binaries. I. Demonstrations and Perspectives}",
      journal = {\apj},
         year = 2005,
        month = jul,
       volume = {628},
       number = {1},
        pages = {426-438},
          doi = {10.1086/430591},
archivePrefix = {arXiv},
       eprint = {astro-ph/0503361},
 primaryClass = {astro-ph},
       adsurl = {https://ui.adsabs.harvard.edu/abs/2005ApJ...628..426P}
}

@ARTICLE{2016ApJS..227...29P,
       author = {{Pr{\v{s}}a}, A. and {Conroy}, K.~E. and {Horvat}, M. and {Pablo}, H. and {Kochoska}, A. and {Bloemen}, S. and {Giammarco}, J. and {Hambleton}, K.~M. and {Degroote}, P.},
        title = "{Physics Of Eclipsing Binaries. II. Toward the Increased Model Fidelity}",
      journal = {\apjs},
         year = 2016,
        month = dec,
       volume = {227},
       number = {2},
          eid = {29},
        pages = {29},
          doi = {10.3847/1538-4365/227/2/29},
archivePrefix = {arXiv},
       eprint = {1609.08135},
 primaryClass = {astro-ph.SR},
       adsurl = {https://ui.adsabs.harvard.edu/abs/2016ApJS..227...29P}
}

@ARTICLE{2020ApJS..250...34C,
       author = {{Conroy}, Kyle E. and {Kochoska}, Angela and {Hey}, Daniel and {Pablo}, Herbert and {Hambleton}, Kelly M. and {Jones}, David and {Giammarco}, Joseph and {Abdul-Masih}, Michael and {Pr{\v{s}}a}, Andrej},
        title = "{Physics of Eclipsing Binaries. V. General Framework for Solving the Inverse Problem}",
      journal = {\apjs},
         year = 2020,
        month = oct,
       volume = {250},
       number = {2},
          eid = {34},
        pages = {34},
          doi = {10.3847/1538-4365/abb4e2},
archivePrefix = {arXiv},
       eprint = {2006.16951},
 primaryClass = {astro-ph.SR},
       adsurl = {https://ui.adsabs.harvard.edu/abs/2020ApJS..250...34C}
}

@ARTICLE{1998PASP..110.1416M,
       author = {{Marchenko}, Sergey V. and {Moffat}, Anthony F.~J. and {Eenens}, Philippe R.~J.},
        title = "{The Wolf-Rayet Binary WR 141 (WN5O + O5 V-III) Revisited}",
      journal = {\pasp},
         year = 1998,
        month = dec,
       volume = {110},
       number = {754},
        pages = {1416-1422},
          doi = {10.1086/316280},
       adsurl = {https://ui.adsabs.harvard.edu/abs/1998PASP..110.1416M}
}

@ARTICLE{2023MNRAS.518.2991S,
       author = {{Shahaf}, S. and {Bashi}, D. and {Mazeh}, T. and {Faigler}, S. and {Arenou}, F. and {El-Badry}, K. and {Rix}, H.~W.},
        title = "{Triage of the Gaia DR3 astrometric orbits - I. A sample of binaries with probable compact companions}",
      journal = {\mnras},
         year = 2023,
        month = jan,
       volume = {518},
       number = {2},
        pages = {2991-3003},
          doi = {10.1093/mnras/stac3290},
archivePrefix = {arXiv},
       eprint = {2209.00828},
 primaryClass = {astro-ph.SR},
       adsurl = {https://ui.adsabs.harvard.edu/abs/2023MNRAS.518.2991S}
}

@ARTICLE{2022A&A...664A.159M,
       author = {{Mahy}, L. and {Sana}, H. and {Shenar}, T. and {Sen}, K. and {Langer}, N. and {Marchant}, P. and {Abdul-Masih}, M. and {Banyard}, G. and {Bodensteiner}, J. and {Bowman}, D.~M. and {Dsilva}, K. and {Fabry}, M. and {Hawcroft}, C. and {Janssens}, S. and {Van Reeth}, T. and {Eldridge}, C.},
        title = "{Identifying quiescent compact objects in massive Galactic single-lined spectroscopic binaries}",
      journal = {\aap},
         year = 2022,
        month = aug,
       volume = {664},
          eid = {A159},
        pages = {A159},
          doi = {10.1051/0004-6361/202243147},
archivePrefix = {arXiv},
       eprint = {2207.07752},
 primaryClass = {astro-ph.SR},
       adsurl = {https://ui.adsabs.harvard.edu/abs/2022A&A...664A.159M}
}

@ARTICLE{1977A&A....57..383Z,
       author = {{Zahn}, J.-P.},
        title = "{Tidal friction in close binary systems.}",
      journal = {\aap},
         year = 1977,
        month = may,
       volume = {57},
        pages = {383-394},
       adsurl = {https://ui.adsabs.harvard.edu/abs/1977A&A....57..383Z}
}

@ARTICLE{1981A&A....99..126H,
       author = {{Hut}, P.},
        title = "{Tidal evolution in close binary systems.}",
      journal = {\aap},
         year = 1981,
        month = jun,
       volume = {99},
        pages = {126-140},
       adsurl = {https://ui.adsabs.harvard.edu/abs/1981A&A....99..126H}
}

@ARTICLE{2019A&A...628A..29C,
       author = {{Claret}, A.},
        title = "{Updating the theoretical tidal evolution constants: Apsidal motion and the moment of inertia}",
      journal = {\aap},
         year = 2019,
        month = aug,
       volume = {628},
          eid = {A29},
        pages = {A29},
          doi = {10.1051/0004-6361/201936007},
archivePrefix = {arXiv},
       eprint = {1907.11538},
 primaryClass = {astro-ph.SR},
       adsurl = {https://ui.adsabs.harvard.edu/abs/2019A&A...628A..29C}
}

@ARTICLE{corral2016,
       author = {{Corral-Santana}, J.~M. and {Casares}, J. and {Mu{\~n}oz-Darias}, T. and {Bauer}, F.~E. and {Mart{\'\i}nez-Pais}, I.~G. and {Russell}, D.~M.},
        title = "{BlackCAT: A catalogue of stellar-mass black holes in X-ray transients}",
      journal = {\aap},
         year = 2016,
        month = mar,
       volume = {587},
          eid = {A61},
        pages = {A61},
          doi = {10.1051/0004-6361/201527130},
archivePrefix = {arXiv},
       eprint = {1510.08869},
 primaryClass = {astro-ph.HE},
       adsurl = {https://ui.adsabs.harvard.edu/abs/2016A&A...587A..61C}
}

@ARTICLE{Althaus2022,
       author = {{Althaus}, Leandro G. and {Camisassa}, Mar{\'\i}a E. and {Torres}, Santiago and {Battich}, Tiara and {C{\'o}rsico}, Alejandro H. and {Rebassa-Mansergas}, Alberto and {Raddi}, Roberto},
        title = "{Structure and evolution of ultra-massive white dwarfs in general relativity}",
      journal = {\aap},
         year = 2022,
        month = dec,
       volume = {668},
          eid = {A58},
        pages = {A58},
          doi = {10.1051/0004-6361/202244604},
archivePrefix = {arXiv},
       eprint = {2208.14144},
 primaryClass = {astro-ph.SR},
       adsurl = {https://ui.adsabs.harvard.edu/abs/2022A&A...668A..58A}
}

@ARTICLE{Paxton2015,
       author = {{Paxton}, Bill and {Marchant}, Pablo and {Schwab}, Josiah and {Bauer}, Evan B. and {Bildsten}, Lars and {Cantiello}, Matteo and {Dessart}, Luc and {Farmer}, R. and {Hu}, H. and {Langer}, N. and {Townsend}, R.~H.~D. and {Townsley}, Dean M. and {Timmes}, F.~X.},
        title = "{Modules for Experiments in Stellar Astrophysics (MESA): Binaries, Pulsations, and Explosions}",
      journal = {\apjs},
         year = 2015,
        month = sep,
       volume = {220},
       number = {1},
          eid = {15},
        pages = {15},
          doi = {10.1088/0067-0049/220/1/15},
archivePrefix = {arXiv},
       eprint = {1506.03146},
 primaryClass = {astro-ph.SR},
       adsurl = {https://ui.adsabs.harvard.edu/abs/2015ApJS..220...15P}
}

@ARTICLE{Hurley2002,
       author = {{Hurley}, Jarrod R. and {Tout}, Christopher A. and {Pols}, Onno R.},
        title = "{Evolution of binary stars and the effect of tides on binary populations}",
      journal = {\mnras},
         year = 2002,
        month = feb,
       volume = {329},
       number = {4},
        pages = {897-928},
          doi = {10.1046/j.1365-8711.2002.05038.x},
archivePrefix = {arXiv},
       eprint = {astro-ph/0201220},
 primaryClass = {astro-ph},
       adsurl = {https://ui.adsabs.harvard.edu/abs/2002MNRAS.329..897H}
}

@ARTICLE{Lindegren2021,
       author = {{Lindegren}, L. and {Klioner}, S.~A. and {Hern{\'a}ndez}, J. and {Bombrun}, A. and {Ramos-Lerate}, M. and {Steidelm{\"u}ller}, H. and {Bastian}, U. and {Biermann}, M. and {de Torres}, A. and {Gerlach}, E. and {Geyer}, R. and {Hilger}, T. and {Hobbs}, D. and {Lammers}, U. and {McMillan}, P.~J. and {Stephenson}, C.~A. and {Casta{\~n}eda}, J. and {Davidson}, M. and {Fabricius}, C. and {Gracia-Abril}, G. and {Portell}, J. and {Rowell}, N. and {Teyssier}, D. and {Torra}, F. and {Bartolom{\'e}}, S. and {Clotet}, M. and {Garralda}, N. and {Gonz{\'a}lez-Vidal}, J.~J. and {Torra}, J. and {Abbas}, U. and {Altmann}, M. and {Anglada Varela}, E. and {Balaguer-N{\'u}{\~n}ez}, L. and {Balog}, Z. and {Barache}, C. and {Becciani}, U. and {Bernet}, M. and {Bertone}, S. and {Bianchi}, L. and {Bouquillon}, S. and {Brown}, A.~G.~A. and {Bucciarelli}, B. and {Busonero}, D. and {Butkevich}, A.~G. and {Buzzi}, R. and {Cancelliere}, R. and {Carlucci}, T. and {Charlot}, P. and {Cioni}, M. -R.~L. and {Crosta}, M. and {Crowley}, C. and {del Peloso}, E.~F. and {del Pozo}, E. and {Drimmel}, R. and {Esquej}, P. and {Fienga}, A. and {Fraile}, E. and {Gai}, M. and {Garcia-Reinaldos}, M. and {Guerra}, R. and {Hambly}, N.~C. and {Hauser}, M. and {Jan{\ss}en}, K. and {Jordan}, S. and {Kostrzewa-Rutkowska}, Z. and {Lattanzi}, M.~G. and {Liao}, S. and {Licata}, E. and {Lister}, T.~A. and {L{\"o}ffler}, W. and {Marchant}, J.~M. and {Masip}, A. and {Mignard}, F. and {Mints}, A. and {Molina}, D. and {Mora}, A. and {Morbidelli}, R. and {Murphy}, C.~P. and {Pagani}, C. and {Panuzzo}, P. and {Pe{\~n}alosa Esteller}, X. and {Poggio}, E. and {Re Fiorentin}, P. and {Riva}, A. and {Sagrist{\`a} Sell{\'e}s}, A. and {Sanchez Gimenez}, V. and {Sarasso}, M. and {Sciacca}, E. and {Siddiqui}, H.~I. and {Smart}, R.~L. and {Souami}, D. and {Spagna}, A. and {Steele}, I.~A. and {Taris}, F. and {Utrilla}, E. and {van Reeven}, W. and {Vecchiato}, A.},
        title = "{Gaia Early Data Release 3. The astrometric solution}",
      journal = {\aap},
         year = 2021,
        month = may,
       volume = {649},
          eid = {A2},
        pages = {A2},
          doi = {10.1051/0004-6361/202039709},
archivePrefix = {arXiv},
       eprint = {2012.03380},
 primaryClass = {astro-ph.IM},
       adsurl = {https://ui.adsabs.harvard.edu/abs/2021A&A...649A...2L}
}

@ARTICLE{Lindegren2018,
       author = {{Lindegren}, L. and {Hern{\'a}ndez}, J. and {Bombrun}, A. and {Klioner}, S. and {Bastian}, U. and {Ramos-Lerate}, M. and {de Torres}, A. and {Steidelm{\"u}ller}, H. and {Stephenson}, C. and {Hobbs}, D. and {Lammers}, U. and {Biermann}, M. and {Geyer}, R. and {Hilger}, T. and {Michalik}, D. and {Stampa}, U. and {McMillan}, P.~J. and {Casta{\~n}eda}, J. and {Clotet}, M. and {Comoretto}, G. and {Davidson}, M. and {Fabricius}, C. and {Gracia}, G. and {Hambly}, N.~C. and {Hutton}, A. and {Mora}, A. and {Portell}, J. and {van Leeuwen}, F. and {Abbas}, U. and {Abreu}, A. and {Altmann}, M. and {Andrei}, A. and {Anglada}, E. and {Balaguer-N{\'u}{\~n}ez}, L. and {Barache}, C. and {Becciani}, U. and {Bertone}, S. and {Bianchi}, L. and {Bouquillon}, S. and {Bourda}, G. and {Br{\"u}semeister}, T. and {Bucciarelli}, B. and {Busonero}, D. and {Buzzi}, R. and {Cancelliere}, R. and {Carlucci}, T. and {Charlot}, P. and {Cheek}, N. and {Crosta}, M. and {Crowley}, C. and {de Bruijne}, J. and {de Felice}, F. and {Drimmel}, R. and {Esquej}, P. and {Fienga}, A. and {Fraile}, E. and {Gai}, M. and {Garralda}, N. and {Gonz{\'a}lez-Vidal}, J.~J. and {Guerra}, R. and {Hauser}, M. and {Hofmann}, W. and {Holl}, B. and {Jordan}, S. and {Lattanzi}, M.~G. and {Lenhardt}, H. and {Liao}, S. and {Licata}, E. and {Lister}, T. and {L{\"o}ffler}, W. and {Marchant}, J. and {Martin-Fleitas}, J. -M. and {Messineo}, R. and {Mignard}, F. and {Morbidelli}, R. and {Poggio}, E. and {Riva}, A. and {Rowell}, N. and {Salguero}, E. and {Sarasso}, M. and {Sciacca}, E. and {Siddiqui}, H. and {Smart}, R.~L. and {Spagna}, A. and {Steele}, I. and {Taris}, F. and {Torra}, J. and {van Elteren}, A. and {van Reeven}, W. and {Vecchiato}, A.},
        title = "{Gaia Data Release 2. The astrometric solution}",
      journal = {\aap},
         year = 2018,
        month = aug,
       volume = {616},
          eid = {A2},
        pages = {A2},
          doi = {10.1051/0004-6361/201832727},
archivePrefix = {arXiv},
       eprint = {1804.09366},
 primaryClass = {astro-ph.IM},
       adsurl = {https://ui.adsabs.harvard.edu/abs/2018A&A...616A...2L}
}

@ARTICLE{Avdeeva2024,
       author = {{Avdeeva}, Aleksandra S. and {Kovaleva}, Dana A. and {Malkov}, Oleg Yu and {Zhao}, Gang},
        title = "{Quality flags for GSP-Phot Gaia DR3 astrophysical parameters with machine learning: effective temperatures case study}",
      journal = {\mnras},
         year = 2024,
        month = jan,
       volume = {527},
       number = {3},
        pages = {7382-7393},
          doi = {10.1093/mnras/stad3601},
archivePrefix = {arXiv},
       eprint = {2310.15671},
 primaryClass = {astro-ph.SR},
       adsurl = {https://ui.adsabs.harvard.edu/abs/2024MNRAS.527.7382A}
}

@ARTICLE{Andrae2023,
       author = {{Andrae}, R. and {Fouesneau}, M. and {Sordo}, R. and {Bailer-Jones}, C.~A.~L. and {Dharmawardena}, T.~E. and {Rybizki}, J. and {De Angeli}, F. and {Lindstr{\o}m}, H.~E.~P. and {Marshall}, D.~J. and {Drimmel}, R. and {Korn}, A.~J. and {Soubiran}, C. and {Brouillet}, N. and {Casamiquela}, L. and {Rix}, H. -W. and {Abreu Aramburu}, A. and {{\'A}lvarez}, M.~A. and {Bakker}, J. and {Bellas-Velidis}, I. and {Bijaoui}, A. and {Brugaletta}, E. and {Burlacu}, A. and {Carballo}, R. and {Chaoul}, L. and {Chiavassa}, A. and {Contursi}, G. and {Cooper}, W.~J. and {Creevey}, O.~L. and {Dafonte}, C. and {Dapergolas}, A. and {de Laverny}, P. and {Delchambre}, L. and {Demouchy}, C. and {Edvardsson}, B. and {Fr{\'e}mat}, Y. and {Garabato}, D. and {Garc{\'\i}a-Lario}, P. and {Garc{\'\i}a-Torres}, M. and {Gavel}, A. and {Gomez}, A. and {Gonz{\'a}lez-Santamar{\'\i}a}, I. and {Hatzidimitriou}, D. and {Heiter}, U. and {Jean-Antoine Piccolo}, A. and {Kontizas}, M. and {Kordopatis}, G. and {Lanzafame}, A.~C. and {Lebreton}, Y. and {Licata}, E.~L. and {Livanou}, E. and {Lobel}, A. and {Lorca}, A. and {Magdaleno Romeo}, A. and {Manteiga}, M. and {Marocco}, F. and {Mary}, N. and {Nicolas}, C. and {Ordenovic}, C. and {Pailler}, F. and {Palicio}, P.~A. and {Pallas-Quintela}, L. and {Panem}, C. and {Pichon}, B. and {Poggio}, E. and {Recio-Blanco}, A. and {Riclet}, F. and {Robin}, C. and {Santove{\~n}a}, R. and {Sarro}, L.~M. and {Schultheis}, M.~S. and {Segol}, M. and {Silvelo}, A. and {Slezak}, I. and {Smart}, R.~L. and {S{\"u}veges}, M. and {Th{\'e}venin}, F. and {Torralba Elipe}, G. and {Ulla}, A. and {Utrilla}, E. and {Vallenari}, A. and {van Dillen}, E. and {Zhao}, H. and {Zorec}, J.},
        title = "{Gaia Data Release 3. Analysis of the Gaia BP/RP spectra using the General Stellar Parameterizer from Photometry}",
      journal = {\aap},
         year = 2023,
        month = jun,
       volume = {674},
          eid = {A27},
        pages = {A27},
          doi = {10.1051/0004-6361/202243462},
archivePrefix = {arXiv},
       eprint = {2206.06138},
 primaryClass = {astro-ph.SR},
       adsurl = {https://ui.adsabs.harvard.edu/abs/2023A&A...674A..27A}
}

@ARTICLE{Collins1995,
       author = {{Collins}, II, George W. and {Truax}, Ryland J.},
        title = "{Classical Rotational Broadening of Spectral Lines}",
      journal = {\apj},
         year = 1995,
        month = feb,
       volume = {439},
        pages = {860},
          doi = {10.1086/175225},
       adsurl = {https://ui.adsabs.harvard.edu/abs/1995ApJ...439..860C}
}

@ARTICLE{Shahbaz2003,
       author = {{Shahbaz}, T.},
        title = "{Determining the spectroscopic mass ratio in interacting binaries: application to X-Ray Nova Sco 1994}",
      journal = {\mnras},
         year = 2003,
        month = mar,
       volume = {339},
       number = {4},
        pages = {1031-1040},
          doi = {10.1046/j.1365-8711.2003.06258.x},
archivePrefix = {arXiv},
       eprint = {astro-ph/0211266},
 primaryClass = {astro-ph},
       adsurl = {https://ui.adsabs.harvard.edu/abs/2003MNRAS.339.1031S}
}

@ARTICLE{2022MNRAS.510.4736K,
       author = {{Knight}, Amy H. and {Ingram}, Adam and {Middleton}, Matthew and {Drake}, Jeremy},
        title = "{Eclipse mapping of EXO 0748-676: evidence for a massive neutron star}",
      journal = {\mnras},
         year = 2022,
        month = mar,
       volume = {510},
       number = {4},
        pages = {4736-4756},
          doi = {10.1093/mnras/stab3722},
archivePrefix = {arXiv},
       eprint = {2201.02188},
 primaryClass = {astro-ph.HE},
       adsurl = {https://ui.adsabs.harvard.edu/abs/2022MNRAS.510.4736K}
}

@ARTICLE{2012RSPTA.370.2765A,
       author = {{Allard}, F. and {Homeier}, D. and {Freytag}, B.},
        title = "{Models of very-low-mass stars, brown dwarfs and exoplanets}",
      journal = {Philosophical Transactions of the Royal Society of London Series A},
         year = 2012,
        month = jun,
       volume = {370},
       number = {1968},
        pages = {2765-2777},
          doi = {10.1098/rsta.2011.0269},
archivePrefix = {arXiv},
       eprint = {1112.3591},
 primaryClass = {astro-ph.SR},
       adsurl = {https://ui.adsabs.harvard.edu/abs/2012RSPTA.370.2765A}
}

@ARTICLE{1993KurCD..13.....K,
       author = {{Kurucz}, Robert},
        title = "{ATLAS9 Stellar Atmosphere Programs and 2 km/s grid.}",
      journal = {Robert Kurucz CD-ROM},
         year = 1993,
        month = jan,
       volume = {13},
       adsurl = {https://ui.adsabs.harvard.edu/abs/1993KurCD..13.....K}
}

@INPROCEEDINGS{2003IAUS..210P.A20C,
       author = {{Castelli}, F. and {Kurucz}, R.~L.},
        title = "{New Grids of ATLAS9 Model Atmospheres}",
    booktitle = {Modelling of Stellar Atmospheres},
         year = 2003,
       editor = {{Piskunov}, N. and {Weiss}, W.~W. and {Gray}, D.~F.},
       series = {IAU Symposium},
       volume = {210},
        month = jan,
        pages = {A20},
          doi = {10.48550/arXiv.astro-ph/0405087},
archivePrefix = {arXiv},
       eprint = {astro-ph/0405087},
 primaryClass = {astro-ph},
       adsurl = {https://ui.adsabs.harvard.edu/abs/2003IAUS..210P.A20C}
}

@ARTICLE{2013A&A...553A...6H,
       author = {{Husser}, T. -O. and {Wende-von Berg}, S. and {Dreizler}, S. and {Homeier}, D. and {Reiners}, A. and {Barman}, T. and {Hauschildt}, P.~H.},
        title = "{A new extensive library of PHOENIX stellar atmospheres and synthetic spectra}",
      journal = {\aap},
         year = 2013,
        month = may,
       volume = {553},
          eid = {A6},
        pages = {A6},
          doi = {10.1051/0004-6361/201219058},
archivePrefix = {arXiv},
       eprint = {1303.5632},
 primaryClass = {astro-ph.SR},
       adsurl = {https://ui.adsabs.harvard.edu/abs/2013A&A...553A...6H}
}

@ARTICLE{Liuchao2020,
       author = {{Liu}, Chao and {Fu}, Jianning and {Shi}, Jianrong and {Wu}, Hong and {Han}, Zhanwen and {Chen}, Li and {Dong}, Subo and {Zhao}, Yongheng and {Chen}, Jian-Jun and {Zhang}, Haotong and {Bai}, Zhong-Rui and {Chen}, Xuefei and {Cui}, Wenyuan and {Du}, Bing and {Hsia}, Chih-Hao and {Jiang}, Deng-Kai and {Hou}, Jinliang and {Hou}, Wen and {Li}, Haining and {Li}, Jiao and {Li}, Lifang and {Liu}, Jiaming and {Liu}, Jifeng and {Luo}, A-Li and {Ren}, Juan-Juan and {Tian}, Hai-Jun and {Tian}, Hao and {Wang}, Jia-Xin and {Wu}, Chao-Jian and {Xie}, Ji-Wei and {Yan}, Hong-Liang and {Yang}, Fan and {Yu}, Jincheng and {Zhang}, Bo and {Zhang}, Huawei and {Zhang}, Li-Yun and {Zhang}, Wei and {Zhao}, Gang and {Zhong}, Jing and {Zong}, Weikai and {Zuo}, Fang},
        title = "{LAMOST Medium-Resolution Spectroscopic Survey (LAMOST-MRS): Scientific goals and survey plan}",
      journal = {arXiv e-prints},
         year = 2020,
        month = may,
          eid = {arXiv:2005.07210},
        pages = {arXiv:2005.07210},
          doi = {10.48550/arXiv.2005.07210},
archivePrefix = {arXiv},
       eprint = {2005.07210},
 primaryClass = {astro-ph.SR},
       adsurl = {https://ui.adsabs.harvard.edu/abs/2020arXiv200507210L}
}

@ARTICLE{Majewski2017,
       author = {{Majewski}, Steven R. and {Schiavon}, Ricardo P. and {Frinchaboy}, Peter M. and {Allende Prieto}, Carlos and {Barkhouser}, Robert and {Bizyaev}, Dmitry and {Blank}, Basil and {Brunner}, Sophia and {Burton}, Adam and {Carrera}, Ricardo and {Chojnowski}, S. Drew and {Cunha}, K{\'a}tia and {Epstein}, Courtney and {Fitzgerald}, Greg and {Garc{\'\i}a P{\'e}rez}, Ana E. and {Hearty}, Fred R. and {Henderson}, Chuck and {Holtzman}, Jon A. and {Johnson}, Jennifer A. and {Lam}, Charles R. and {Lawler}, James E. and {Maseman}, Paul and {M{\'e}sz{\'a}ros}, Szabolcs and {Nelson}, Matthew and {Nguyen}, Duy Coung and {Nidever}, David L. and {Pinsonneault}, Marc and {Shetrone}, Matthew and {Smee}, Stephen and {Smith}, Verne V. and {Stolberg}, Todd and {Skrutskie}, Michael F. and {Walker}, Eric and {Wilson}, John C. and {Zasowski}, Gail and {Anders}, Friedrich and {Basu}, Sarbani and {Beland}, Stephane and {Blanton}, Michael R. and {Bovy}, Jo and {Brownstein}, Joel R. and {Carlberg}, Joleen and {Chaplin}, William and {Chiappini}, Cristina and {Eisenstein}, Daniel J. and {Elsworth}, Yvonne and {Feuillet}, Diane and {Fleming}, Scott W. and {Galbraith-Frew}, Jessica and {Garc{\'\i}a}, Rafael A. and {Garc{\'\i}a-Hern{\'a}ndez}, D. An{\'\i}bal and {Gillespie}, Bruce A. and {Girardi}, L{\'e}o and {Gunn}, James E. and {Hasselquist}, Sten and {Hayden}, Michael R. and {Hekker}, Saskia and {Ivans}, Inese and {Kinemuchi}, Karen and {Klaene}, Mark and {Mahadevan}, Suvrath and {Mathur}, Savita and {Mosser}, Beno{\^\i}t and {Muna}, Demitri and {Munn}, Jeffrey A. and {Nichol}, Robert C. and {O'Connell}, Robert W. and {Parejko}, John K. and {Robin}, A.~C. and {Rocha-Pinto}, Helio and {Schultheis}, Matthias and {Serenelli}, Aldo M. and {Shane}, Neville and {Silva Aguirre}, Victor and {Sobeck}, Jennifer S. and {Thompson}, Benjamin and {Troup}, Nicholas W. and {Weinberg}, David H. and {Zamora}, Olga},
        title = "{The Apache Point Observatory Galactic Evolution Experiment (APOGEE)}",
      journal = {\aj},
         year = 2017,
        month = sep,
       volume = {154},
       number = {3},
          eid = {94},
        pages = {94},
          doi = {10.3847/1538-3881/aa784d},
archivePrefix = {arXiv},
       eprint = {1509.05420},
 primaryClass = {astro-ph.IM},
       adsurl = {https://ui.adsabs.harvard.edu/abs/2017AJ....154...94M}
}

@ARTICLE{gaia2016,
       author = {{Gaia Collaboration} and {Prusti}, T. and {de Bruijne}, J.~H.~J. and {Brown}, A.~G.~A. and {Vallenari}, A. and {Babusiaux}, C. and {Bailer-Jones}, C.~A.~L. and {Bastian}, U. and {Biermann}, M. and {Evans}, D.~W. and {Eyer}, L. and {Jansen}, F. and {Jordi}, C. and {Klioner}, S.~A. and {Lammers}, U. and {Lindegren}, L. and {Luri}, X. and {Mignard}, F. and {Milligan}, D.~J. and {Panem}, C. and {Poinsignon}, V. and {Pourbaix}, D. and {Randich}, S. and {Sarri}, G. and {Sartoretti}, P. and {Siddiqui}, H.~I. and {Soubiran}, C. and {Valette}, V. and {van Leeuwen}, F. and {Walton}, N.~A. and {Aerts}, C. and {Arenou}, F. and {Cropper}, M. and {Drimmel}, R. and {H{\o}g}, E. and {Katz}, D. and {Lattanzi}, M.~G. and {O'Mullane}, W. and {Grebel}, E.~K. and {Holland}, A.~D. and {Huc}, C. and {Passot}, X. and {Bramante}, L. and {Cacciari}, C. and {Casta{\~n}eda}, J. and {Chaoul}, L. and {Cheek}, N. and {De Angeli}, F. and {Fabricius}, C. and {Guerra}, R. and {Hern{\'a}ndez}, J. and {Jean-Antoine-Piccolo}, A. and {Masana}, E. and {Messineo}, R. and {Mowlavi}, N. and {Nienartowicz}, K. and {Ord{\'o}{\~n}ez-Blanco}, D. and {Panuzzo}, P. and {Portell}, J. and {Richards}, P.~J. and {Riello}, M. and {Seabroke}, G.~M. and {Tanga}, P. and {Th{\'e}venin}, F. and {Torra}, J. and {Els}, S.~G. and {Gracia-Abril}, G. and {Comoretto}, G. and {Garcia-Reinaldos}, M. and {Lock}, T. and {Mercier}, E. and {Altmann}, M. and {Andrae}, R. and {Astraatmadja}, T.~L. and {Bellas-Velidis}, I. and {Benson}, K. and {Berthier}, J. and {Blomme}, R. and {Busso}, G. and {Carry}, B. and {Cellino}, A. and {Clementini}, G. and {Cowell}, S. and {Creevey}, O. and {Cuypers}, J. and {Davidson}, M. and {De Ridder}, J. and {de Torres}, A. and {Delchambre}, L. and {Dell'Oro}, A. and {Ducourant}, C. and {Fr{\'e}mat}, Y. and {Garc{\'\i}a-Torres}, M. and {Gosset}, E. and {Halbwachs}, J. -L. and {Hambly}, N.~C. and {Harrison}, D.~L. and {Hauser}, M. and {Hestroffer}, D. and {Hodgkin}, S.~T. and {Huckle}, H.~E. and {Hutton}, A. and {Jasniewicz}, G. and {Jordan}, S. and {Kontizas}, M. and {Korn}, A.~J. and {Lanzafame}, A.~C. and {Manteiga}, M. and {Moitinho}, A. and {Muinonen}, K. and {Osinde}, J. and {Pancino}, E. and {Pauwels}, T. and {Petit}, J. -M. and {Recio-Blanco}, A. and {Robin}, A.~C. and {Sarro}, L.~M. and {Siopis}, C. and {Smith}, M. and {Smith}, K.~W. and {Sozzetti}, A. and {Thuillot}, W. and {van Reeven}, W. and {Viala}, Y. and {Abbas}, U. and {Abreu Aramburu}, A. and {Accart}, S. and {Aguado}, J.~J. and {Allan}, P.~M. and {Allasia}, W. and {Altavilla}, G. and {{\'A}lvarez}, M.~A. and {Alves}, J. and {Anderson}, R.~I. and {Andrei}, A.~H. and {Anglada Varela}, E. and {Antiche}, E. and {Antoja}, T. and {Ant{\'o}n}, S. and {Arcay}, B. and {Atzei}, A. and {Ayache}, L. and {Bach}, N. and {Baker}, S.~G. and {Balaguer-N{\'u}{\~n}ez}, L. and {Barache}, C. and {Barata}, C. and {Barbier}, A. and {Barblan}, F. and {Baroni}, M. and {Barrado y Navascu{\'e}s}, D. and {Barros}, M. and {Barstow}, M.~A. and {Becciani}, U. and {Bellazzini}, M. and {Bellei}, G. and {Bello Garc{\'\i}a}, A. and {Belokurov}, V. and {Bendjoya}, P. and {Berihuete}, A. and {Bianchi}, L. and {Bienaym{\'e}}, O. and {Billebaud}, F. and {Blagorodnova}, N. and {Blanco-Cuaresma}, S. and {Boch}, T. and {Bombrun}, A. and {Borrachero}, R. and {Bouquillon}, S. and {Bourda}, G. and {Bouy}, H. and {Bragaglia}, A. and {Breddels}, M.~A. and {Brouillet}, N. and {Br{\"u}semeister}, T. and {Bucciarelli}, B. and {Budnik}, F. and {Burgess}, P. and {Burgon}, R. and {Burlacu}, A. and {Busonero}, D. and {Buzzi}, R. and {Caffau}, E. and {Cambras}, J. and {Campbell}, H. and {Cancelliere}, R. and {Cantat-Gaudin}, T. and {Carlucci}, T. and {Carrasco}, J.~M. and {Castellani}, M. and {Charlot}, P. and {Charnas}, J. and {Charvet}, P. and {Chassat}, F. and {Chiavassa}, A. and {Clotet}, M. and {Cocozza}, G. and {Collins}, R.~S. and {Collins}, P. and {Costigan}, G.},
        title = "{The Gaia mission}",
      journal = {\aap},
         year = 2016,
        month = nov,
       volume = {595},
          eid = {A1},
        pages = {A1},
          doi = {10.1051/0004-6361/201629272},
archivePrefix = {arXiv},
       eprint = {1609.04153},
 primaryClass = {astro-ph.IM},
       adsurl = {https://ui.adsabs.harvard.edu/abs/2016A&A...595A...1G}
}

@ARTICLE{2022ApJ...933..193Z,
       author = {{Zhang}, Zhi-Xiang and {Zheng}, Ling-Lin and {Gu}, Wei-Min and {Sun}, Mouyuan and {Yi}, Tuan and {Shi}, Jian-Rong and {Wang}, Song and {Bai}, Zhong-Rui and {Zhang}, Hao-Tong and {Cui}, Wen-Yuan and {Wang}, Junfeng and {Wu}, Jianfeng and {Li}, Xiang-Dong and {Shao}, Yong and {Lu}, Kai-Xing and {Bai}, Yu and {Li}, Chunqian and {Fu}, Jin-Bo and {Liu}, Jifeng},
        title = "{A Long-period Pre-ELM System Discovered from the LAMOST Medium-resolution Survey}",
      journal = {\apj},
         year = 2022,
        month = jul,
       volume = {933},
       number = {2},
          eid = {193},
        pages = {193},
          doi = {10.3847/1538-4357/ac75b6},
archivePrefix = {arXiv},
       eprint = {2209.13889},
 primaryClass = {astro-ph.SR},
       adsurl = {https://ui.adsabs.harvard.edu/abs/2022ApJ...933..193Z}
}

@ARTICLE{zhang2024,
       author = {{Zhang}, Zhi-Xiang and {Liu}, Hao-Bin and {Yi}, Tuan and {Sun}, Mouyuan and {Gu}, Wei-Min},
        title = "{The Hidden Companion in J1527: A 0.69 Solar-mass White Dwarf?}",
      journal = {\apjl},
         year = 2024,
        month = feb,
       volume = {961},
       number = {2},
          eid = {L48},
        pages = {L48},
          doi = {10.3847/2041-8213/ad1eef},
archivePrefix = {arXiv},
       eprint = {2401.08289},
 primaryClass = {astro-ph.SR},
       adsurl = {https://ui.adsabs.harvard.edu/abs/2024ApJ...961L..48Z}
}

@ARTICLE{2024ApJ...969..114L,
       author = {{Liu}, Hao-Bin and {Gu}, Wei-Min and {Zhang}, Zhi-Xiang and {Yi}, Tuan and {Liu}, Jin-Zhong and {Sun}, Mouyuan},
        title = "{A Sample of Compact Object Candidates in Single-lined Spectroscopic Binaries from LAMOST Medium-resolution Survey}",
      journal = {\apj},
         year = 2024,
        month = jul,
       volume = {969},
       number = {2},
          eid = {114},
        pages = {114},
          doi = {10.3847/1538-4357/ad4c6f},
archivePrefix = {arXiv},
       eprint = {2405.09825},
 primaryClass = {astro-ph.SR},
       adsurl = {https://ui.adsabs.harvard.edu/abs/2024ApJ...969..114L}
}

@software{spectool2025,
  author       = {Zhang, Zhixiang},
  title        = {zhang-zhixiang/spectool: spectool version 1.0.1},
  month        = feb,
  year         = 2025,
  publisher    = {Zenodo},
  version      = {v1.0.1},
  doi          = {10.5281/zenodo.14947417},
  url          = {https://doi.org/10.5281/zenodo.14947417},
}

@software{2015ascl.soft03010M,
       author = {{Morton}, Timothy D.},
        title = "{isochrones: Stellar model grid package}",
 howpublished = {Astrophysics Source Code Library, record ascl:1503.010},
         year = 2015,
        month = mar,
          eid = {ascl:1503.010},
       adsurl = {https://ui.adsabs.harvard.edu/abs/2015ascl.soft03010M}
}

@article{Tauris2017,
  author = {Tauris, T. M. and Kramer, M. and Freire, P. C. C. and Wex, N. and Janka, H.-T. and Langer, N. and Podsiadlowski, Ph. and Bozzo, E. and Chaty, S. and Kruckow, M. and van den Heuvel, E. P. J. and Antoniadis, J. and Breton, R. P. and Champion, D. J. and Costa, J. E. S. and Demorest, P. B. and Desvignes, G. and Ghosh, T. and Janssen, G. H. and Lazarus, P. and Liu, K. and Lyne, A. G. and O'Brien, A. N. and Palfreyman, J. L. and Stairs, I. H. and van Leeuwen, J. and Weltevrede, P.},
  title = {Formation of Double Neutron Star Systems},
  journal = {ApJ},
  year = {2017},
  volume = {846},
  number = {2},
  pages = {170},
  doi = {10.3847/1538-4357/aa7e89}
}

@article{Manchester2005,
  author = {Manchester, R. N. and Hobbs, G. B. and Teoh, A. and Hobbs, M.},
  title = {The Australia Telescope National Facility Pulsar Catalogue},
  journal = {AJ},
  year = {2005},
  volume = {129},
  pages = {1993-2006},
  doi = {10.1086/428488}
}

@ARTICLE{2023MNRAS.521.4323E,
       author = {{El-Badry}, Kareem and {Rix}, Hans-Walter and {Cendes}, Yvette and {Rodriguez}, Antonio C. and {Conroy}, Charlie and {Quataert}, Eliot and {Hawkins}, Keith and {Zari}, Eleonora and {Hobson}, Melissa and {Breivik}, Katelyn and {Rau}, Arne and {Berger}, Edo and {Shahaf}, Sahar and {Seeburger}, Rhys and {Burdge}, Kevin B. and {Latham}, David W. and {Buchhave}, Lars A. and {Bieryla}, Allyson and {Bashi}, Dolev and {Mazeh}, Tsevi and {Faigler}, Simchon},
        title = "{A red giant orbiting a black hole}",
      journal = {\mnras},
         year = 2023,
        month = may,
       volume = {521},
       number = {3},
        pages = {4323-4348},
          doi = {10.1093/mnras/stad799},
archivePrefix = {arXiv},
       eprint = {2302.07880},
 primaryClass = {astro-ph.SR},
       adsurl = {https://ui.adsabs.harvard.edu/abs/2023MNRAS.521.4323E}
}

@ARTICLE{2020A&A...638A..94O,
       author = {{Olejak}, A. and {Belczynski}, K. and {Bulik}, T. and {Sobolewska}, M.},
        title = "{Synthetic catalog of black holes in the Milky Way}",
      journal = {\aap},
         year = 2020,
        month = jun,
       volume = {638},
          eid = {A94},
        pages = {A94},
          doi = {10.1051/0004-6361/201936557},
archivePrefix = {arXiv},
       eprint = {1908.08775},
 primaryClass = {astro-ph.SR},
       adsurl = {https://ui.adsabs.harvard.edu/abs/2020A&A...638A..94O}
}

@ARTICLE{2010A&A...510A..23S,
       author = {{Sartore}, N. and {Ripamonti}, E. and {Treves}, A. and {Turolla}, R.},
        title = "{Galactic neutron stars. I. Space and velocity distributions in the disk and in the halo}",
      journal = {\aap},
         year = 2010,
        month = feb,
       volume = {510},
          eid = {A23},
        pages = {A23},
          doi = {10.1051/0004-6361/200912222},
archivePrefix = {arXiv},
       eprint = {0908.3182},
 primaryClass = {astro-ph.GA},
       adsurl = {https://ui.adsabs.harvard.edu/abs/2010A&A...510A..23S}
}

@ARTICLE{2022NatAs...6.1203Y,
       author = {{Yi}, Tuan and {Gu}, Wei-Min and {Zhang}, Zhi-Xiang and {Zheng}, Ling-Lin and {Sun}, Mouyuan and {Wang}, Junfeng and {Bai}, Zhongrui and {Wang}, Pei and {Wu}, Jianfeng and {Bai}, Yu and {Wang}, Song and {Zhang}, Haotong and {Dong}, Yize and {Shao}, Yong and {Li}, Xiang-Dong and {Zhang}, Jia and {Huang}, Yang and {Yang}, Fan and {Yu}, Qingzheng and {Mu}, Hui-Jun and {Fu}, Jin-Bo and {Qi}, Senyu and {Guo}, Jing and {Fang}, Xuan and {Zheng}, Chuanjie and {Li}, Chun-Qian and {Shi}, Jian-Rong and {Chen}, Huanyang and {Liu}, Jifeng},
        title = "{A dynamically discovered and characterized non-accreting neutron star-M dwarf binary candidate}",
      journal = {Nature Astronomy},
         year = 2022,
        month = sep,
       volume = {6},
        pages = {1203-1212},
          doi = {10.1038/s41550-022-01766-0},
archivePrefix = {arXiv},
       eprint = {2209.12141},
 primaryClass = {astro-ph.SR},
       adsurl = {https://ui.adsabs.harvard.edu/abs/2022NatAs...6.1203Y}
}

@ARTICLE{2023AJ....165..187Q,
       author = {{Qi}, Senyu and {Gu}, Wei-Min and {Yi}, Tuan and {Zhang}, Zhi-Xiang and {Wang}, Song and {Liu}, Jifeng},
        title = "{Searching for Compact Object Candidates from LAMOST Time-domain Survey of Four K2 Plates}",
      journal = {\aj},
         year = 2023,
        month = may,
       volume = {165},
       number = {5},
          eid = {187},
        pages = {187},
          doi = {10.3847/1538-3881/acc389},
archivePrefix = {arXiv},
       eprint = {2303.06083},
 primaryClass = {astro-ph.SR},
       adsurl = {https://ui.adsabs.harvard.edu/abs/2023AJ....165..187Q}
}

@ARTICLE{2022ApJ...936...33Z,
       author = {{Zheng}, Ling-Lin and {Gu}, Wei-Min and {Sun}, Mouyuan and {Zhang}, Zhi-Xiang and {Yi}, Tuan and {Wu}, Jianfeng and {Wang}, Junfeng and {Fu}, Jin-Bo and {Qi}, Sen-Yu and {Yang}, Fan and {Wang}, Song and {Wang}, Liang and {Bai}, Zhong-Rui and {Zhang}, Haotong and {Li}, Chun-Qian and {Shi}, Jian-Rong and {Zong}, Weikai and {Bai}, Yu and {Liu}, Jifeng},
        title = "{A White Dwarf-Main-sequence Binary Unveiled by Time-domain Observations from LAMOST and TESS}",
      journal = {\apj},
         year = 2022,
        month = sep,
       volume = {936},
       number = {1},
          eid = {33},
        pages = {33},
          doi = {10.3847/1538-4357/ac853f},
archivePrefix = {arXiv},
       eprint = {2209.13924},
 primaryClass = {astro-ph.SR},
       adsurl = {https://ui.adsabs.harvard.edu/abs/2022ApJ...936...33Z}
}

@ARTICLE{2023SCPMA..6629512Z,
       author = {{Zheng}, Ling-Lin and {Sun}, Mouyuan and {Gu}, Wei-Min and {Yi}, Tuan and {Zhang}, Zhi-Xiang and {Wang}, Pei and {Wang}, Junfeng and {Wu}, Jianfeng and {Weng}, Shan-Shan and {Wang}, Song and {Qi}, Sen-Yu and {Zhang}, Jia and {Li}, Chun-Qian and {Shi}, Jian-Rong and {Shao}, Yong and {Li}, Xiang-Dong and {Fu}, Jin-Bo and {Yang}, Fan and {Bai}, Zhongrui and {Bai}, Yu and {Zhang}, Haotong and {Liu}, Jifeng},
        title = "{The nearest neutron star candidate in a binary revealed by optical time-domain surveys}",
      journal = {Science China Physics, Mechanics, and Astronomy},
         year = 2023,
        month = dec,
       volume = {66},
       number = {12},
          eid = {129512},
        pages = {129512},
          doi = {10.1007/s11433-023-2247-x},
archivePrefix = {arXiv},
       eprint = {2210.04685},
 primaryClass = {astro-ph.HE},
       adsurl = {https://ui.adsabs.harvard.edu/abs/2023SCPMA..6629512Z}
}

@ARTICLE{2011AA...529A..75C,
       author = {{Claret}, A. and {Bloemen}, S.},
        title = "{Gravity and limb-darkening coefficients for the Kepler, CoRoT, Spitzer, uvby, UBVRIJHK, and Sloan photometric systems}",
      journal = {\aap},
         year = 2011,
        month = may,
       volume = {529},
          eid = {A75},
        pages = {A75},
          doi = {10.1051/0004-6361/201116451},
       adsurl = {https://ui.adsabs.harvard.edu/abs/2011A&A...529A..75C}
}

@ARTICLE{2017AA...600A..30C,
       author = {{Claret}, A.},
        title = "{Limb and gravity-darkening coefficients for the TESS satellite at several metallicities, surface gravities, and microturbulent velocities}",
      journal = {\aap},
         year = 2017,
        month = apr,
       volume = {600},
          eid = {A30},
        pages = {A30},
          doi = {10.1051/0004-6361/201629705},
archivePrefix = {arXiv},
       eprint = {1804.10295},
 primaryClass = {astro-ph.SR},
       adsurl = {https://ui.adsabs.harvard.edu/abs/2017A&A...600A..30C}
}

@ARTICLE{2022MNRAS.513.2719V,
       author = {{Vines}, Jose I. and {Jenkins}, James S.},
        title = "{ARIADNE: measuring accurate and precise stellar parameters through SED fitting}",
      journal = {\mnras},
         year = 2022,
        month = jun,
       volume = {513},
       number = {2},
        pages = {2719-2731},
          doi = {10.1093/mnras/stac956},
archivePrefix = {arXiv},
       eprint = {2204.03769},
 primaryClass = {astro-ph.SR},
       adsurl = {https://ui.adsabs.harvard.edu/abs/2022MNRAS.513.2719V}
}

@ARTICLE{2023AA...674A...1G,
       author = {{Gaia Collaboration} and {Vallenari}, A. and {Brown}, A.~G.~A. and {Prusti}, T. and {de Bruijne}, J.~H.~J. and {Arenou}, F. and {Babusiaux}, C. and {Biermann}, M. and {Creevey}, O.~L. and {Ducourant}, C. and {Evans}, D.~W. and {Eyer}, L. and {Guerra}, R. and {Hutton}, A. and {Jordi}, C. and {Klioner}, S.~A. and {Lammers}, U.~L. and {Lindegren}, L. and {Luri}, X. and {Mignard}, F. and {Panem}, C. and {Pourbaix}, D. and {Randich}, S. and {Sartoretti}, P. and {Soubiran}, C. and {Tanga}, P. and {Walton}, N.~A. and {Bailer-Jones}, C.~A.~L. and {Bastian}, U. and {Drimmel}, R. and {Jansen}, F. and {Katz}, D. and {Lattanzi}, M.~G. and {van Leeuwen}, F. and {Bakker}, J. and {Cacciari}, C. and {Casta{\~n}eda}, J. and {De Angeli}, F. and {Fabricius}, C. and {Fouesneau}, M. and {Fr{\'e}mat}, Y. and {Galluccio}, L. and {Guerrier}, A. and {Heiter}, U. and {Masana}, E. and {Messineo}, R. and {Mowlavi}, N. and {Nicolas}, C. and {Nienartowicz}, K. and {Pailler}, F. and {Panuzzo}, P. and {Riclet}, F. and {Roux}, W. and {Seabroke}, G.~M. and {Sordo}, R. and {Th{\'e}venin}, F. and {Gracia-Abril}, G. and {Portell}, J. and {Teyssier}, D. and {Altmann}, M. and {Andrae}, R. and {Audard}, M. and {Bellas-Velidis}, I. and {Benson}, K. and {Berthier}, J. and {Blomme}, R. and {Burgess}, P.~W. and {Busonero}, D. and {Busso}, G. and {C{\'a}novas}, H. and {Carry}, B. and {Cellino}, A. and {Cheek}, N. and {Clementini}, G. and {Damerdji}, Y. and {Davidson}, M. and {de Teodoro}, P. and {Nu{\~n}ez Campos}, M. and {Delchambre}, L. and {Dell'Oro}, A. and {Esquej}, P. and {Fern{\'a}ndez-Hern{\'a}ndez}, J. and {Fraile}, E. and {Garabato}, D. and {Garc{\'\i}a-Lario}, P. and {Gosset}, E. and {Haigron}, R. and {Halbwachs}, J. -L. and {Hambly}, N.~C. and {Harrison}, D.~L. and {Hern{\'a}ndez}, J. and {Hestroffer}, D. and {Hodgkin}, S.~T. and {Holl}, B. and {Jan{\ss}en}, K. and {Jevardat de Fombelle}, G. and {Jordan}, S. and {Krone-Martins}, A. and {Lanzafame}, A.~C. and {L{\"o}ffler}, W. and {Marchal}, O. and {Marrese}, P.~M. and {Moitinho}, A. and {Muinonen}, K. and {Osborne}, P. and {Pancino}, E. and {Pauwels}, T. and {Recio-Blanco}, A. and {Reyl{\'e}}, C. and {Riello}, M. and {Rimoldini}, L. and {Roegiers}, T. and {Rybizki}, J. and {Sarro}, L.~M. and {Siopis}, C. and {Smith}, M. and {Sozzetti}, A. and {Utrilla}, E. and {van Leeuwen}, M. and {Abbas}, U. and {{\'A}brah{\'a}m}, P. and {Abreu Aramburu}, A. and {Aerts}, C. and {Aguado}, J.~J. and {Ajaj}, M. and {Aldea-Montero}, F. and {Altavilla}, G. and {{\'A}lvarez}, M.~A. and {Alves}, J. and {Anders}, F. and {Anderson}, R.~I. and {Anglada Varela}, E. and {Antoja}, T. and {Baines}, D. and {Baker}, S.~G. and {Balaguer-N{\'u}{\~n}ez}, L. and {Balbinot}, E. and {Balog}, Z. and {Barache}, C. and {Barbato}, D. and {Barros}, M. and {Barstow}, M.~A. and {Bartolom{\'e}}, S. and {Bassilana}, J. -L. and {Bauchet}, N. and {Becciani}, U. and {Bellazzini}, M. and {Berihuete}, A. and {Bernet}, M. and {Bertone}, S. and {Bianchi}, L. and {Binnenfeld}, A. and {Blanco-Cuaresma}, S. and {Blazere}, A. and {Boch}, T. and {Bombrun}, A. and {Bossini}, D. and {Bouquillon}, S. and {Bragaglia}, A. and {Bramante}, L. and {Breedt}, E. and {Bressan}, A. and {Brouillet}, N. and {Brugaletta}, E. and {Bucciarelli}, B. and {Burlacu}, A. and {Butkevich}, A.~G. and {Buzzi}, R. and {Caffau}, E. and {Cancelliere}, R. and {Cantat-Gaudin}, T. and {Carballo}, R. and {Carlucci}, T. and {Carnerero}, M.~I. and {Carrasco}, J.~M. and {Casamiquela}, L. and {Castellani}, M. and {Castro-Ginard}, A. and {Chaoul}, L. and {Charlot}, P. and {Chemin}, L. and {Chiaramida}, V. and {Chiavassa}, A. and {Chornay}, N. and {Comoretto}, G. and {Contursi}, G. and {Cooper}, W.~J. and {Cornez}, T. and {Cowell}, S. and {Crifo}, F. and {Cropper}, M. and {Crosta}, M. and {Crowley}, C. and {Dafonte}, C. and {Dapergolas}, A. and {David}, M. and {David}, P. and {de Laverny}, P. and {De Luise}, F. and {De March}, R.},
        title = "{Gaia Data Release 3. Summary of the content and survey properties}",
      journal = {\aap},
         year = 2023,
        month = jun,
       volume = {674},
          eid = {A1},
        pages = {A1},
          doi = {10.1051/0004-6361/202243940},
archivePrefix = {arXiv},
       eprint = {2208.00211},
 primaryClass = {astro-ph.GA},
       adsurl = {https://ui.adsabs.harvard.edu/abs/2023A&A...674A...1G}
}

@ARTICLE{2023A&A...675A.199A,
       author = {{Avakyan}, A. and {Neumann}, M. and {Zainab}, A. and {Doroshenko}, V. and {Wilms}, J. and {Santangelo}, A.},
        title = "{XRBcats: Galactic low-mass X-ray binary catalogue}",
      journal = {\aap},
         year = 2023,
        month = jul,
       volume = {675},
          eid = {A199},
        pages = {A199},
          doi = {10.1051/0004-6361/202346522},
archivePrefix = {arXiv},
       eprint = {2303.16168},
 primaryClass = {astro-ph.HE},
       adsurl = {https://ui.adsabs.harvard.edu/abs/2023A&A...675A.199A}
}

@ARTICLE{2023A&A...677A.134N,
       author = {{Neumann}, M. and {Avakyan}, A. and {Doroshenko}, V. and {Santangelo}, A.},
        title = "{XRBcats: Galactic High Mass X-ray Binary Catalogue★}",
      journal = {\aap},
         year = 2023,
        month = sep,
       volume = {677},
          eid = {A134},
        pages = {A134},
          doi = {10.1051/0004-6361/202245728},
archivePrefix = {arXiv},
       eprint = {2303.16137},
 primaryClass = {astro-ph.HE},
       adsurl = {https://ui.adsabs.harvard.edu/abs/2023A&A...677A.134N}
}

@ARTICLE{2022NatAs...6.1085S,
       author = {{Shenar}, Tomer and {Sana}, Hugues and {Mahy}, Laurent and {El-Badry}, Kareem and {Marchant}, Pablo and {Langer}, Norbert and {Hawcroft}, Calum and {Fabry}, Matthias and {Sen}, Koushik and {Almeida}, Leonardo A. and {Abdul-Masih}, Michael and {Bodensteiner}, Julia and {Crowther}, Paul A. and {Gieles}, Mark and {Gromadzki}, Mariusz and {H{\'e}nault-Brunet}, Vincent and {Herrero}, Artemio and {de Koter}, Alex and {Iwanek}, Patryk and {Koz{\l}owski}, Szymon and {Lennon}, Daniel J. and {Ma{\'\i}z Apell{\'a}niz}, Jes{\'u}s and {Mr{\'o}z}, Przemys{\l}aw and {Moffat}, Anthony F.~J. and {Picco}, Annachiara and {Pietrukowicz}, Pawe{\l} and {Poleski}, Rados{\l}aw and {Rybicki}, Krzysztof and {Schneider}, Fabian R.~N. and {Skowron}, Dorota M. and {Skowron}, Jan and {Soszy{\'n}ski}, Igor and {Szyma{\'n}ski}, Micha{\l} K. and {Toonen}, Silvia and {Udalski}, Andrzej and {Ulaczyk}, Krzysztof and {Vink}, Jorick S. and {Wrona}, Marcin},
        title = "{An X-ray-quiet black hole born with a negligible kick in a massive binary within the Large Magellanic Cloud}",
      journal = {Nature Astronomy},
         year = 2022,
        month = jul,
       volume = {6},
        pages = {1085-1092},
          doi = {10.1038/s41550-022-01730-y},
archivePrefix = {arXiv},
       eprint = {2207.07675},
 primaryClass = {astro-ph.HE},
       adsurl = {https://ui.adsabs.harvard.edu/abs/2022NatAs...6.1085S}
}

@ARTICLE{2019ApJ...872L..20G,
       author = {{Gu}, Wei-Min and {Mu}, Hui-Jun and {Fu}, Jin-Bo and {Zheng}, Ling-Lin and {Yi}, Tuan and {Bai}, Zhong-Rui and {Wang}, Song and {Zhang}, Hao-Tong and {Lei}, Ya-Juan and {Bai}, Yu and {Wu}, Jianfeng and {Wang}, Junfeng and {Liu}, Jifeng},
        title = "{A Method to Search for Black Hole Candidates with Giant Companions by LAMOST}",
      journal = {\apjl},
         year = 2019,
        month = feb,
       volume = {872},
       number = {2},
          eid = {L20},
        pages = {L20},
          doi = {10.3847/2041-8213/ab04f0},
archivePrefix = {arXiv},
       eprint = {1902.02813},
 primaryClass = {astro-ph.SR},
       adsurl = {https://ui.adsabs.harvard.edu/abs/2019ApJ...872L..20G}
}

@ARTICLE{2021MNRAS.501.2822G,
       author = {{Gomel}, Roy and {Faigler}, Simchon and {Mazeh}, Tsevi},
        title = "{Search for dormant black holes in ellipsoidal variables I. Revisiting the expected amplitudes of the photometric modulation}",
      journal = {\mnras},
         year = 2021,
        month = feb,
       volume = {501},
       number = {2},
        pages = {2822-2832},
          doi = {10.1093/mnras/staa3305},
archivePrefix = {arXiv},
       eprint = {2008.11209},
 primaryClass = {astro-ph.SR},
       adsurl = {https://ui.adsabs.harvard.edu/abs/2021MNRAS.501.2822G}
}

@ARTICLE{2020AJ....159...81M,
       author = {{Masuda}, Kento and {Winn}, Joshua N.},
        title = "{On the Inference of a Star's Inclination Angle from its Rotation Velocity and Projected Rotation Velocity}",
      journal = {\aj},
         year = 2020,
        month = mar,
       volume = {159},
       number = {3},
          eid = {81},
        pages = {81},
          doi = {10.3847/1538-3881/ab65be},
archivePrefix = {arXiv},
       eprint = {2001.04973},
 primaryClass = {astro-ph.IM},
       adsurl = {https://ui.adsabs.harvard.edu/abs/2020AJ....159...81M}
}

@ARTICLE{2024Ap&SS.369...16L,
       author = {{Ludlam}, Renee M.},
        title = "{Reflecting on accretion in neutron star low-mass X-ray binaries}",
      journal = {\apss},
         year = 2024,
        month = jan,
       volume = {369},
       number = {1},
          eid = {16},
        pages = {16},
          doi = {10.1007/s10509-024-04281-y},
archivePrefix = {arXiv},
       eprint = {2401.15787},
 primaryClass = {astro-ph.HE},
       adsurl = {https://ui.adsabs.harvard.edu/abs/2024Ap&SS.369...16L}
}

@ARTICLE{2002ApJ...565.1107P,
       author = {{Podsiadlowski}, Ph. and {Rappaport}, S. and {Pfahl}, E.~D.},
        title = "{Evolutionary Sequences for Low- and Intermediate-Mass X-Ray Binaries}",
      journal = {\apj},
         year = 2002,
        month = feb,
       volume = {565},
       number = {2},
        pages = {1107-1133},
          doi = {10.1086/324686},
archivePrefix = {arXiv},
       eprint = {astro-ph/0107261},
 primaryClass = {astro-ph},
       adsurl = {https://ui.adsabs.harvard.edu/abs/2002ApJ...565.1107P}
}

@ARTICLE{2019ApJ...878L..38T,
       author = {{Townsley}, Dean M. and {Miles}, Broxton J. and {Shen}, Ken J. and {Kasen}, Daniel},
        title = "{Double Detonations with Thin, Modestly Enriched Helium Layers can Make Normal Type Ia Supernovae}",
      journal = {\apjl},
         year = 2019,
        month = jun,
       volume = {878},
       number = {2},
          eid = {L38},
        pages = {L38},
          doi = {10.3847/2041-8213/ab27cd},
archivePrefix = {arXiv},
       eprint = {1903.10960},
 primaryClass = {astro-ph.SR},
       adsurl = {https://ui.adsabs.harvard.edu/abs/2019ApJ...878L..38T}
}

@ARTICLE{2023Natur.615..605G,
       author = {{Greiner}, J. and {Maitra}, C. and {Haberl}, F. and {Willer}, R. and {Burgess}, J.~M. and {Langer}, N. and {Bodensteiner}, J. and {Buckley}, D.~A.~H. and {Monageng}, I.~M. and {Udalski}, A. and {Ritter}, H. and {Werner}, K. and {Maggi}, P. and {Jayaraman}, R. and {Vanderspek}, R.},
        title = "{A helium-burning white dwarf binary as a supersoft X-ray source}",
      journal = {\nat},
         year = 2023,
        month = mar,
       volume = {615},
       number = {7953},
        pages = {605-609},
          doi = {10.1038/s41586-023-05714-4},
archivePrefix = {arXiv},
       eprint = {2303.13338},
 primaryClass = {astro-ph.HE},
       adsurl = {https://ui.adsabs.harvard.edu/abs/2023Natur.615..605G}
}

@ARTICLE{2020NatMe..17..261V,
       author = {{Virtanen}, Pauli and {Gommers}, Ralf and {Oliphant}, Travis E. and {Haberland}, Matt and {Reddy}, Tyler and {Cournapeau}, David and {Burovski}, Evgeni and {Peterson}, Pearu and {Weckesser}, Warren and {Bright}, Jonathan and {van der Walt}, St{\'e}fan J. and {Brett}, Matthew and {Wilson}, Joshua and {Millman}, K. Jarrod and {Mayorov}, Nikolay and {Nelson}, Andrew R.~J. and {Jones}, Eric and {Kern}, Robert and {Larson}, Eric and {Carey}, C.~J. and {Polat}, {\.I}lhan and {Feng}, Yu and {Moore}, Eric W. and {VanderPlas}, Jake and {Laxalde}, Denis and {Perktold}, Josef and {Cimrman}, Robert and {Henriksen}, Ian and {Quintero}, E.~A. and {Harris}, Charles R. and {Archibald}, Anne M. and {Ribeiro}, Ant{\^o}nio H. and {Pedregosa}, Fabian and {van Mulbregt}, Paul and {SciPy 1. 0 Contributors}},
        title = "{SciPy 1.0: fundamental algorithms for scientific computing in Python}",
      journal = {Nature Methods},
         year = 2020,
        month = feb,
       volume = {17},
        pages = {261-272},
          doi = {10.1038/s41592-019-0686-2},
archivePrefix = {arXiv},
       eprint = {1907.10121},
 primaryClass = {cs.MS},
       adsurl = {https://ui.adsabs.harvard.edu/abs/2020NatMe..17..261V}
}

@ARTICLE{2013PASP..125..306F,
       author = {{Foreman-Mackey}, Daniel and {Hogg}, David W. and {Lang}, Dustin and {Goodman}, Jonathan},
        title = "{emcee: The MCMC Hammer}",
      journal = {\pasp},
         year = 2013,
        month = mar,
       volume = {125},
       number = {925},
        pages = {306},
          doi = {10.1086/670067},
archivePrefix = {arXiv},
       eprint = {1202.3665},
 primaryClass = {astro-ph.IM},
       adsurl = {https://ui.adsabs.harvard.edu/abs/2013PASP..125..306F}
}

@ARTICLE{2018AJ....156..123A,
       author = {{Astropy Collaboration} and {Price-Whelan}, A.~M. and {Sip{\H{o}}cz}, B.~M. and {G{\"u}nther}, H.~M. and {Lim}, P.~L. and {Crawford}, S.~M. and {Conseil}, S. and {Shupe}, D.~L. and {Craig}, M.~W. and {Dencheva}, N. and {Ginsburg}, A. and {VanderPlas}, J.~T. and {Bradley}, L.~D. and {P{\'e}rez-Su{\'a}rez}, D. and {de Val-Borro}, M. and {Aldcroft}, T.~L. and {Cruz}, K.~L. and {Robitaille}, T.~P. and {Tollerud}, E.~J. and {Ardelean}, C. and {Babej}, T. and {Bach}, Y.~P. and {Bachetti}, M. and {Bakanov}, A.~V. and {Bamford}, S.~P. and {Barentsen}, G. and {Barmby}, P. and {Baumbach}, A. and {Berry}, K.~L. and {Biscani}, F. and {Boquien}, M. and {Bostroem}, K.~A. and {Bouma}, L.~G. and {Brammer}, G.~B. and {Bray}, E.~M. and {Breytenbach}, H. and {Buddelmeijer}, H. and {Burke}, D.~J. and {Calderone}, G. and {Cano Rodr{\'\i}guez}, J.~L. and {Cara}, M. and {Cardoso}, J.~V.~M. and {Cheedella}, S. and {Copin}, Y. and {Corrales}, L. and {Crichton}, D. and {D'Avella}, D. and {Deil}, C. and {Depagne}, {\'E}. and {Dietrich}, J.~P. and {Donath}, A. and {Droettboom}, M. and {Earl}, N. and {Erben}, T. and {Fabbro}, S. and {Ferreira}, L.~A. and {Finethy}, T. and {Fox}, R.~T. and {Garrison}, L.~H. and {Gibbons}, S.~L.~J. and {Goldstein}, D.~A. and {Gommers}, R. and {Greco}, J.~P. and {Greenfield}, P. and {Groener}, A.~M. and {Grollier}, F. and {Hagen}, A. and {Hirst}, P. and {Homeier}, D. and {Horton}, A.~J. and {Hosseinzadeh}, G. and {Hu}, L. and {Hunkeler}, J.~S. and {Ivezi{\'c}}, {\v{Z}}. and {Jain}, A. and {Jenness}, T. and {Kanarek}, G. and {Kendrew}, S. and {Kern}, N.~S. and {Kerzendorf}, W.~E. and {Khvalko}, A. and {King}, J. and {Kirkby}, D. and {Kulkarni}, A.~M. and {Kumar}, A. and {Lee}, A. and {Lenz}, D. and {Littlefair}, S.~P. and {Ma}, Z. and {Macleod}, D.~M. and {Mastropietro}, M. and {McCully}, C. and {Montagnac}, S. and {Morris}, B.~M. and {Mueller}, M. and {Mumford}, S.~J. and {Muna}, D. and {Murphy}, N.~A. and {Nelson}, S. and {Nguyen}, G.~H. and {Ninan}, J.~P. and {N{\"o}the}, M. and {Ogaz}, S. and {Oh}, S. and {Parejko}, J.~K. and {Parley}, N. and {Pascual}, S. and {Patil}, R. and {Patil}, A.~A. and {Plunkett}, A.~L. and {Prochaska}, J.~X. and {Rastogi}, T. and {Reddy Janga}, V. and {Sabater}, J. and {Sakurikar}, P. and {Seifert}, M. and {Sherbert}, L.~E. and {Sherwood-Taylor}, H. and {Shih}, A.~Y. and {Sick}, J. and {Silbiger}, M.~T. and {Singanamalla}, S. and {Singer}, L.~P. and {Sladen}, P.~H. and {Sooley}, K.~A. and {Sornarajah}, S. and {Streicher}, O. and {Teuben}, P. and {Thomas}, S.~W. and {Tremblay}, G.~R. and {Turner}, J.~E.~H. and {Terr{\'o}n}, V. and {van Kerkwijk}, M.~H. and {de la Vega}, A. and {Watkins}, L.~L. and {Weaver}, B.~A. and {Whitmore}, J.~B. and {Woillez}, J. and {Zabalza}, V. and {Astropy Contributors}},
        title = "{The Astropy Project: Building an Open-science Project and Status of the v2.0 Core Package}",
      journal = {\aj},
         year = 2018,
        month = sep,
       volume = {156},
       number = {3},
          eid = {123},
        pages = {123},
          doi = {10.3847/1538-3881/aabc4f},
archivePrefix = {arXiv},
       eprint = {1801.02634},
 primaryClass = {astro-ph.IM},
       adsurl = {https://ui.adsabs.harvard.edu/abs/2018AJ....156..123A}
}

@ARTICLE{2013A&A...558A..33A,
       author = {{Astropy Collaboration} and {Robitaille}, Thomas P. and {Tollerud}, Erik J. and {Greenfield}, Perry and {Droettboom}, Michael and {Bray}, Erik and {Aldcroft}, Tom and {Davis}, Matt and {Ginsburg}, Adam and {Price-Whelan}, Adrian M. and {Kerzendorf}, Wolfgang E. and {Conley}, Alexander and {Crighton}, Neil and {Barbary}, Kyle and {Muna}, Demitri and {Ferguson}, Henry and {Grollier}, Fr{\'e}d{\'e}ric and {Parikh}, Madhura M. and {Nair}, Prasanth H. and {Unther}, Hans M. and {Deil}, Christoph and {Woillez}, Julien and {Conseil}, Simon and {Kramer}, Roban and {Turner}, James E.~H. and {Singer}, Leo and {Fox}, Ryan and {Weaver}, Benjamin A. and {Zabalza}, Victor and {Edwards}, Zachary I. and {Azalee Bostroem}, K. and {Burke}, D.~J. and {Casey}, Andrew R. and {Crawford}, Steven M. and {Dencheva}, Nadia and {Ely}, Justin and {Jenness}, Tim and {Labrie}, Kathleen and {Lim}, Pey Lian and {Pierfederici}, Francesco and {Pontzen}, Andrew and {Ptak}, Andy and {Refsdal}, Brian and {Servillat}, Mathieu and {Streicher}, Ole},
        title = "{Astropy: A community Python package for astronomy}",
      journal = {\aap},
         year = 2013,
        month = oct,
       volume = {558},
          eid = {A33},
        pages = {A33},
          doi = {10.1051/0004-6361/201322068},
archivePrefix = {arXiv},
       eprint = {1307.6212},
 primaryClass = {astro-ph.IM},
       adsurl = {https://ui.adsabs.harvard.edu/abs/2013A&A...558A..33A}
}

@ARTICLE{2023A&A...674A...8F,
       author = {{Fr{\'e}mat}, Y. and {Royer}, F. and {Marchal}, O. and {Blomme}, R. and {Sartoretti}, P. and {Guerrier}, A. and {Panuzzo}, P. and {Katz}, D. and {Seabroke}, G.~M. and {Th{\'e}venin}, F. and {Cropper}, M. and {Benson}, K. and {Damerdji}, Y. and {Haigron}, R. and {Lobel}, A. and {Smith}, M. and {Baker}, S.~G. and {Chemin}, L. and {David}, M. and {Dolding}, C. and {Gosset}, E. and {Jan{\ss}en}, K. and {Jasniewicz}, G. and {Plum}, G. and {Samaras}, N. and {Snaith}, O. and {Soubiran}, C. and {Vanel}, O. and {Zorec}, J. and {Zwitter}, T. and {Brouillet}, N. and {Caffau}, E. and {Crifo}, F. and {Fabre}, C. and {Fragkoudi}, F. and {Huckle}, H.~E. and {Lasne}, Y. and {Leclerc}, N. and {Mastrobuono-Battisti}, A. and {Jean-Antoine Piccolo}, A. and {Viala}, Y.},
        title = "{Gaia Data Release 3. Properties of the line-broadening parameter derived with the Radial Velocity Spectrometer (RVS)}",
      journal = {\aap},
         year = 2023,
        month = jun,
       volume = {674},
          eid = {A8},
        pages = {A8},
          doi = {10.1051/0004-6361/202243809},
archivePrefix = {arXiv},
       eprint = {2206.10986},
 primaryClass = {astro-ph.SR},
       adsurl = {https://ui.adsabs.harvard.edu/abs/2023A&A...674A...8F}
}

@software{stellarSpecModel2025,
  author       = {Zhang, Zhixiang},
  title        = {zhang-zhixiang/stellarSpecModel: stellarSpecModel
                   release
                  },
  month        = mar,
  year         = 2025,
  publisher    = {Zenodo},
  version      = {v1.0.0},
  doi          = {10.5281/zenodo.15109587},
  url          = {https://doi.org/10.5281/zenodo.15109587},
  swhid        = {swh:1:dir:16c37119e5f7c2eb18a4c17d5fd49655658a31fc
                   ;origin=https://doi.org/10.5281/zenodo.15109586;vi
                   sit=swh:1:snp:b0422bac54352f0670435603beccaa55bcbb
                   3b1c;anchor=swh:1:rel:8ae122b78007860c8ef386f08a73
                   7e7a60927af6;path=zhang-zhixiang-
                   stellarSpecModel-9748279
                  },
}

@ARTICLE{2008A&A...486..951G,
       author = {{Gustafsson}, B. and {Edvardsson}, B. and {Eriksson}, K. and {J{\o}rgensen}, U.~G. and {Nordlund}, {\r{A}}. and {Plez}, B.},
        title = "{A grid of MARCS model atmospheres for late-type stars. I. Methods and general properties}",
      journal = {\aap},
         year = 2008,
        month = aug,
       volume = {486},
       number = {3},
        pages = {951-970},
          doi = {10.1051/0004-6361:200809724},
archivePrefix = {arXiv},
       eprint = {0805.0554},
 primaryClass = {astro-ph},
       adsurl = {https://ui.adsabs.harvard.edu/abs/2008A&A...486..951G}
}

@ARTICLE{2001ApJ...556..357A,
       author = {{Allard}, France and {Hauschildt}, Peter H. and {Alexander}, David R. and {Tamanai}, Akemi and {Schweitzer}, Andreas},
        title = "{The Limiting Effects of Dust in Brown Dwarf Model Atmospheres}",
      journal = {\apj},
         year = 2001,
        month = jul,
       volume = {556},
       number = {1},
        pages = {357-372},
          doi = {10.1086/321547},
archivePrefix = {arXiv},
       eprint = {astro-ph/0104256},
 primaryClass = {astro-ph},
       adsurl = {https://ui.adsabs.harvard.edu/abs/2001ApJ...556..357A}
}

@ARTICLE{2016ApJS..222....8D,
       author = {{Dotter}, Aaron},
        title = "{MESA Isochrones and Stellar Tracks (MIST) 0: Methods for the Construction of Stellar Isochrones}",
      journal = {\apjs},
         year = 2016,
        month = jan,
       volume = {222},
       number = {1},
          eid = {8},
        pages = {8},
          doi = {10.3847/0067-0049/222/1/8},
archivePrefix = {arXiv},
       eprint = {1601.05144},
 primaryClass = {astro-ph.SR},
       adsurl = {https://ui.adsabs.harvard.edu/abs/2016ApJS..222....8D}
}

\appendix
\section{Spectral Fitting for the Sample}\label{specfit}
\begin{figure}[ht!]
    \centering
    \includegraphics[width=0.473\linewidth]{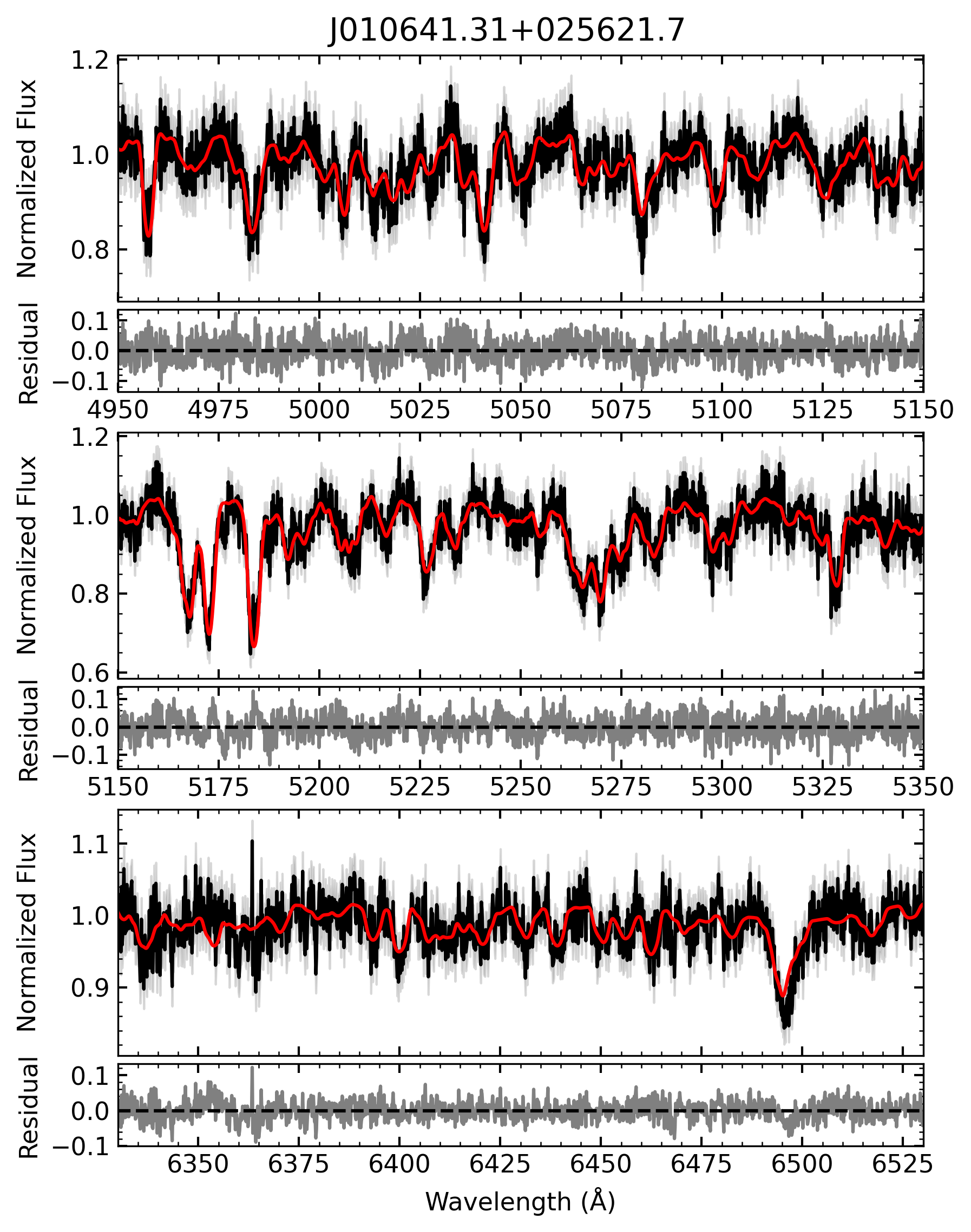}
    \includegraphics[width=0.49\linewidth]{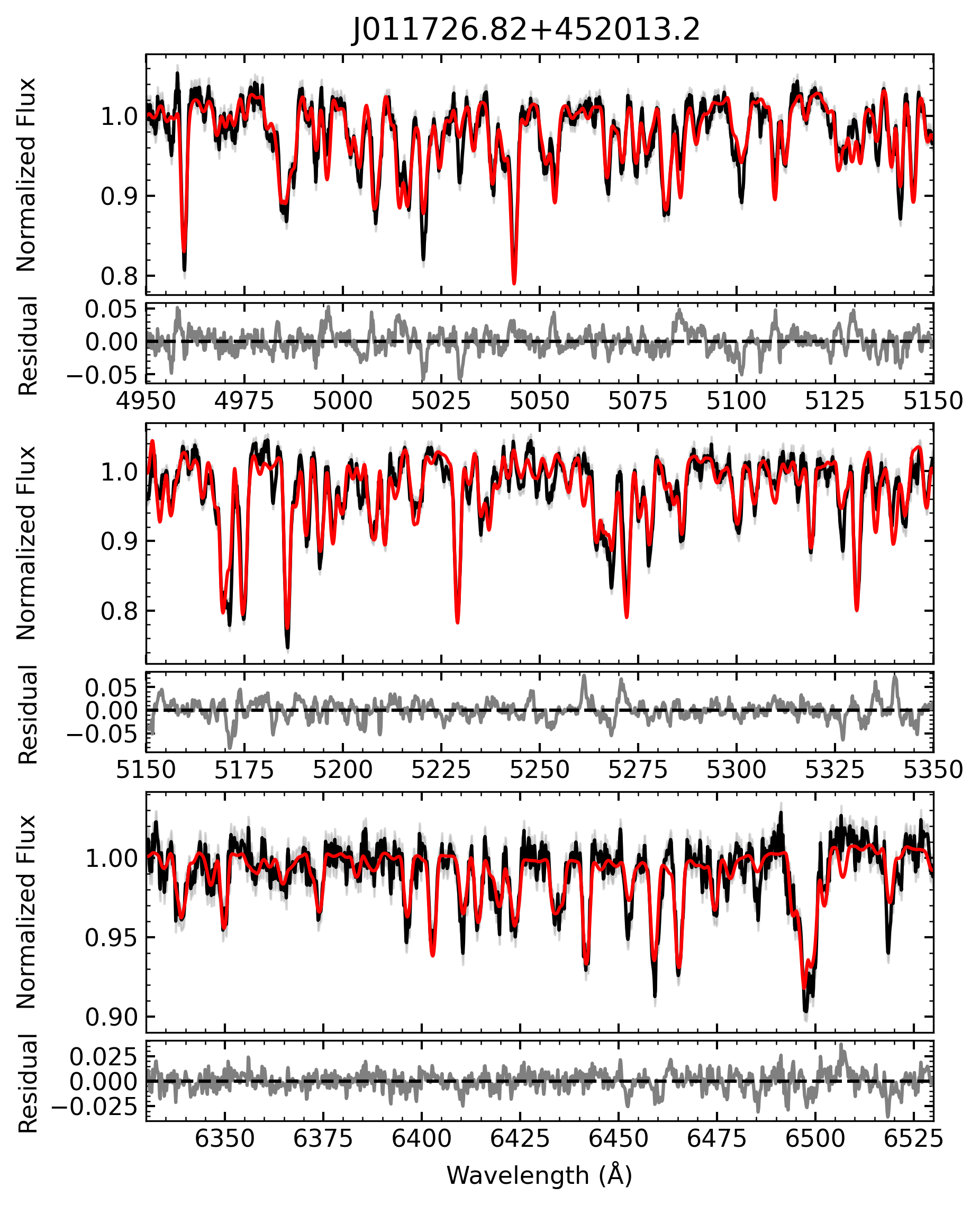}
    \caption{Spectral fitting results for J0106 and J0117. }
    \label{fig:fit_page1}
\end{figure}

\begin{figure}
    \centering
    \includegraphics[width=0.47\linewidth]{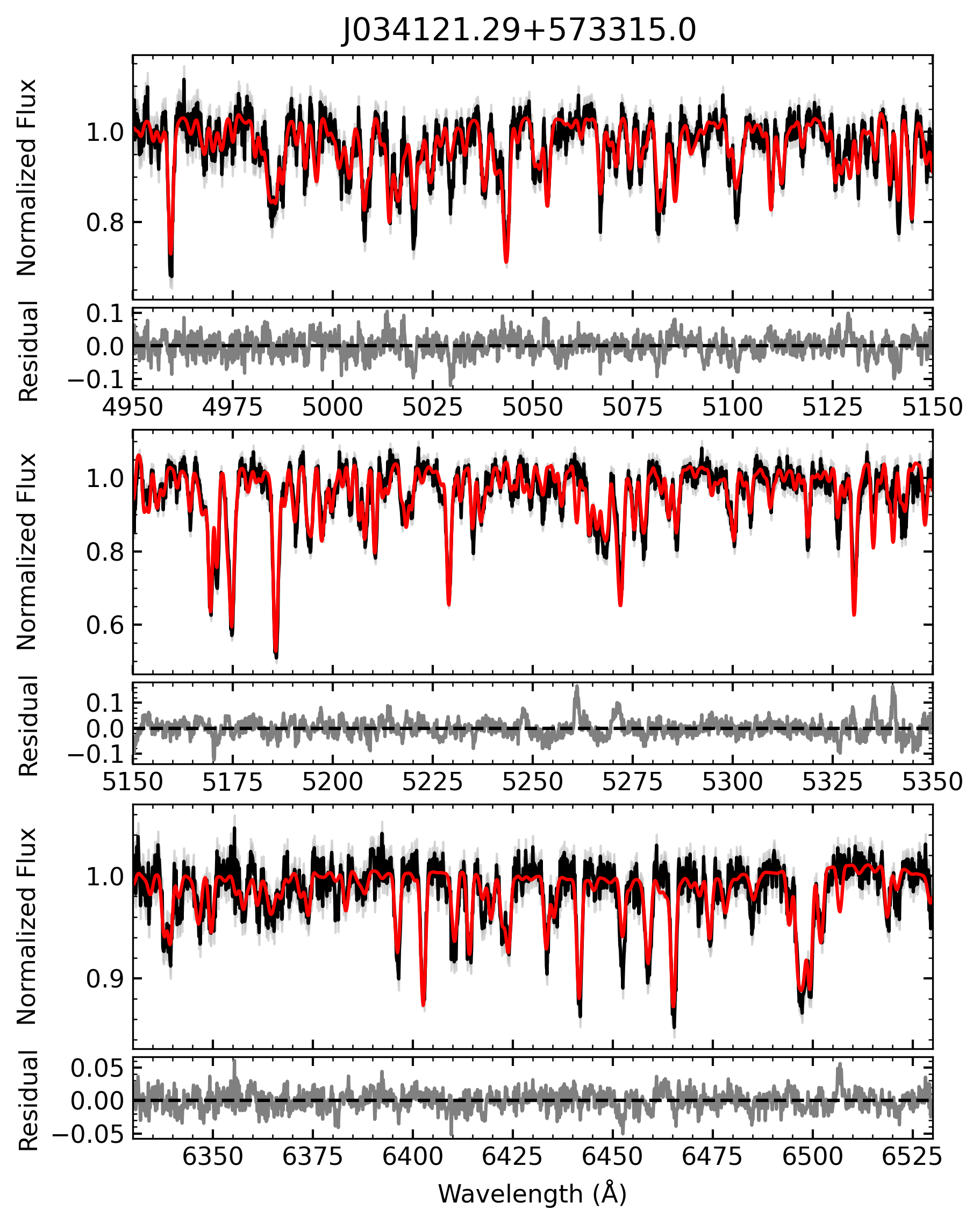}
    \includegraphics[width=0.48\linewidth]{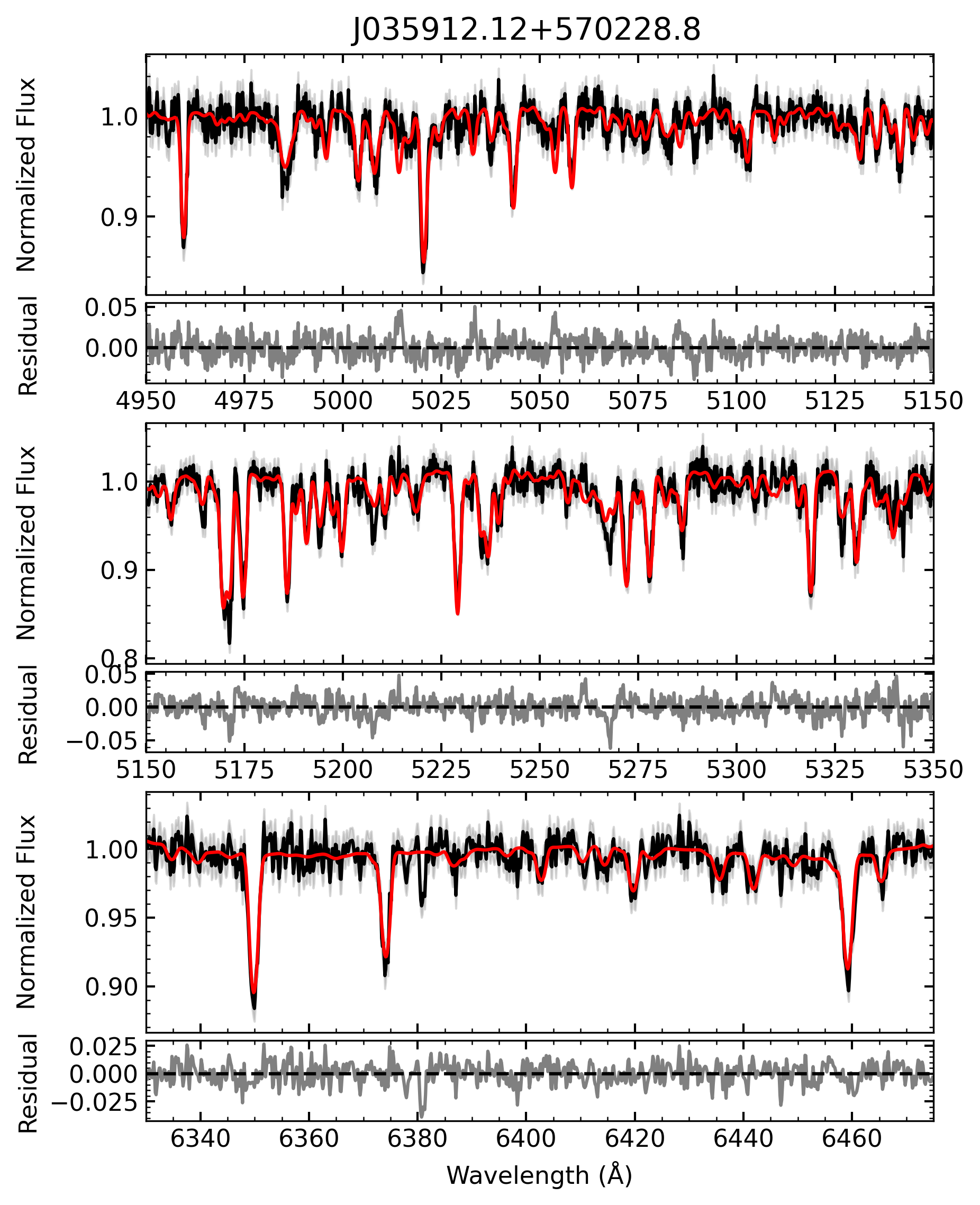}
    \includegraphics[width=0.485\linewidth]{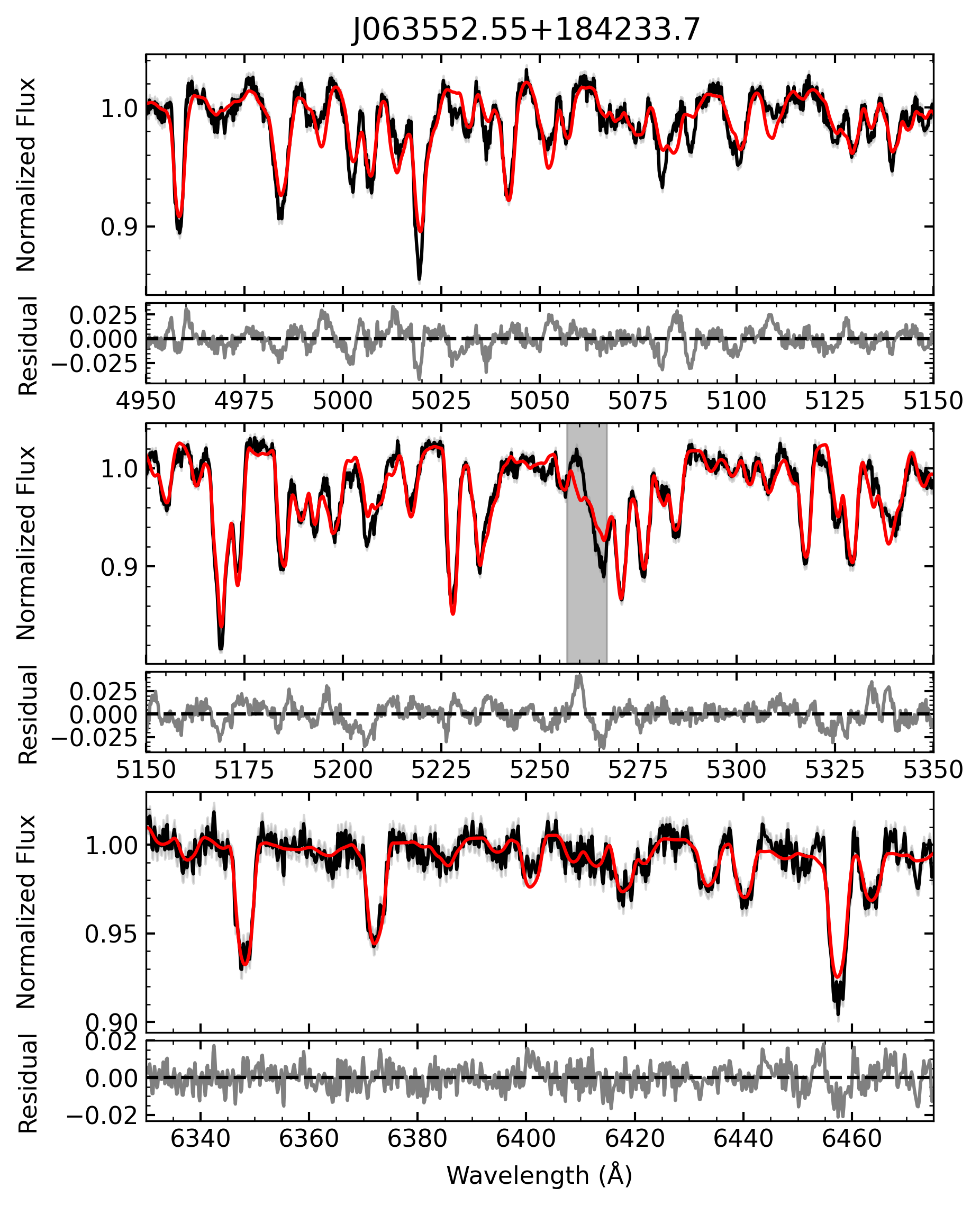}
    \includegraphics[width=0.47\linewidth]{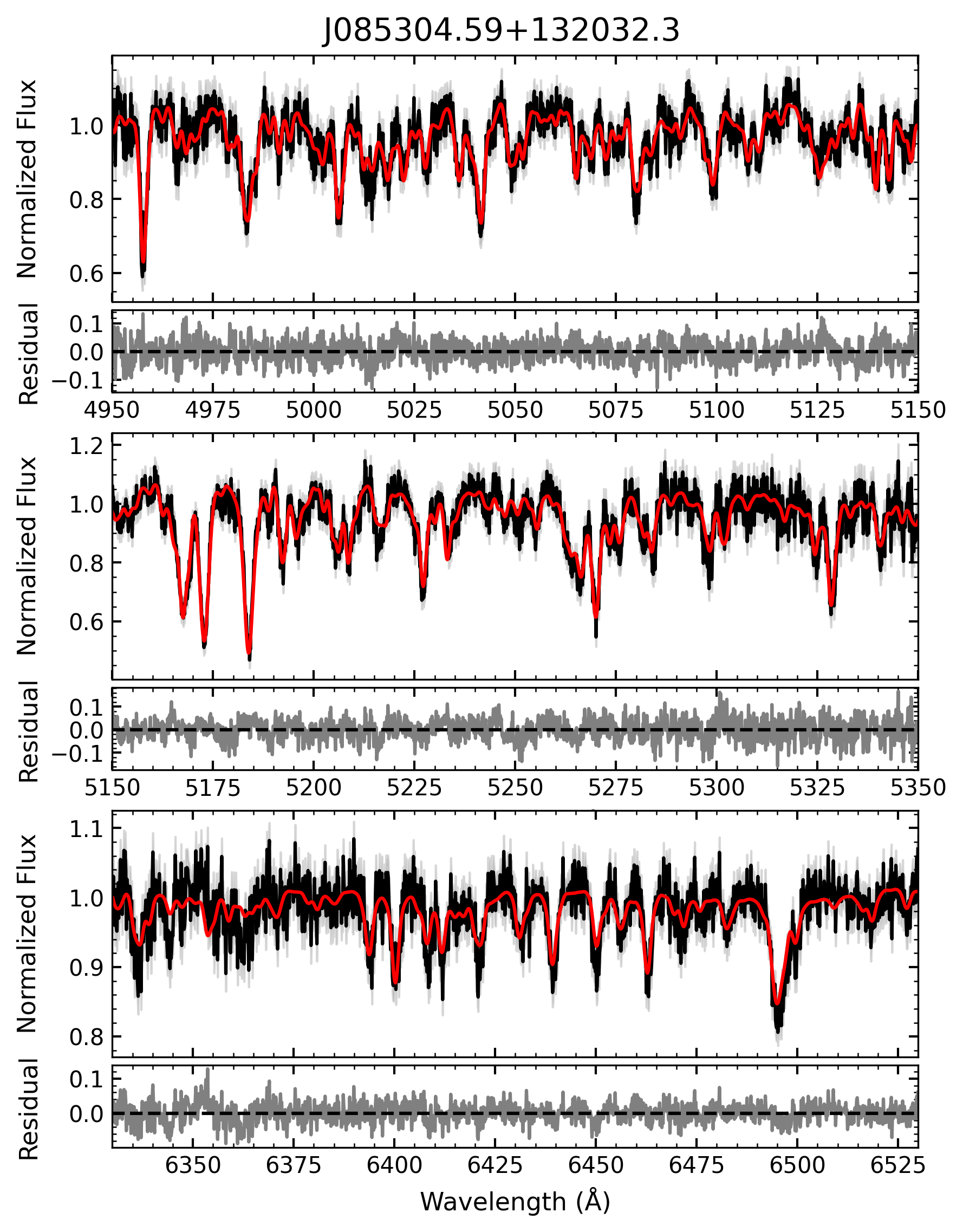}
    \caption{Spectral fitting results for J0341, J0359, J0635 and J0853.}
    \label{fig:fit_page2}
\end{figure}

\begin{figure}
    \centering
    \includegraphics[width=0.49\linewidth]{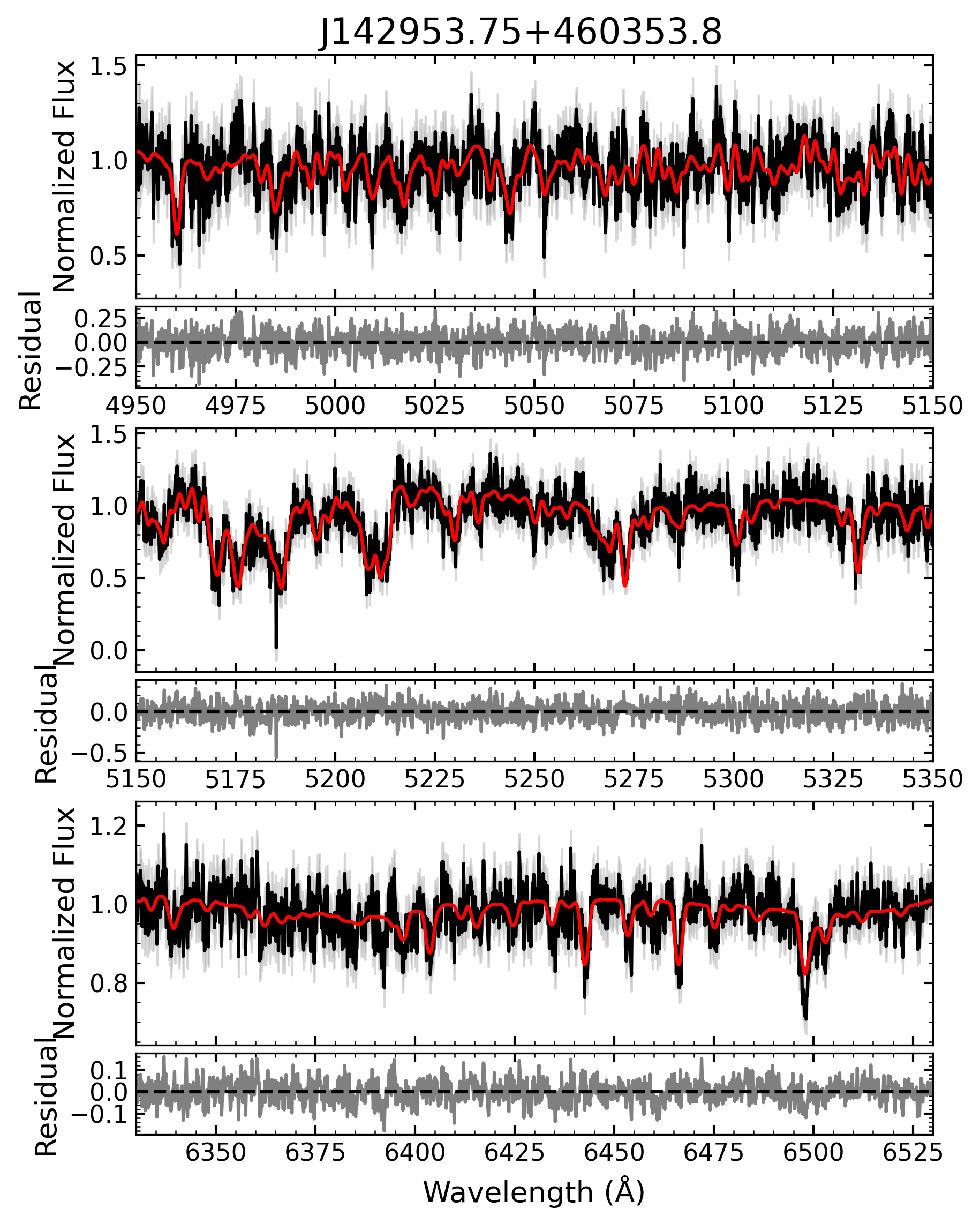}
    \includegraphics[width=0.48\linewidth]{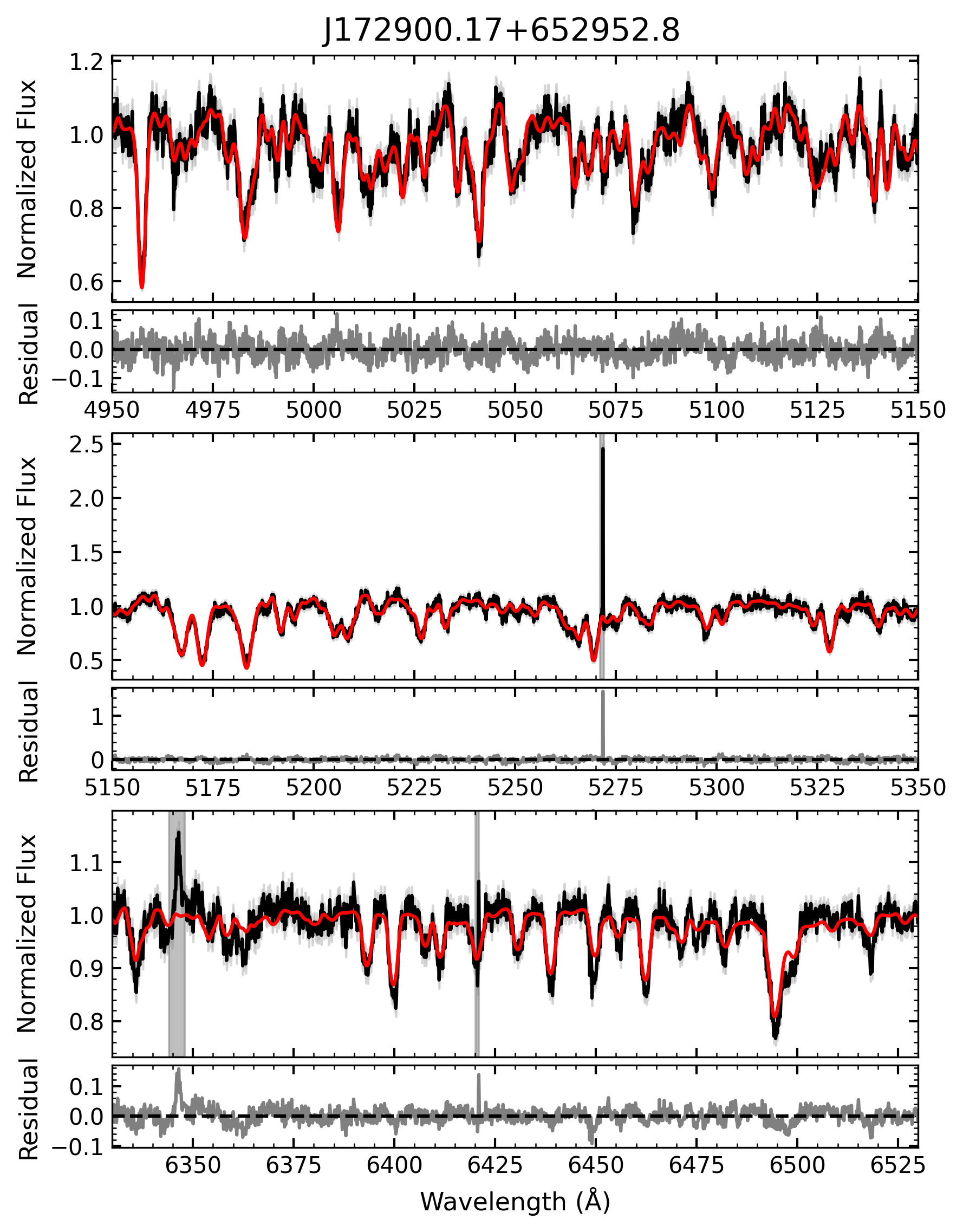}
    \includegraphics[width=0.49\linewidth]{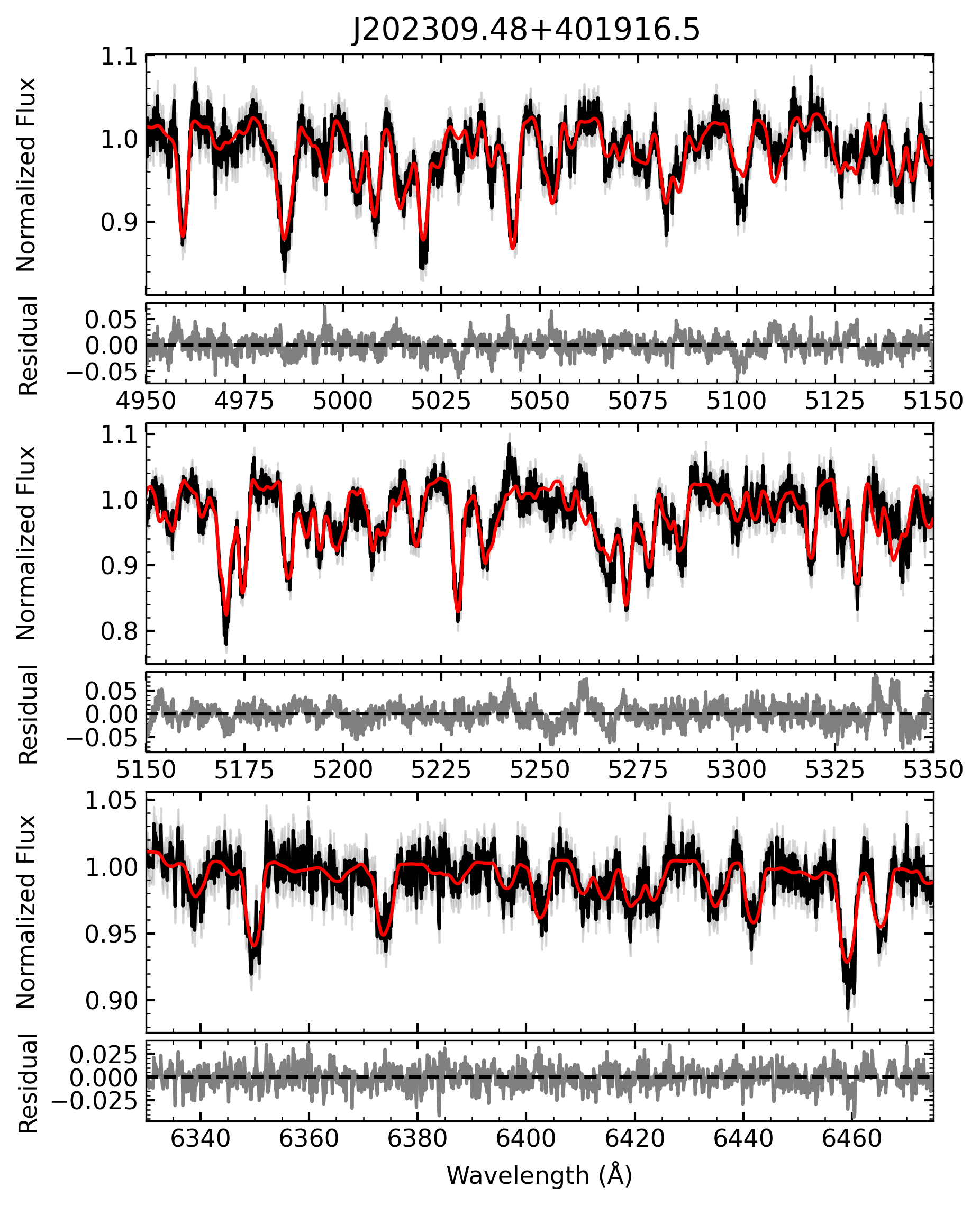}
    \includegraphics[width=0.47\linewidth]{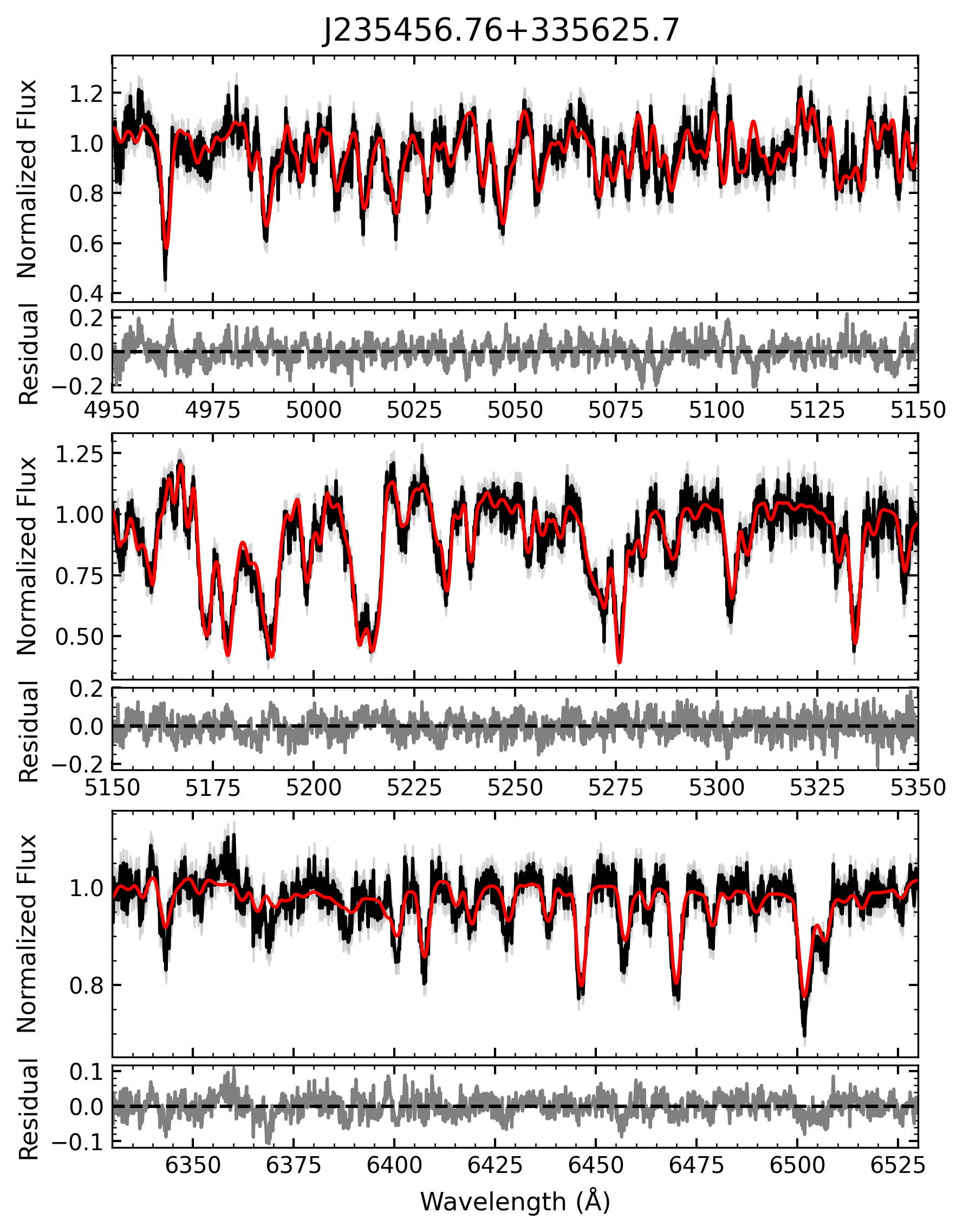}
    \caption{Spectral fitting results for J1429, J1729, J2023 and J2354.}
    \label{fig:fit_page3}
\end{figure}

Figures~\ref{fig:fit_page1}–\ref{fig:fit_page3} present the spectral fitting results for the 10 compact object candidates in our sample. Each figure shows the LAMOST MRS blue and red spectra fitted simultaneously with MARCS/BT-Cond model templates, including rotational and instrumental broadening. Observed spectra are shown in black, best-fit models in red, and residuals in gray. Shaded gray regions indicate masked wavelength ranges excluded from the fitting. These fits are used to determine $T_{\text{spec}}$, $v\sin i$  and Limb as listed in Table~\ref{tab:parameter}.


\clearpage
\section{Posterior Distributions from Spectral Fitting for the Sample}\label{corner_spec}
The posterior distributions of fitting parameters for all targets are presented in this section. These distributions were obtained from the MCMC sampling described in Section~\ref{measure} and reflect the full posterior probability space explored for each source. The median value and the uncertainties corresponding to the 16th–84th percentile ranges of the posterior distributions are shown in each histogram.
\begin{figure}[ht!]
    \centering
    \includegraphics[width=0.49\linewidth]{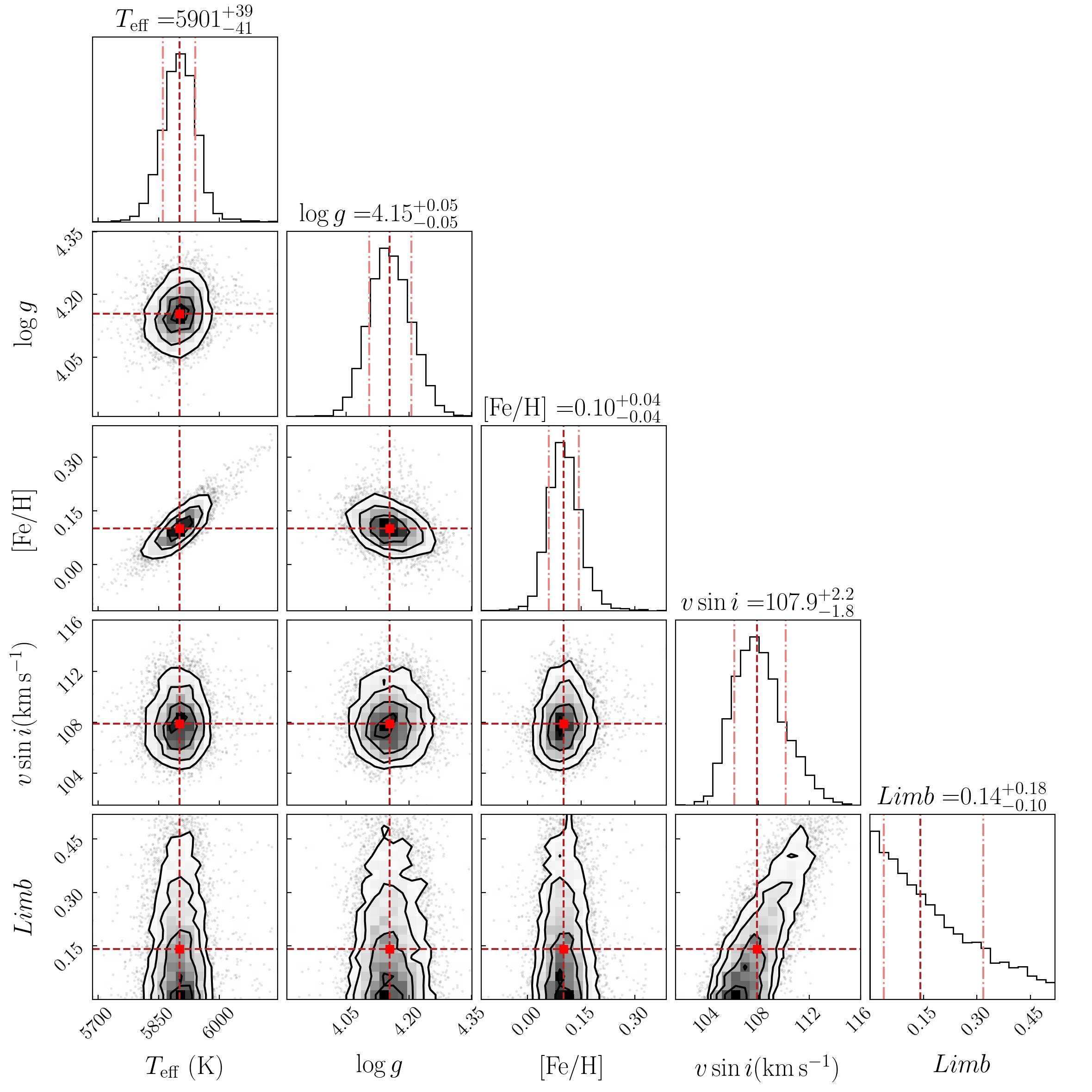}
    \includegraphics[width=0.49\linewidth]{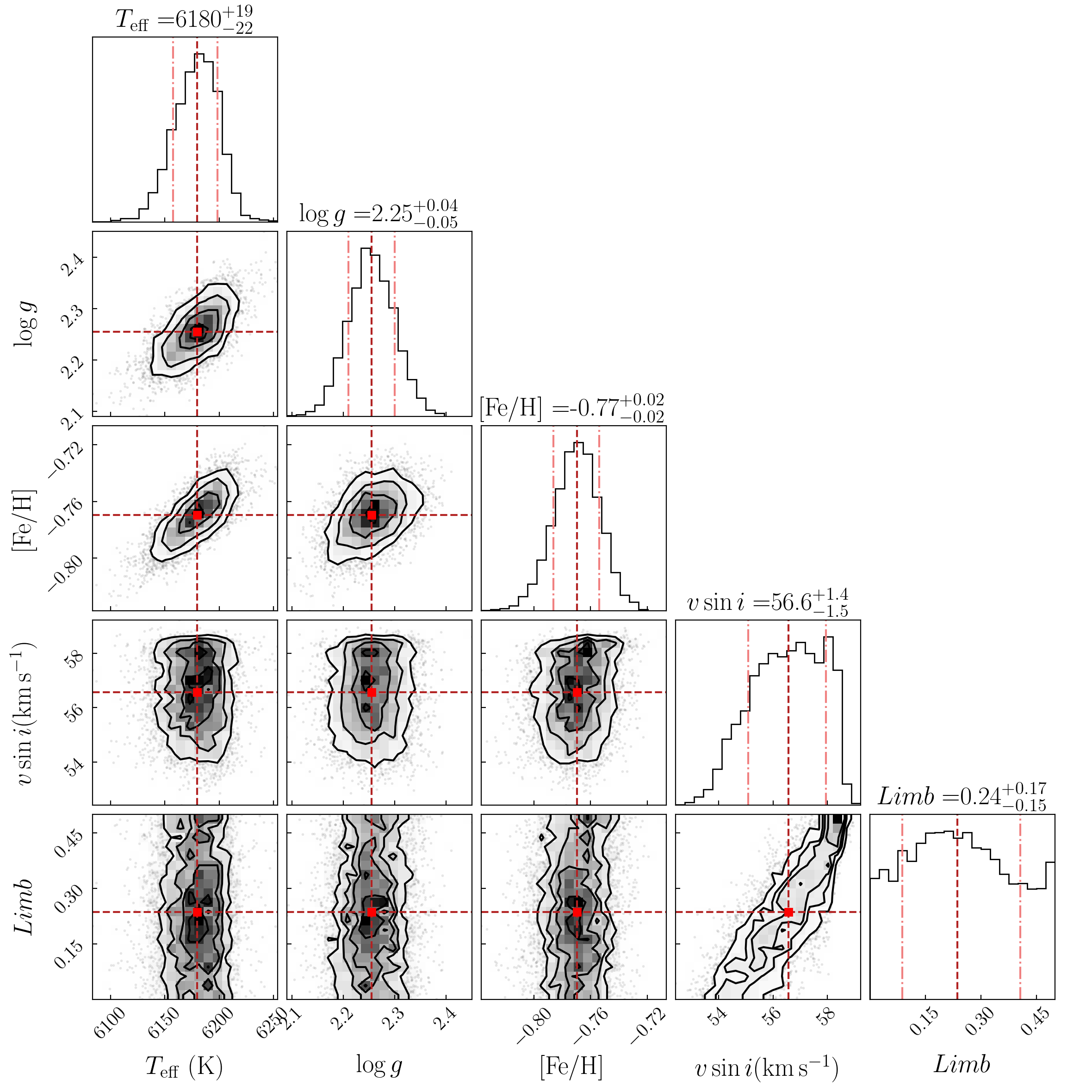}
    \caption{The posterior distributions of $v \sin i$ for J0106 and J0117. }
    \label{fig_dis_1}
\end{figure}

\begin{figure}[ht!]
    \centering
    \includegraphics[width=0.49\linewidth]{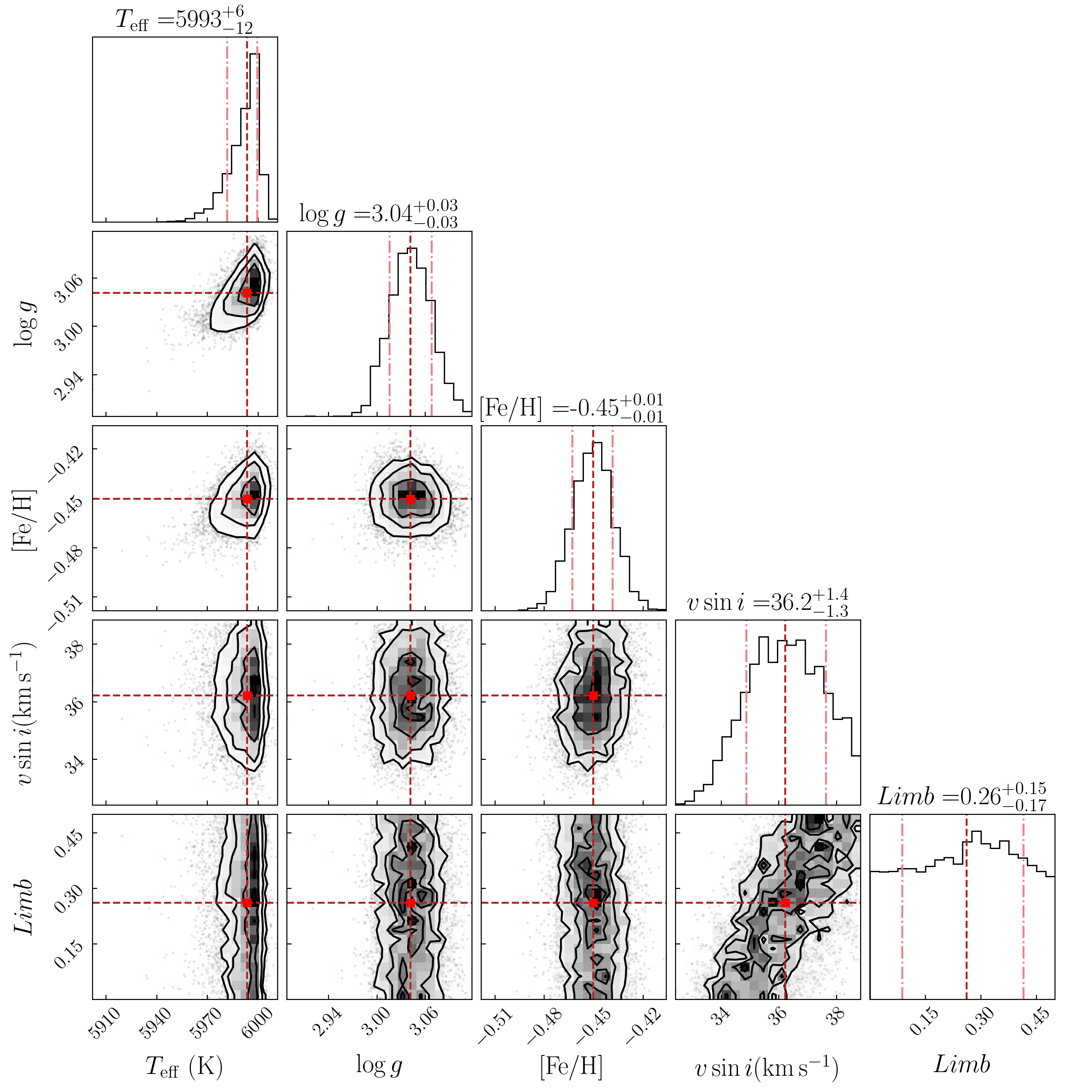}
    \includegraphics[width=0.49\linewidth]{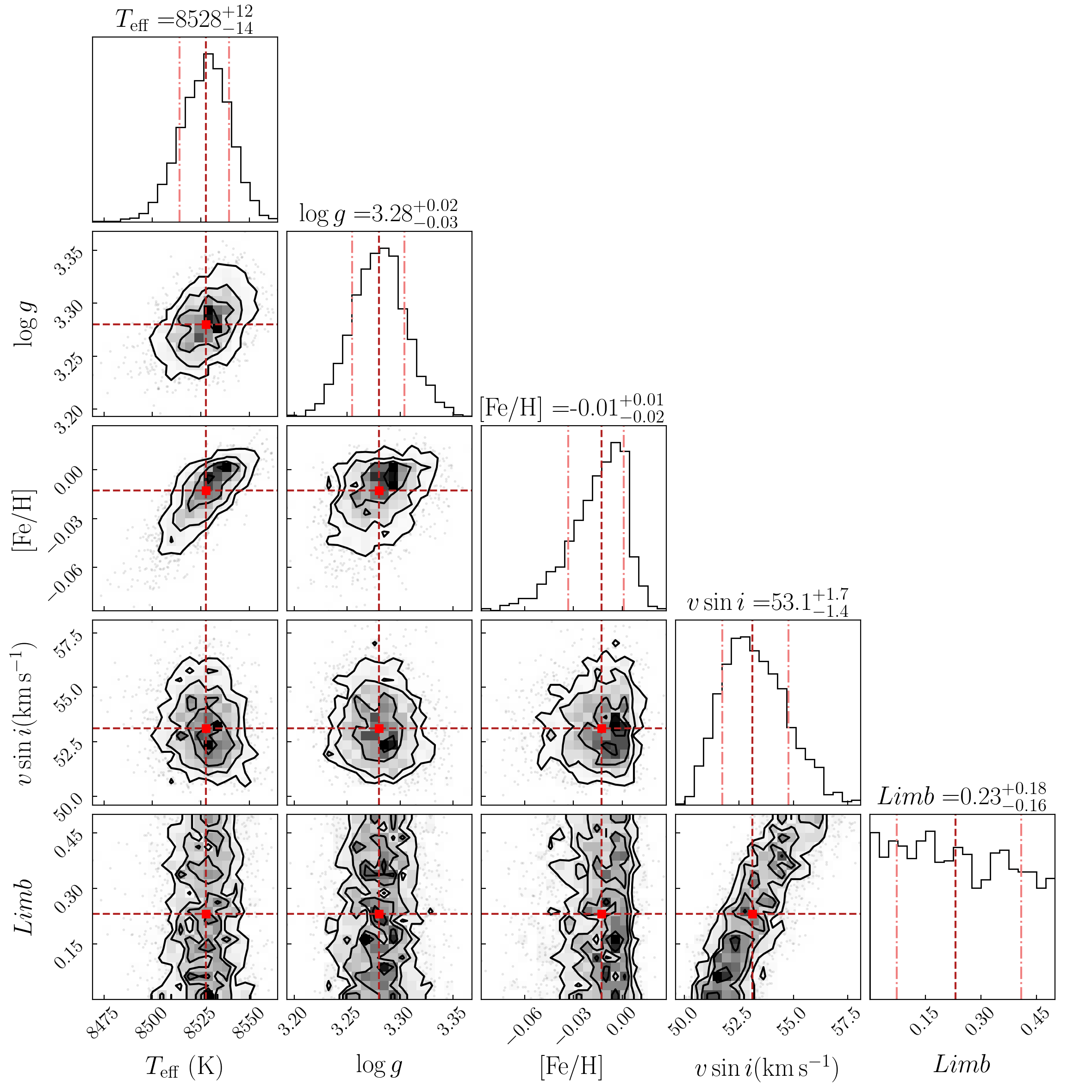}
    \caption{The posterior distributions of $v \sin i$ for J0341 and J0359.}
    \label{fig_dis_2}
\end{figure}

\begin{figure}[ht!]
    \centering
    \includegraphics[width=0.49\linewidth]{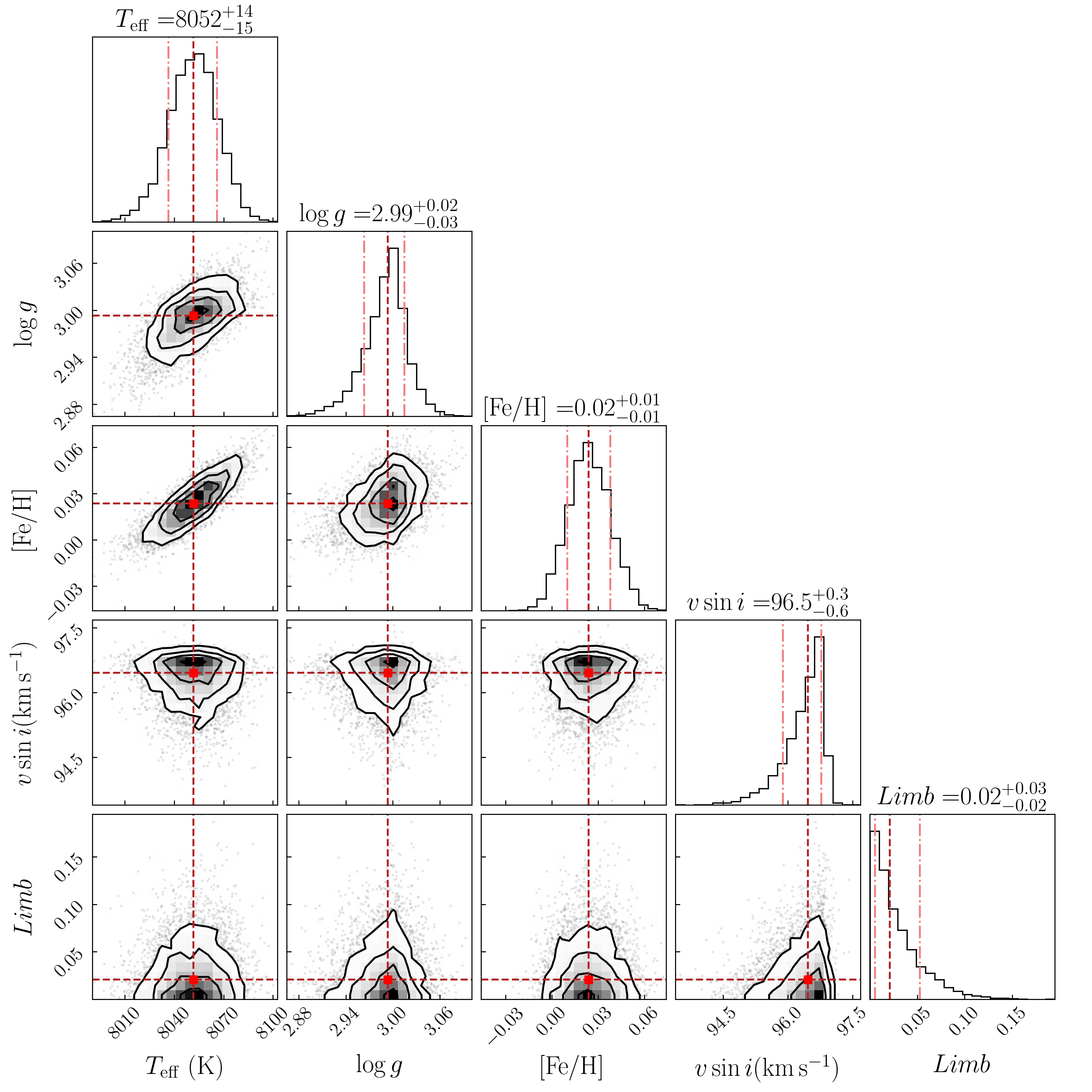}
    \includegraphics[width=0.49\linewidth]{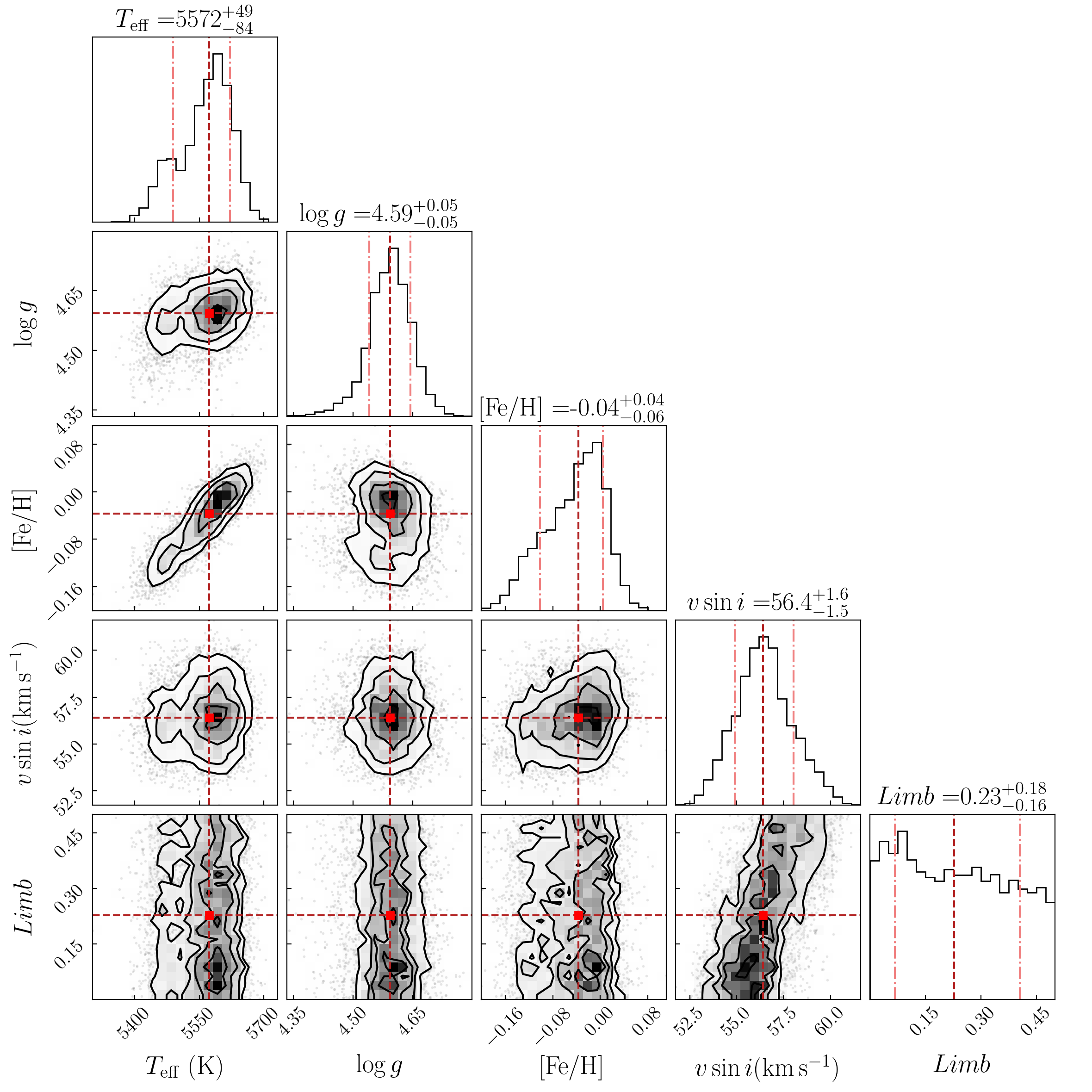}
    \caption{The posterior distributions of $v \sin i$ for J0635 and J0853.}
    \label{fig_dis_3}
\end{figure}

\begin{figure}[ht!]
    \centering
    \includegraphics[width=0.49\linewidth]{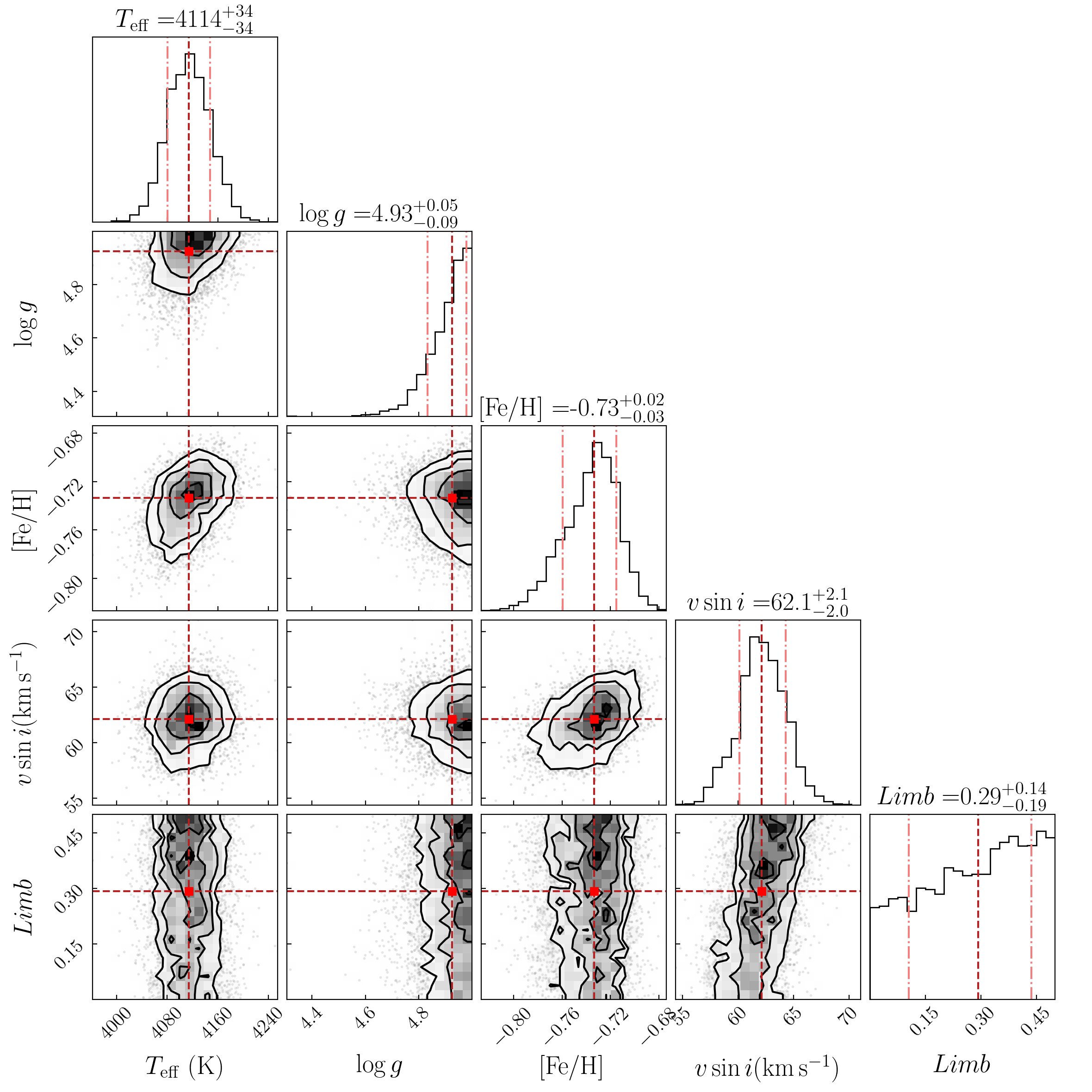}
    \includegraphics[width=0.49\linewidth]{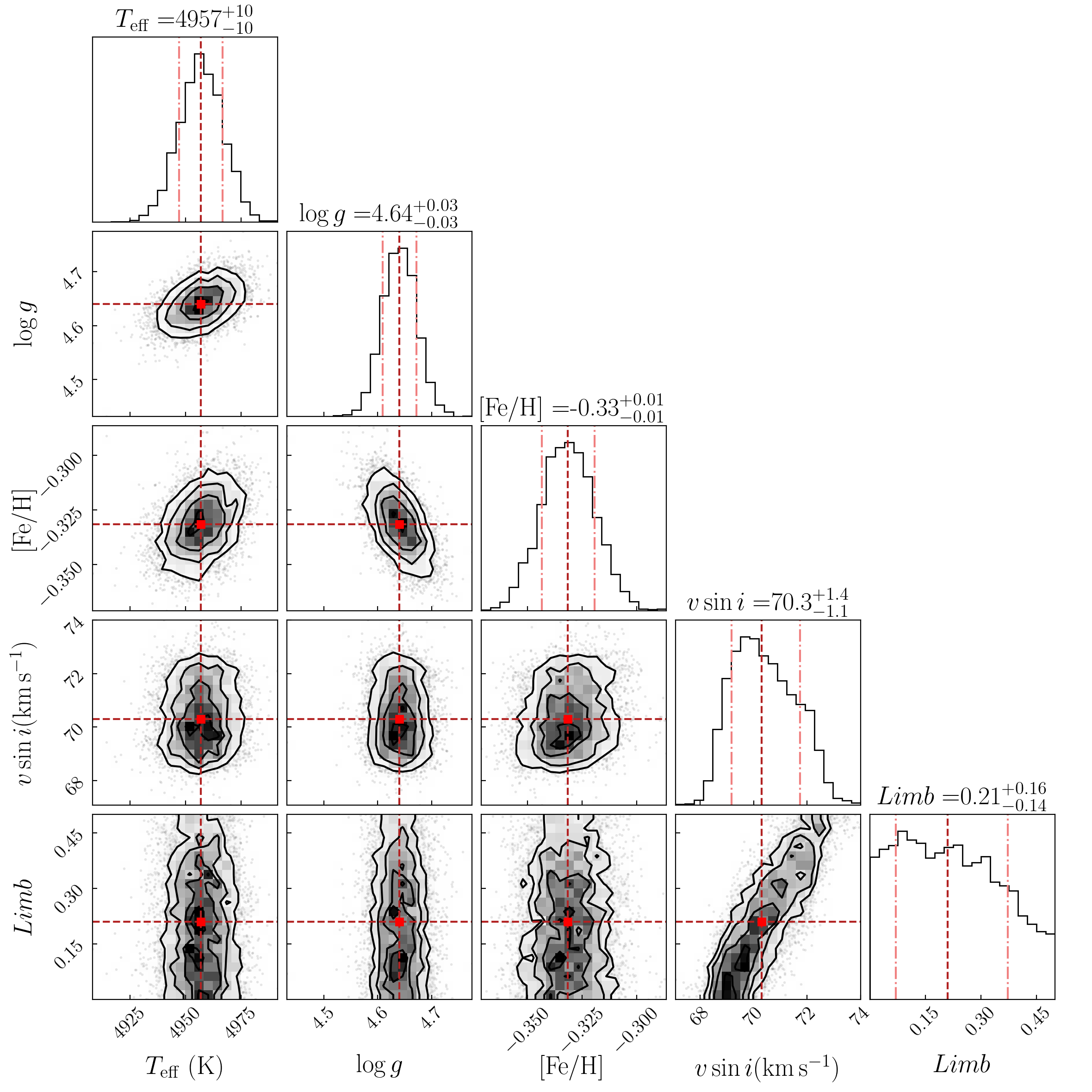}
    \caption{The posterior distributions of $v \sin i$ for J1429 and J1729.}
    \label{fig_dis_4}
\end{figure}

\begin{figure}[ht!]
    \centering
    \includegraphics[width=0.49\linewidth]{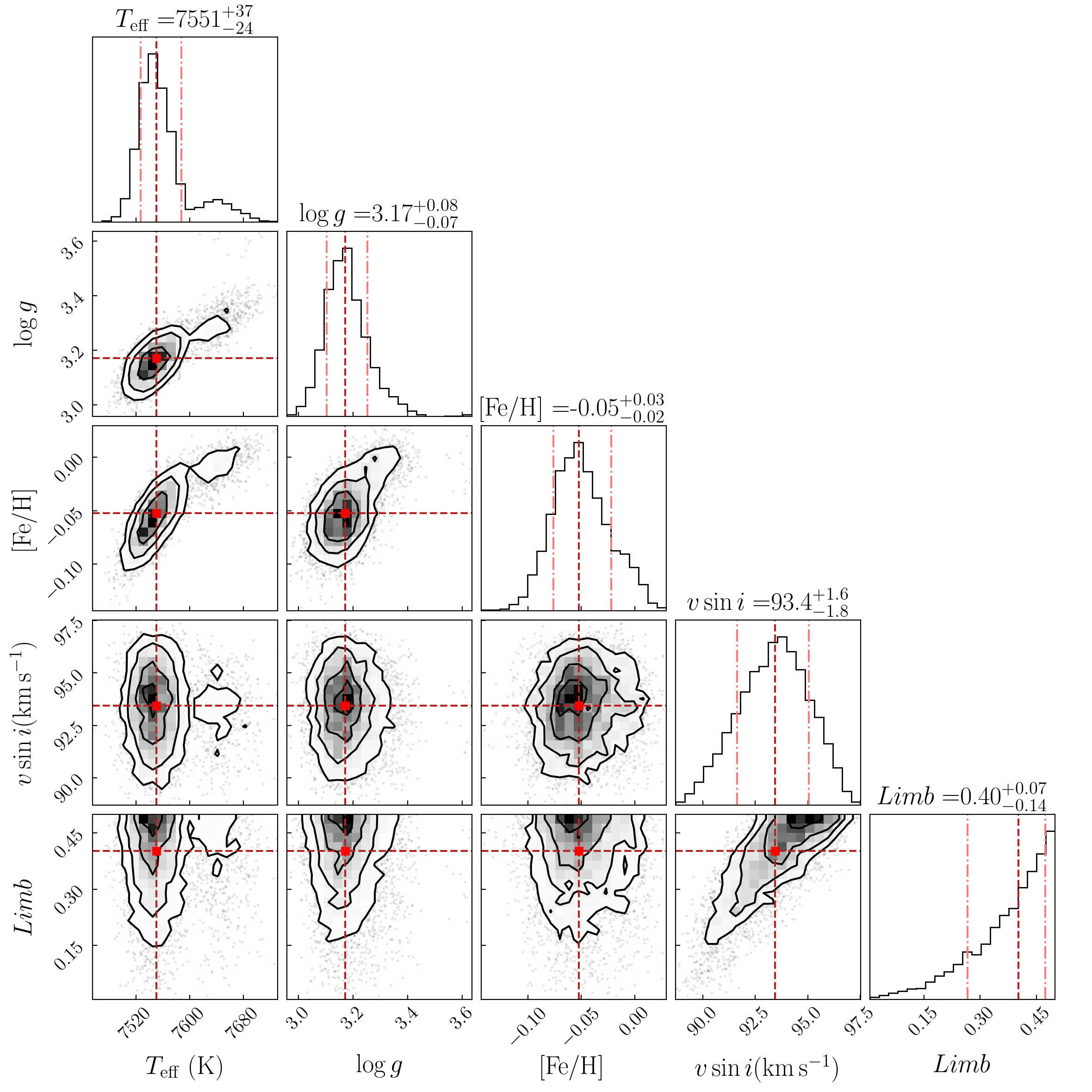}
    \includegraphics[width=0.49\linewidth]{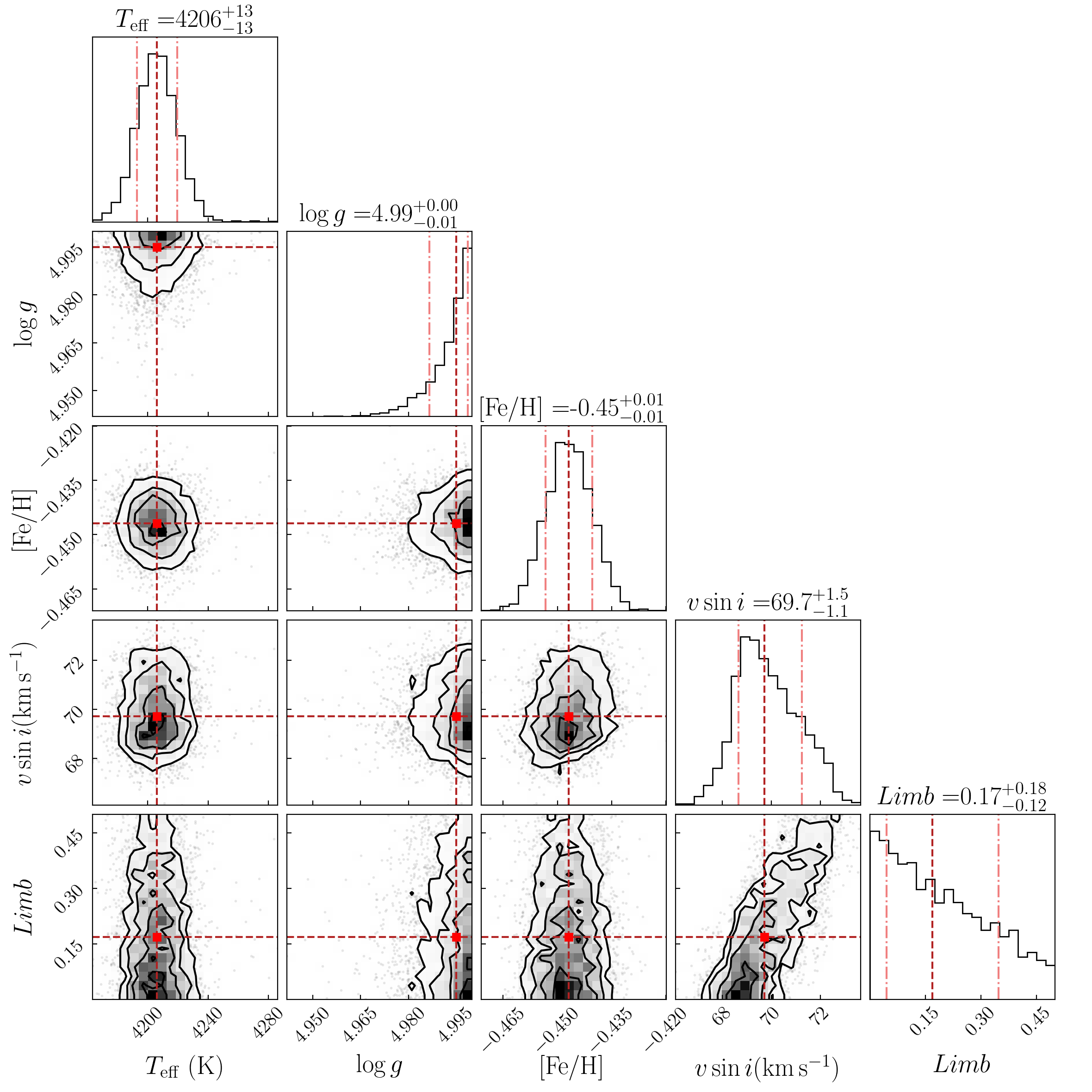}
    \caption{The posterior distributions of $v \sin i$ for J2023 and J2354.}
    \label{fig_dis_5}
\end{figure}

\clearpage
\section{Spectral Energy Distributions for the Sample}\label{SEDfit}
\begin{figure}[ht!]
    \centering
    \includegraphics[width=0.45\linewidth]{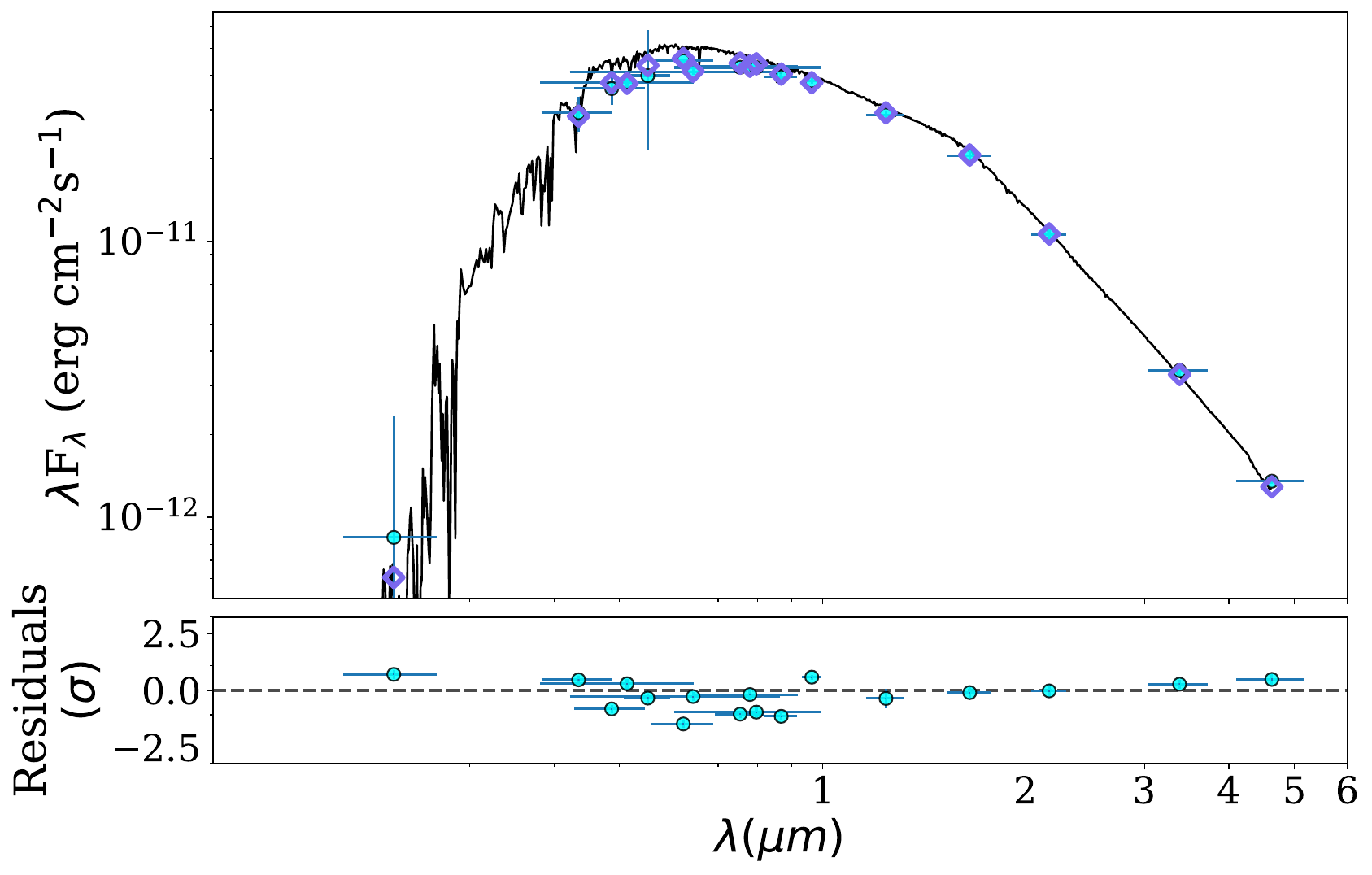}
    \includegraphics[width=0.45\linewidth]{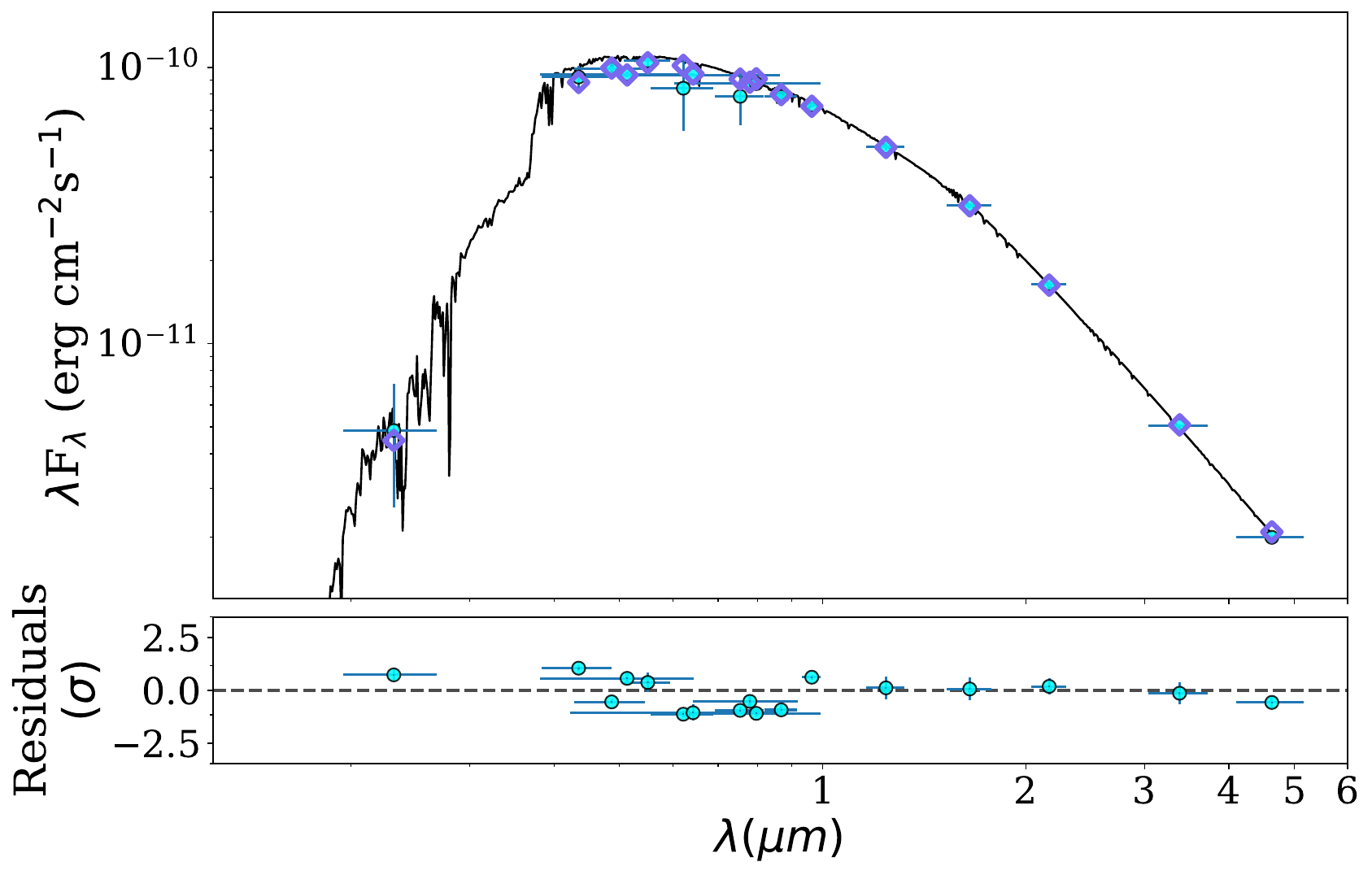}
    \caption{Spectral Energy Distributions for J0106 and J0117. }
    \label{fig:SED1}
\end{figure}

\begin{figure}[ht!]
    \centering
    \includegraphics[width=0.45\linewidth]{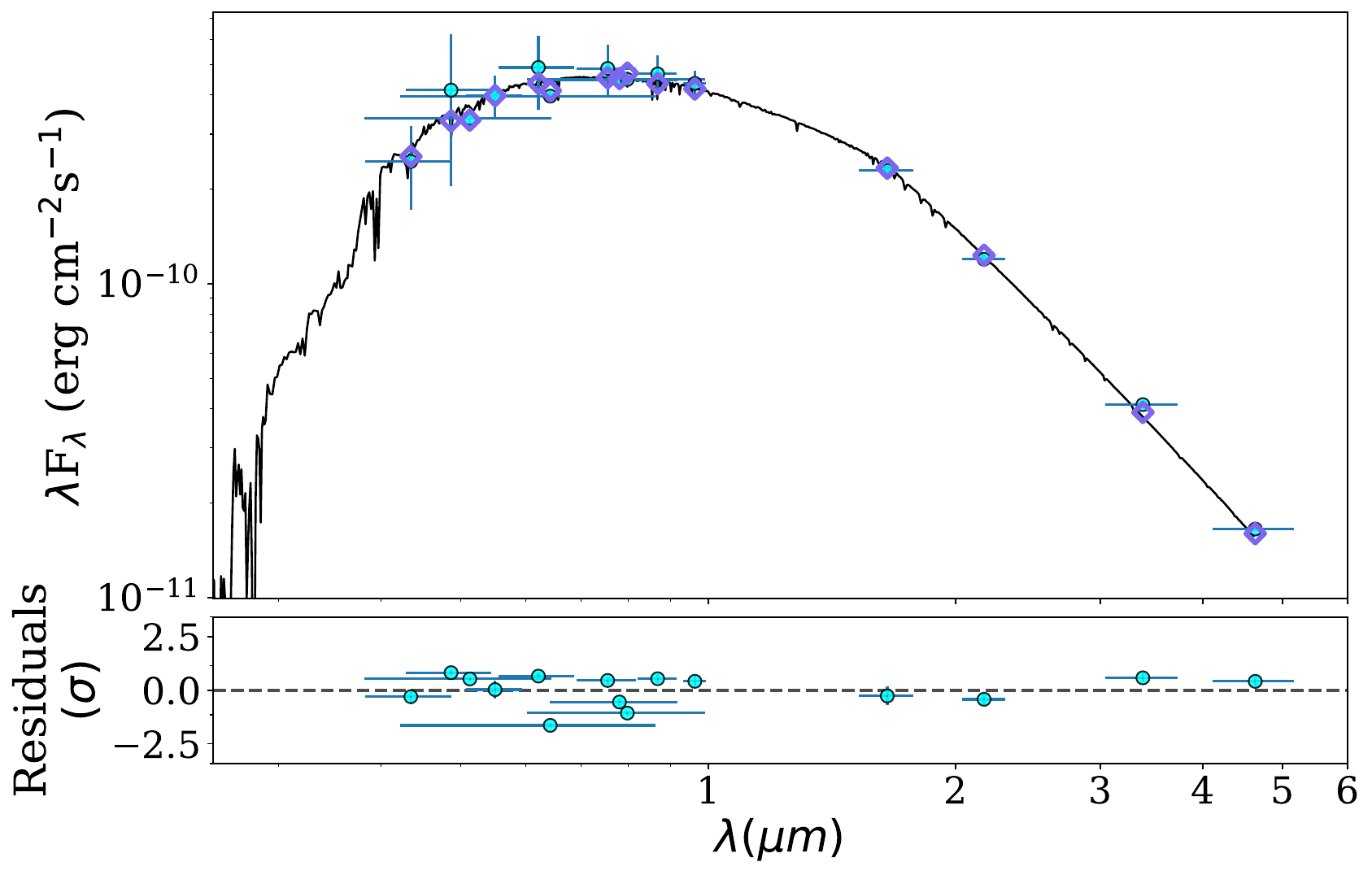}
    \includegraphics[width=0.45\linewidth]{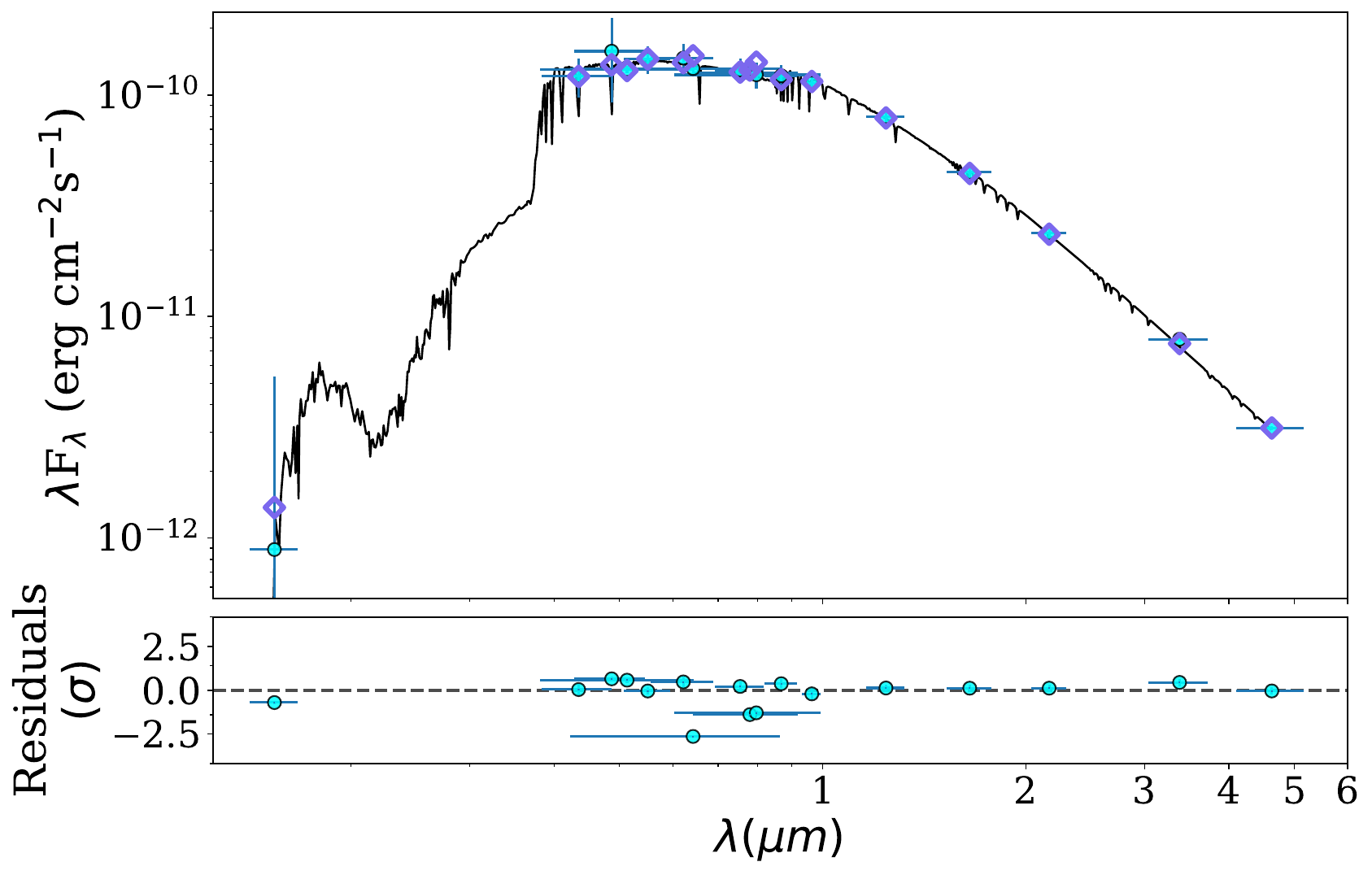}
    \caption{Spectral Energy Distributions for J0341 and J0359. }
    \label{fig:SED2}
\end{figure}

\begin{figure}[ht!]
    \centering
    \includegraphics[width=0.45\linewidth]{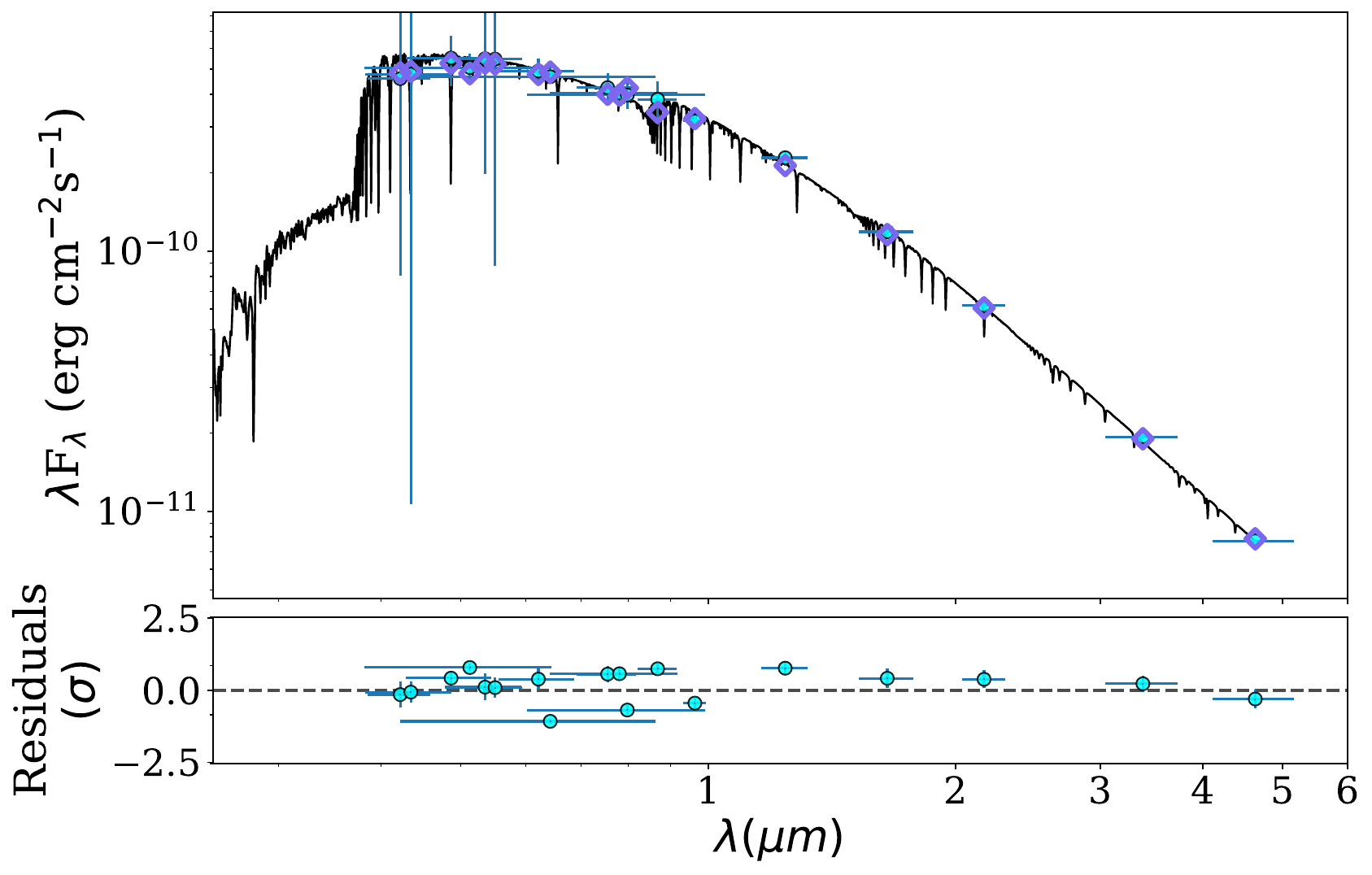}
    \includegraphics[width=0.45\linewidth]{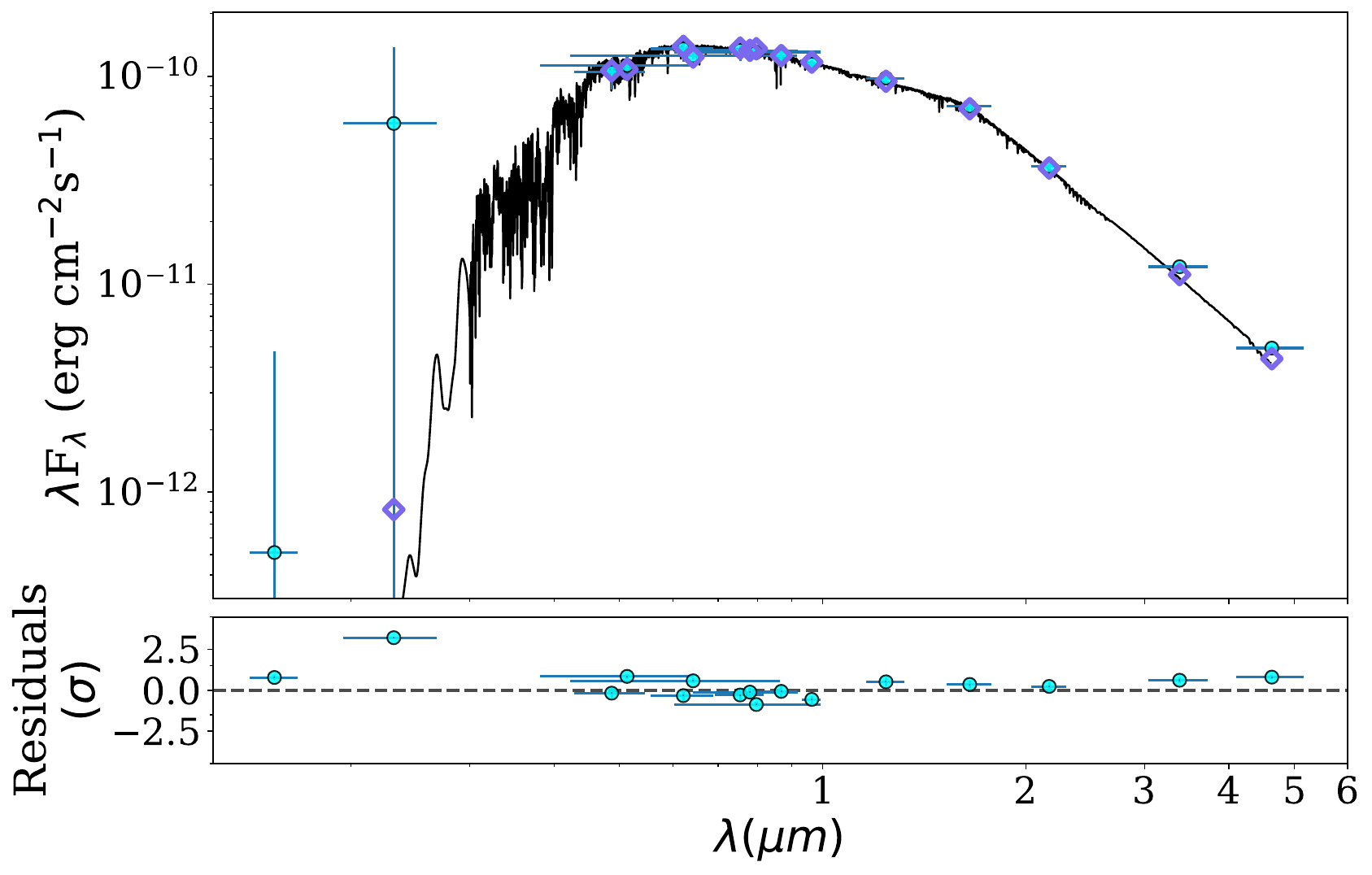}
    \caption{Spectral Energy Distributions for J0635 and J0853. }
    \label{fig:SED3}
\end{figure}

\begin{figure}[ht!]
    \centering
    \includegraphics[width=0.45\linewidth]{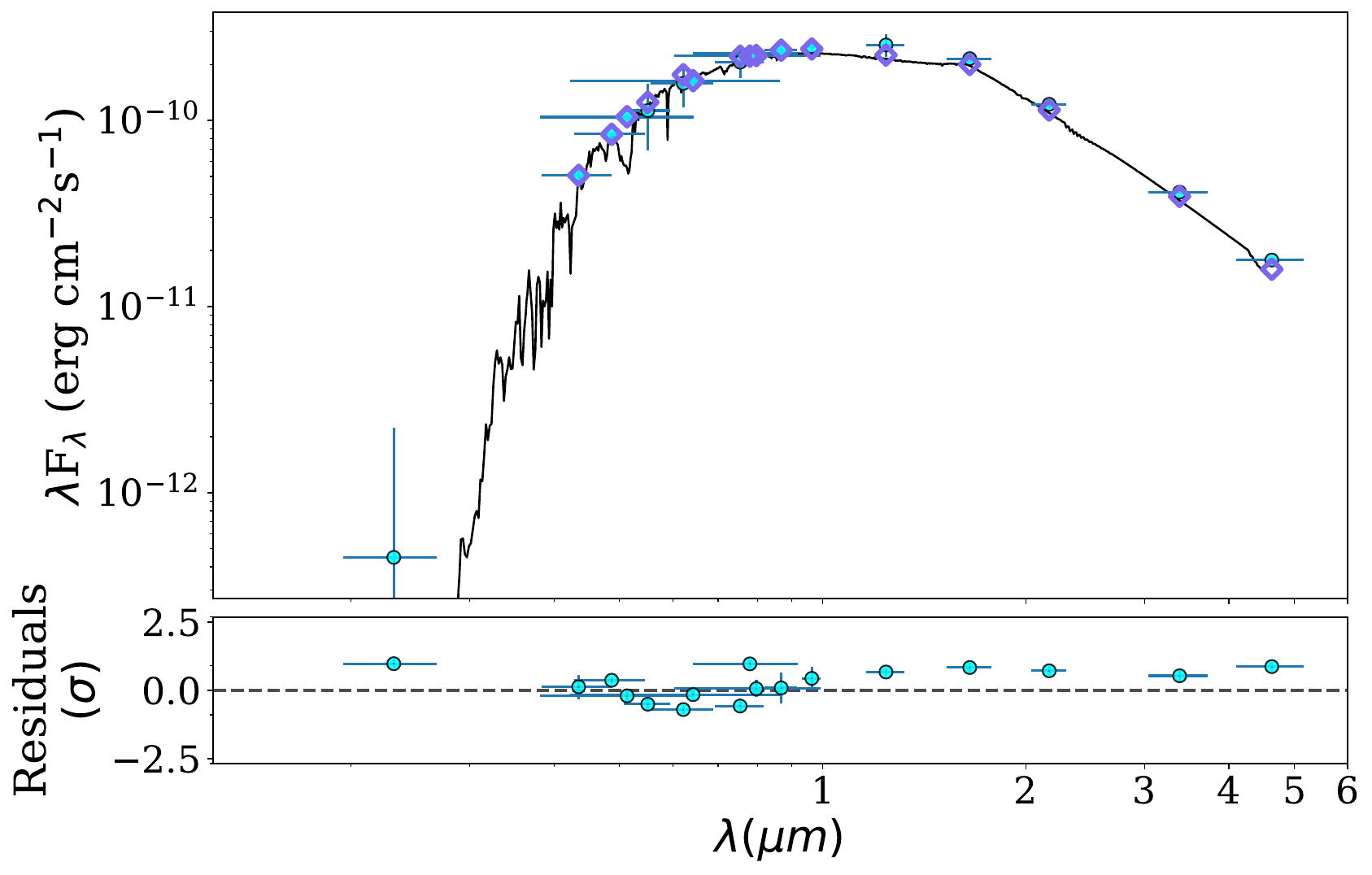}
    \includegraphics[width=0.45\linewidth]{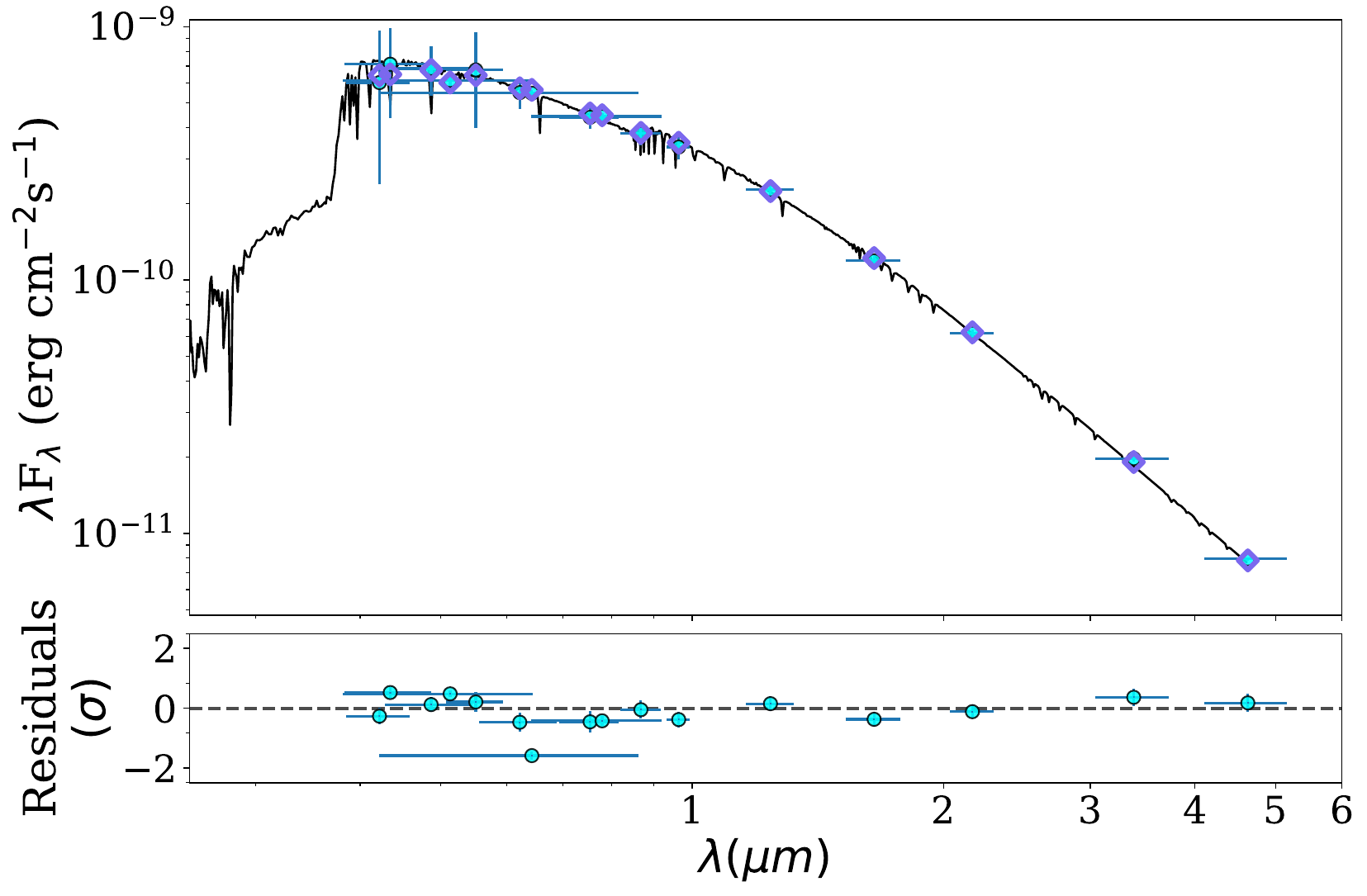}
    \caption{Spectral Energy Distributions for J1429 and J2023. }
    \label{fig:SED4}
\end{figure}

In this appendix, we provide SEDs for the sample of compact object candidates discussed in the main text. For the sources J2354 and J1729, which have already been certified in previous studies, we did not perform new SED fitting. Instead, we directly used the visible star radii in the literature. Figures~\ref{fig:SED1} to \ref{fig:SED4} display the SEDs for the remaining objects. Blue points denote observed fluxes, purple diamonds indicate synthetic fluxes, and the black curve represents the best-fitting SED model.

\clearpage
\section{TESS Light Curves of the Sample}
\begin{figure}[ht!]
    \centering
    \includegraphics[width=0.44\linewidth]{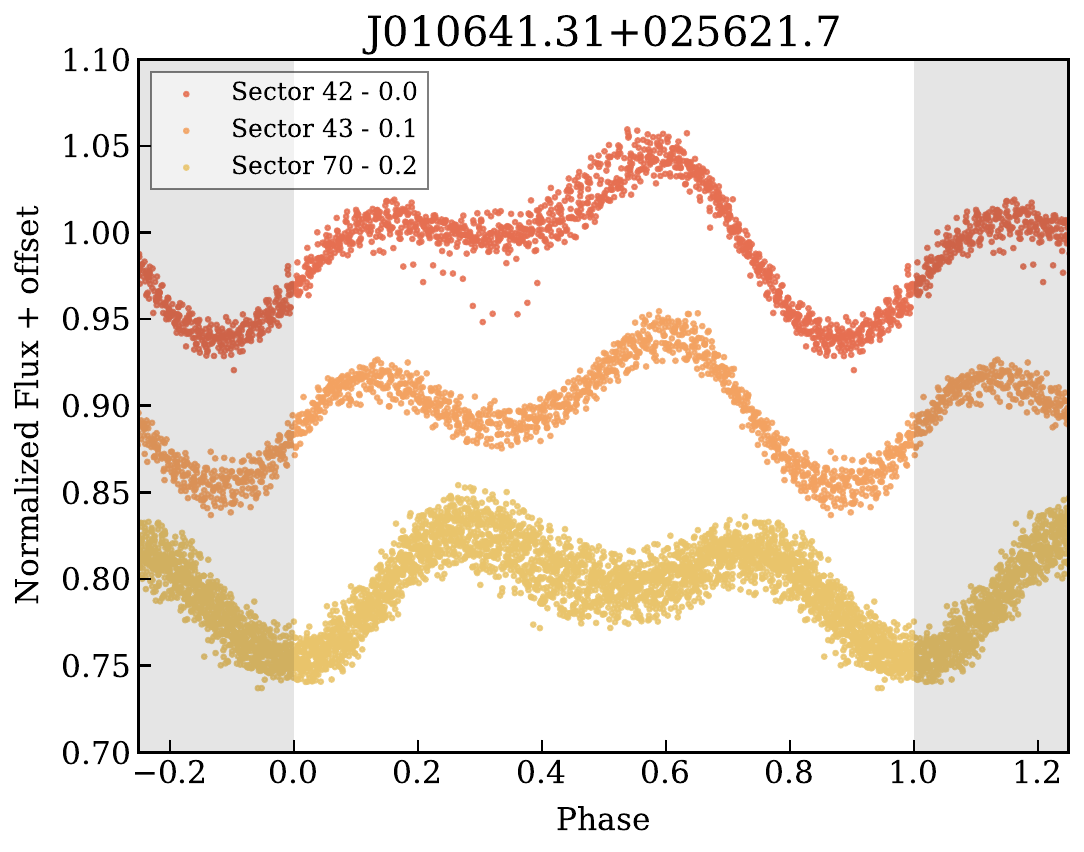}
    \includegraphics[width=0.45\linewidth]{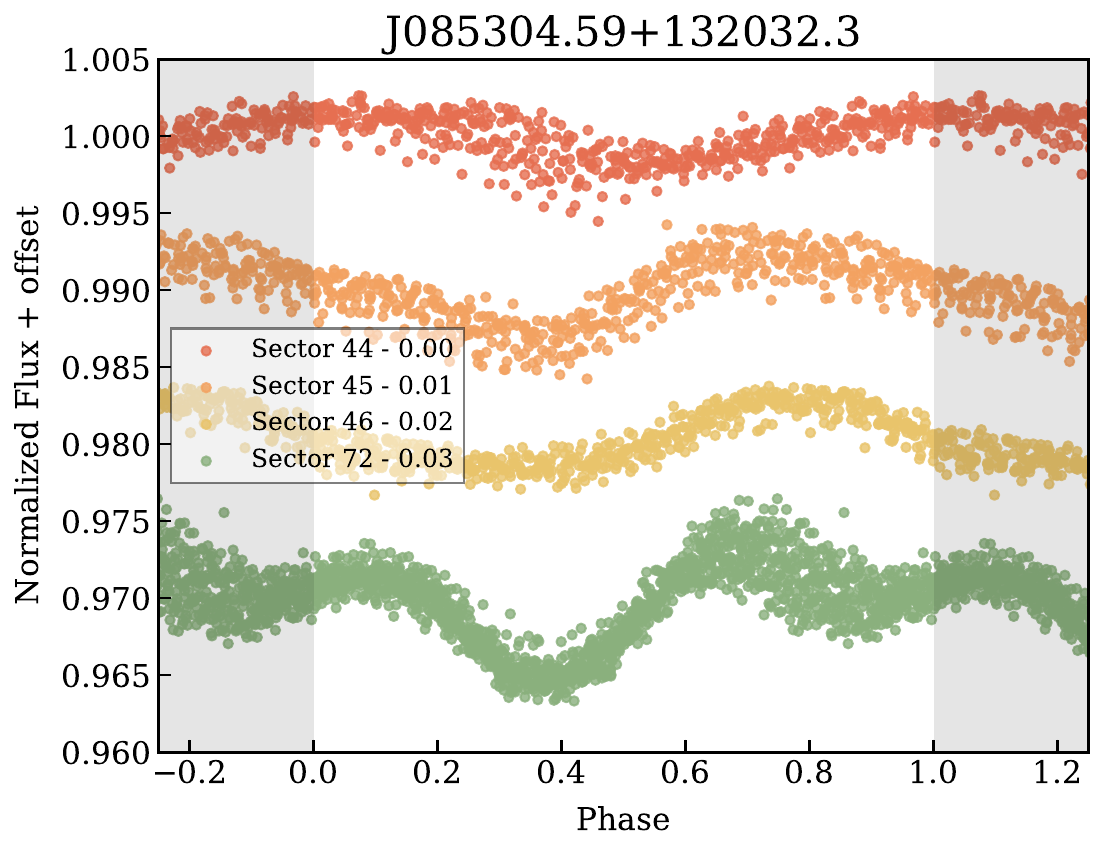}
    \caption{TESS light curves of J0106 (left) and J0853 (right) from multiple observing sectors. The light curve observed in different sectors showed significant changes in the structure.}
    \label{fig:lc}
\end{figure}

\begin{figure}[ht!]
    \centering
    \includegraphics[width=0.8\linewidth]{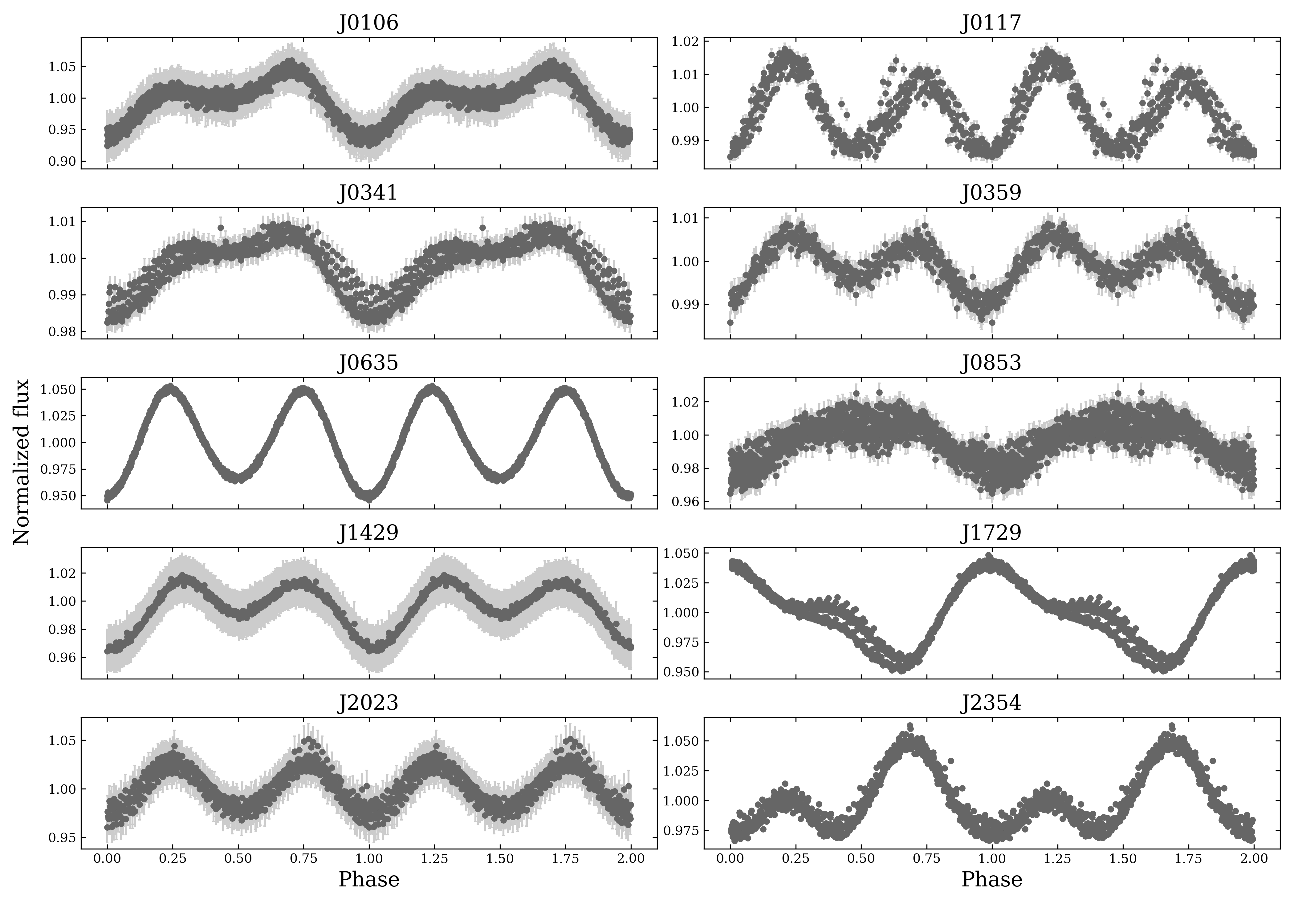}
    \caption{Phase-folded TESS light curves of the eight targets. The photometric data are folded on their respective orbital periods.}
    \label{fig:lc_all}
\end{figure}

Figure~\ref{fig:lc} presents the TESS light curves of J0106 and J0853, compiled from multiple observing sectors. Figure~\ref{fig:lc_all} shows the phase-folded TESS light curves of all targets in our sample, each folded on its known orbital period.

\clearpage
\section{Spectral Disentangling for the Sample}\label{disen}
The spectral disentangling results for all unclassified sources in our sample are presented in this appendix.
Each figure shows the disentangled spectra of the primary and secondary components. The secondary spectra have been rescaled according to the expected flux ratio based on the estimated secondary mass to enhance visibility. For comparison, representative stellar templates corresponding to the secondary masses (black dashed lines) are overplotted to illustrate the expected spectral features. The top panel in each figure displays the original observed spectrum (blue), the reconstructed spectrum from the disentangling procedure (red), and the individual spectra of the primary (green) and secondary (orange) components. The bottom panel shows the residuals between the reconstructed and observed spectra.

\begin{figure}[ht!]
    \centering
    \includegraphics[width=0.49\linewidth]{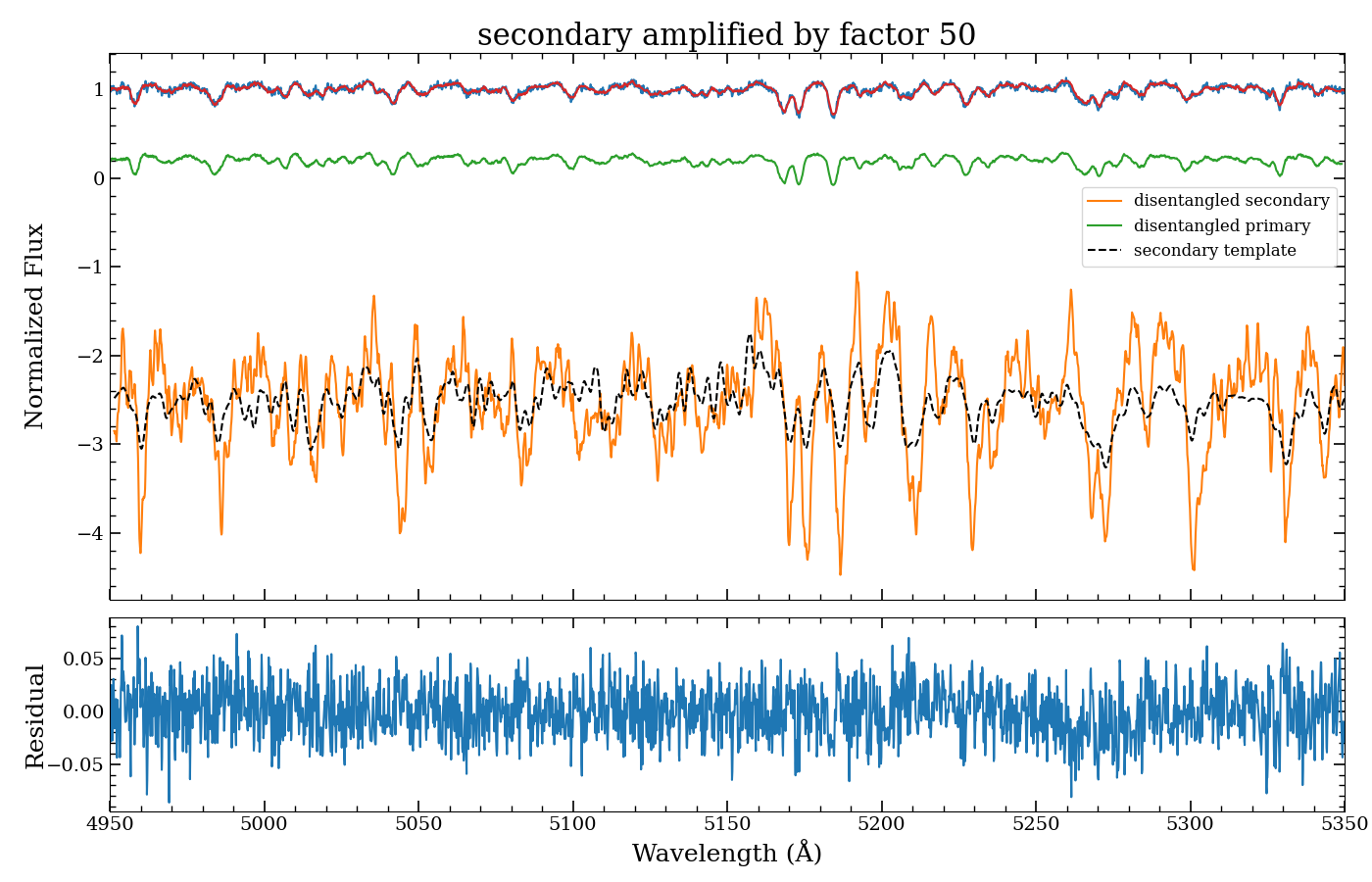}
    \includegraphics[width=0.49\linewidth]{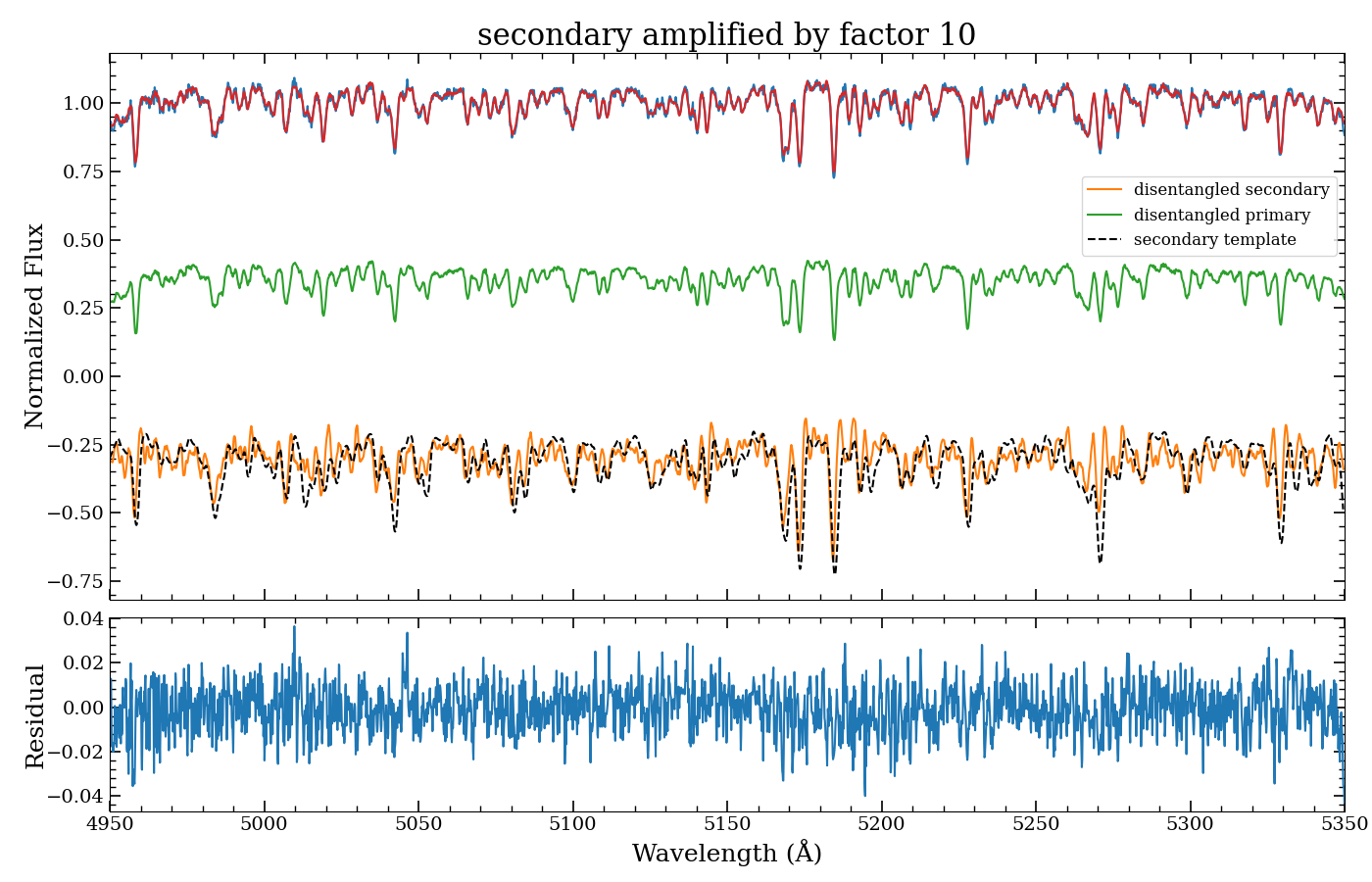}
    \caption{Disentangled spectra for the systems J0106 and J0117. 
\textbf{J0106:} Secondary spectrum rescaled by a factor of 50 and binned at $\Delta\lambda = 1$~\text{\AA}. Stellar template (black dashed line) corresponds to 0.57~$M_\odot$ ($T_{\rm eff}$ = 4424~K, $\log g$ = 4.0, $v \sin i = 80~\mathrm{km\,s^{-1}}$). \textbf{J0117:} Secondary spectrum rescaled by a factor of 10 and binned at $\Delta\lambda = 1$~\text{\AA}. Stellar template (black dashed line) corresponds to 1.08~$M_\odot$ ($T_{\rm eff}$ = 6048~K, $\log g$ = 4.0, $v \sin i = 80~\mathrm{km\,s^{-1}}$).}
    \label{fig_spec_1}
\end{figure}

\begin{figure}[ht!]
    \centering
    \includegraphics[width=0.49\linewidth]{J0341_distangle.png}
    \includegraphics[width=0.49\linewidth]{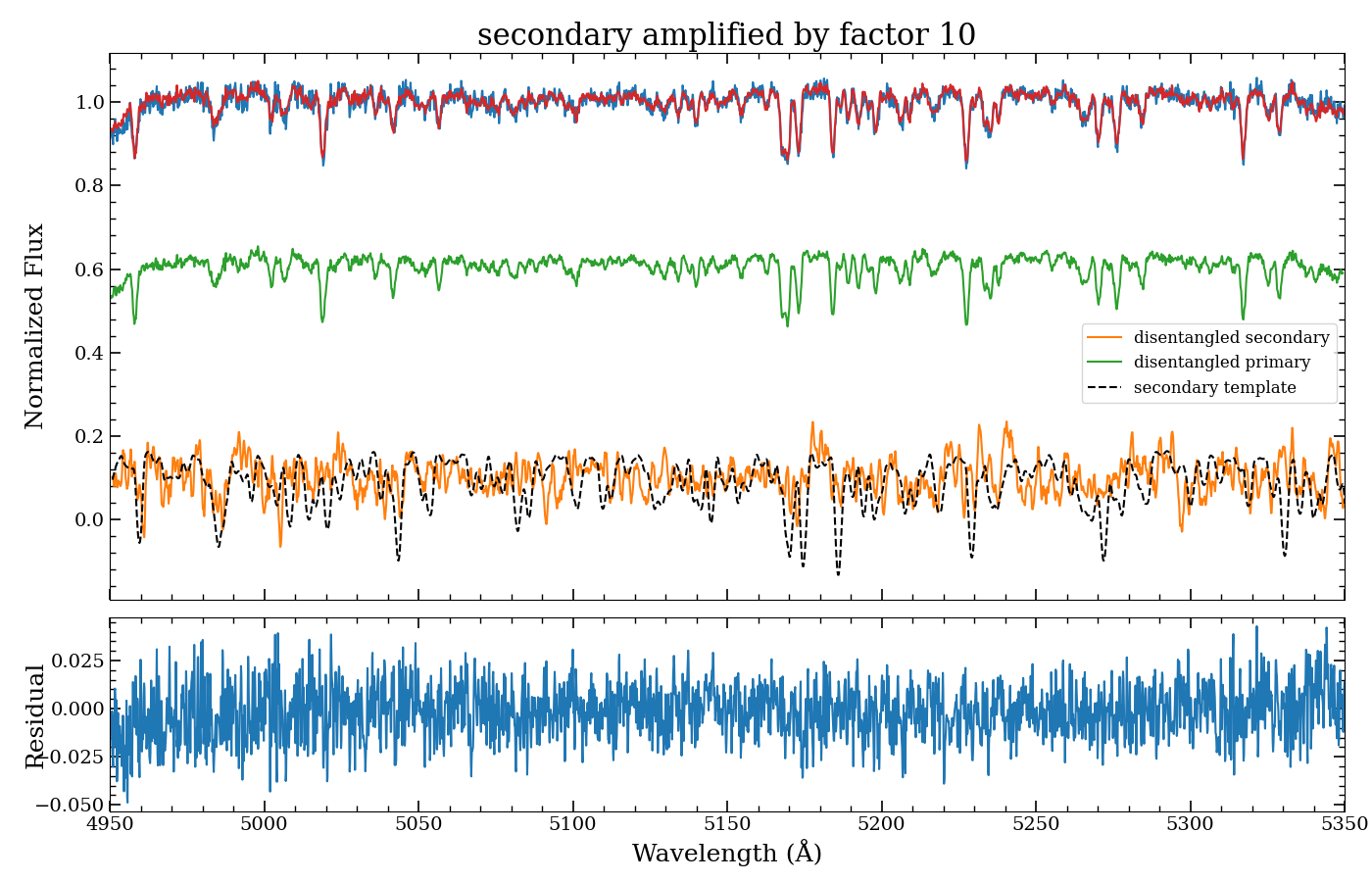}
    \caption{Disentangled spectra for the systems J0341 and J0359. 
\textbf{J0341:} Secondary spectrum rescaled by a factor of 2. Stellar template (black dashed line) corresponds to 1.34~$M_\odot$ ($T_{\rm eff}$ = 7017~K, $\log g$ = 4.0, $v \sin i = 80~\mathrm{km\,s^{-1}}$). \textbf{J0359:} Secondary spectrum rescaled by a factor of 10 and binned at $\Delta\lambda = 1$~\text{\AA}. Stellar template (black dashed line) corresponds to 1.39~$M_\odot$ ($T_{\rm eff}$ = 6867~K, $\log g$ = 4.0, $v \sin i = 80~\mathrm{km\,s^{-1}}$).}
    \label{fig_spec_2}
\end{figure}

\begin{figure}[ht!]
    \centering
    \includegraphics[width=0.49\linewidth]{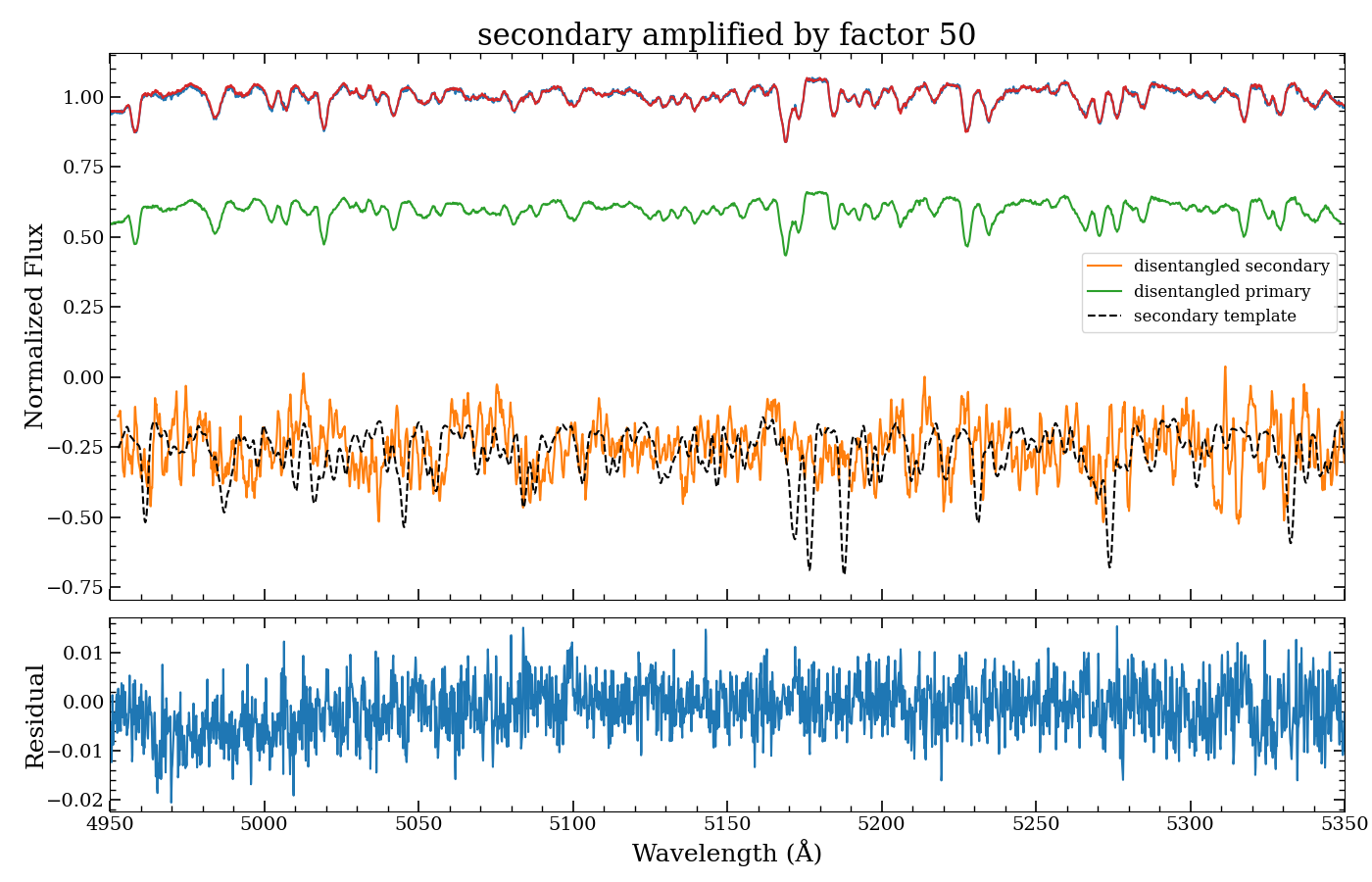}
    \includegraphics[width=0.49\linewidth]{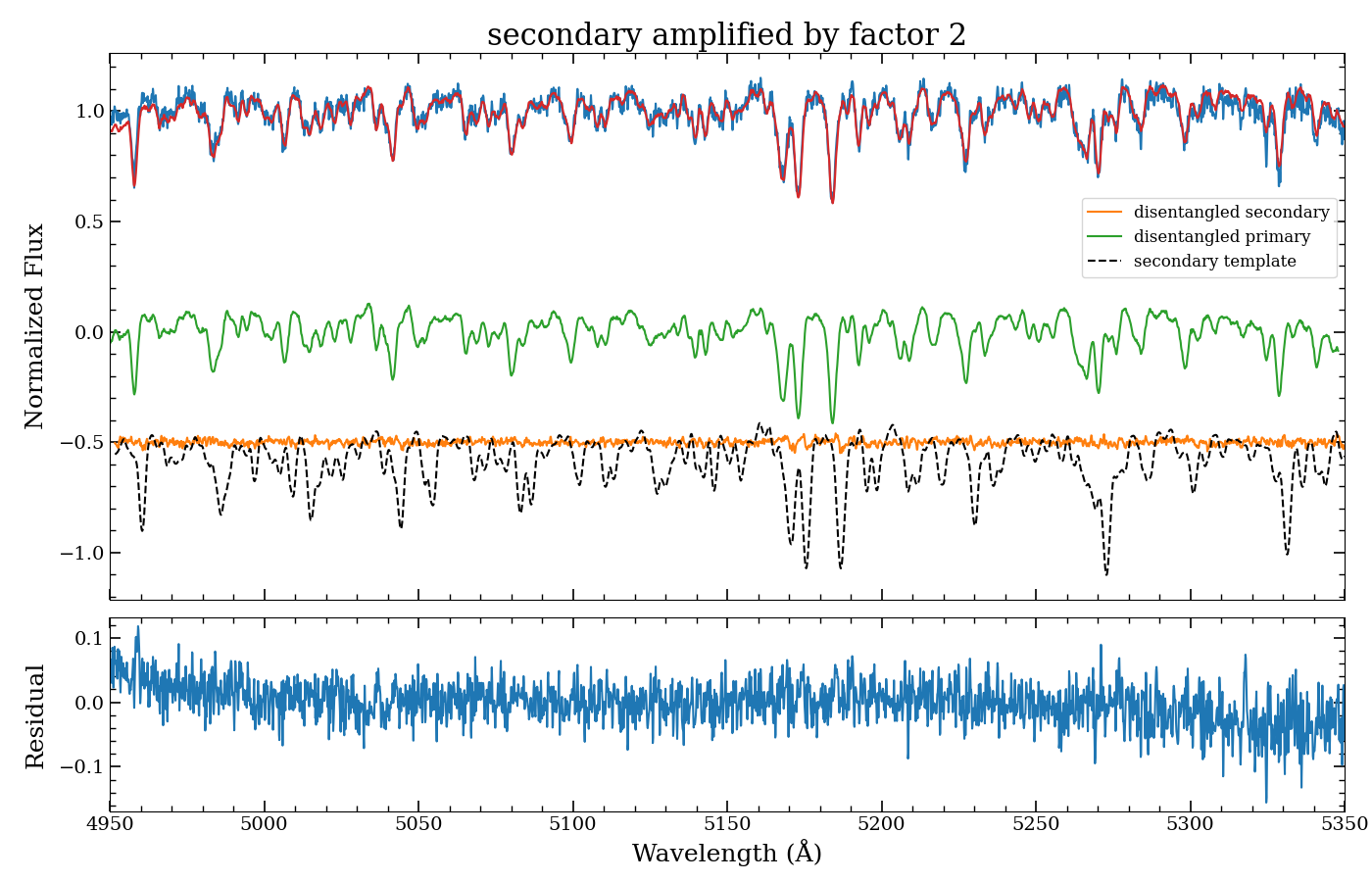}
    \caption{Disentangled spectra for the systems J0635 and J0853. 
\textbf{J0635:} Secondary spectrum rescaled by a factor of 50 and binned at $\Delta\lambda = 1$~\text{\AA}. Stellar template (black dashed line) corresponds to 1.04~$M_\odot$ ($T_{\rm eff}$ = 5466~K, $\log g$ = 4.0, $v \sin i = 80~\mathrm{km\,s^{-1}}$). \textbf{J0853:} Secondary spectrum rescaled by a factor of 2. Stellar template (black dashed line) corresponds to 0.89~$M_\odot$ ($T_{\rm eff}$ = 4157~K, $\log g$ = 4.0, $v \sin i = 80~\mathrm{km\,s^{-1}}$).}
    \label{fig_spec_3}
\end{figure}

\begin{figure}[ht!]
    \centering
    \includegraphics[width=0.49\linewidth]{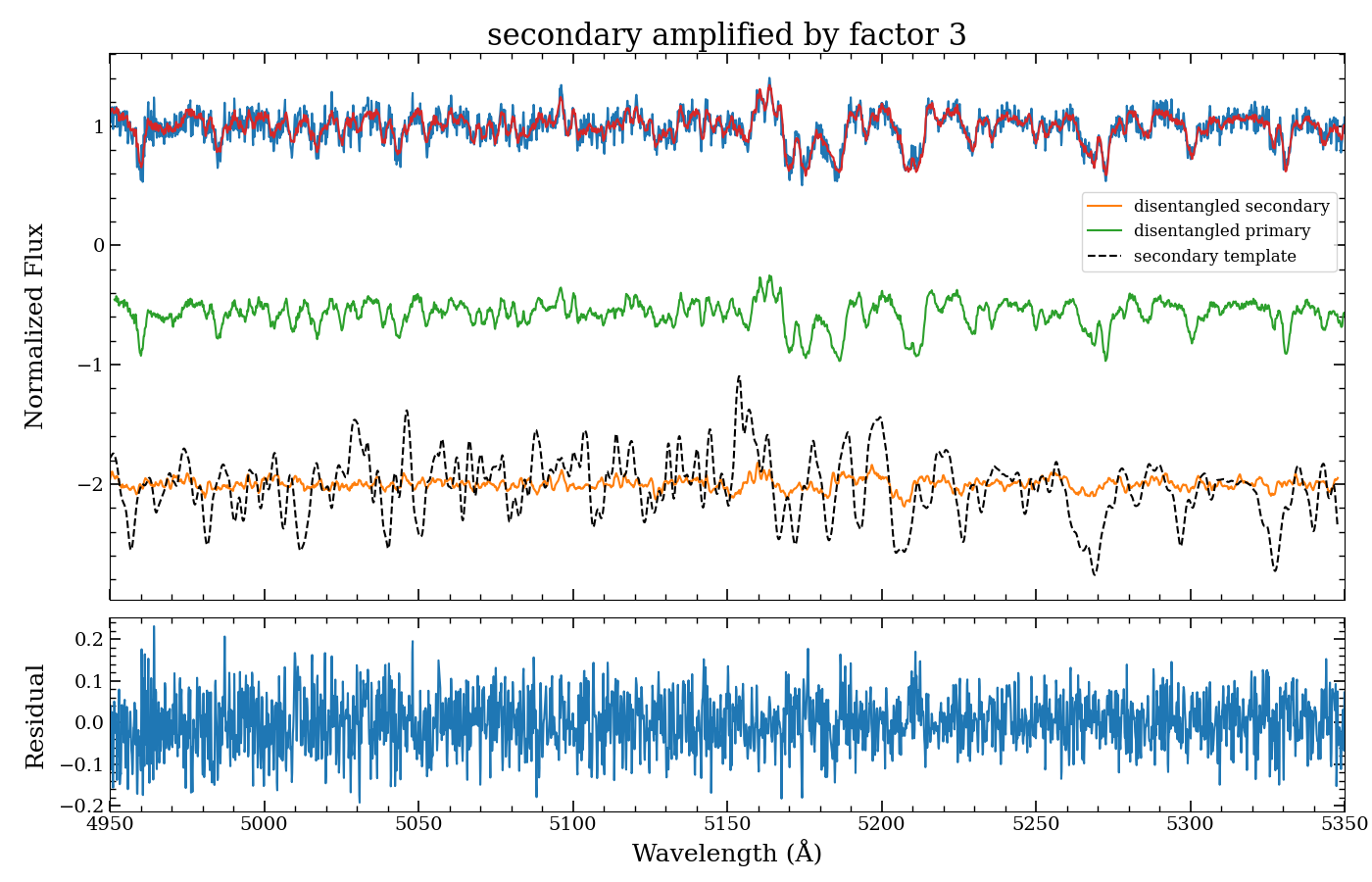}
    \includegraphics[width=0.49\linewidth]{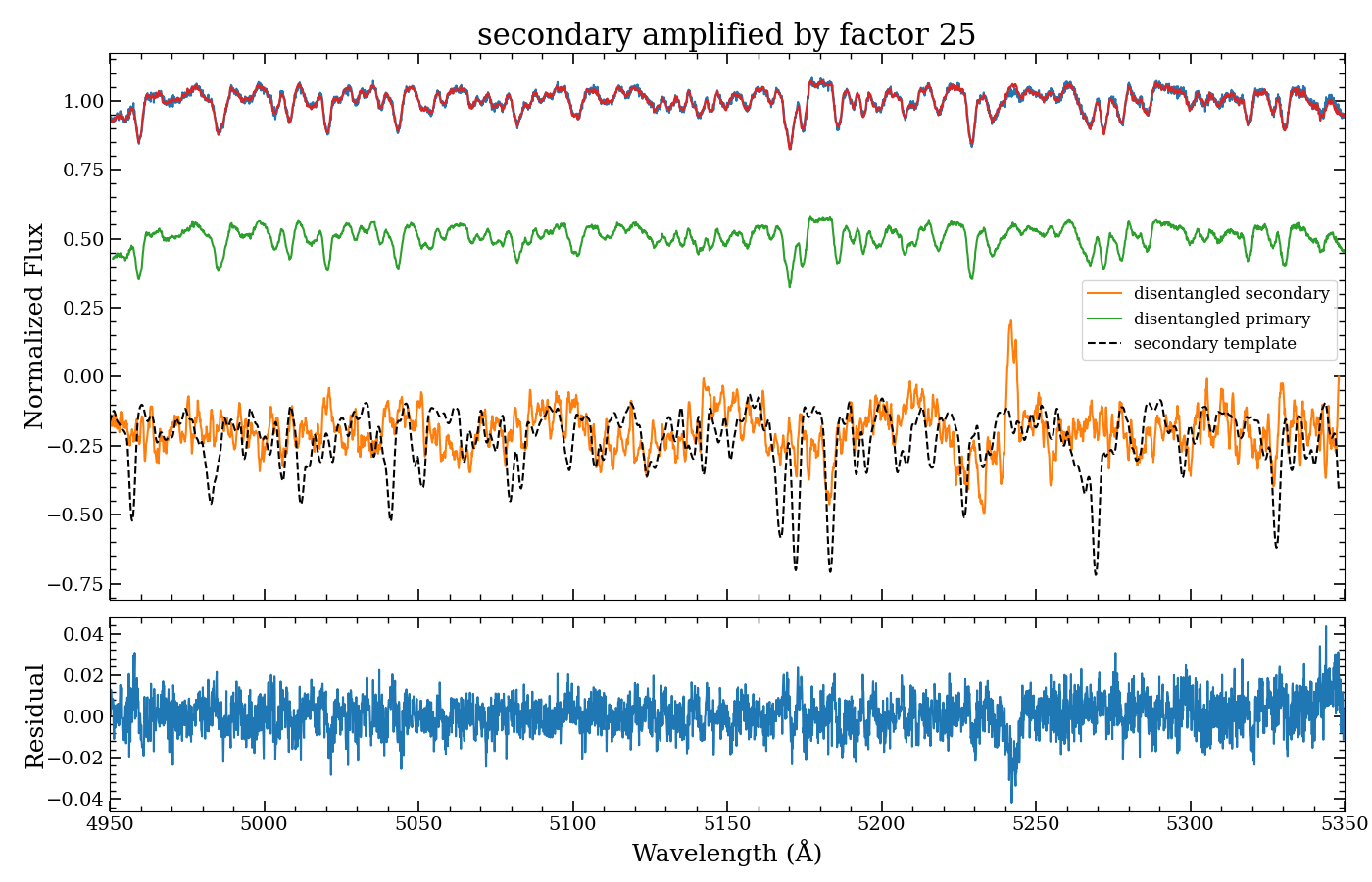}
    \caption{Disentangled spectra for the systems J1429 and J2023. 
\textbf{J1429:} Secondary spectrum rescaled by a factor of 3 and binned at $\Delta\lambda = 1$~\text{\AA}. Stellar template (black dashed line) corresponds to 0.50~$M_\odot$ ($T_{\rm eff}$ = 5638~K, $\log g$ = 4.0, $v \sin i = 80~\mathrm{km\,s^{-1}}$). \textbf{J2023:} Secondary spectrum rescaled by a factor of 25 and binned at $\Delta\lambda = 1$~\text{\AA}. Stellar template (black dashed line) corresponds to 0.95~$M_\odot$ ($T_{\rm eff}$ = 6867~K, $\log g$ = 4.0, $v \sin i = 80~\mathrm{km\,s^{-1}}$).}
    \label{fig_spec_4}
\end{figure}

\clearpage
\section{Ellipsoidal light curve Modeling of J2023}\label{phoebe}
\begin{figure}[ht!]
    \centering
    \includegraphics[width=0.45\linewidth]{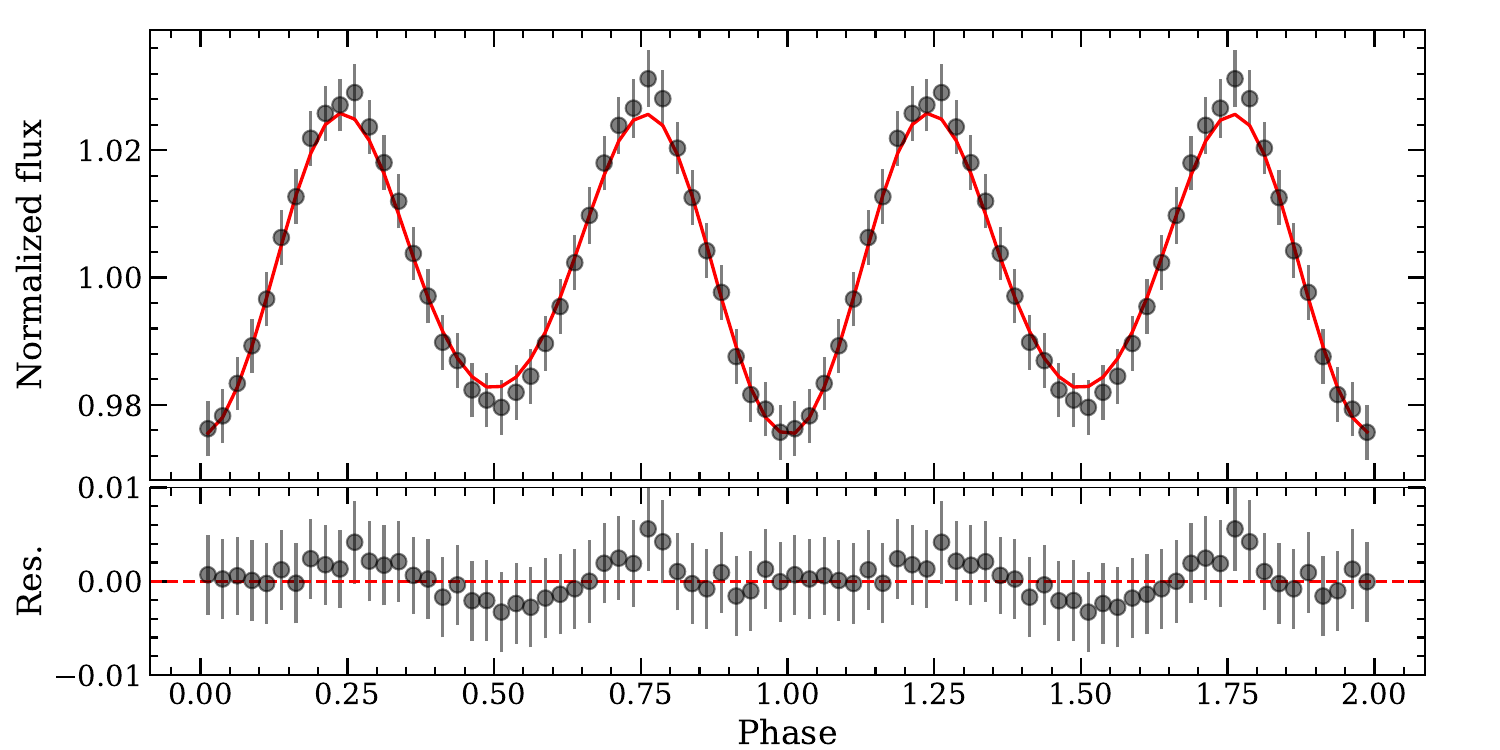}
    \includegraphics[width=0.45\linewidth]{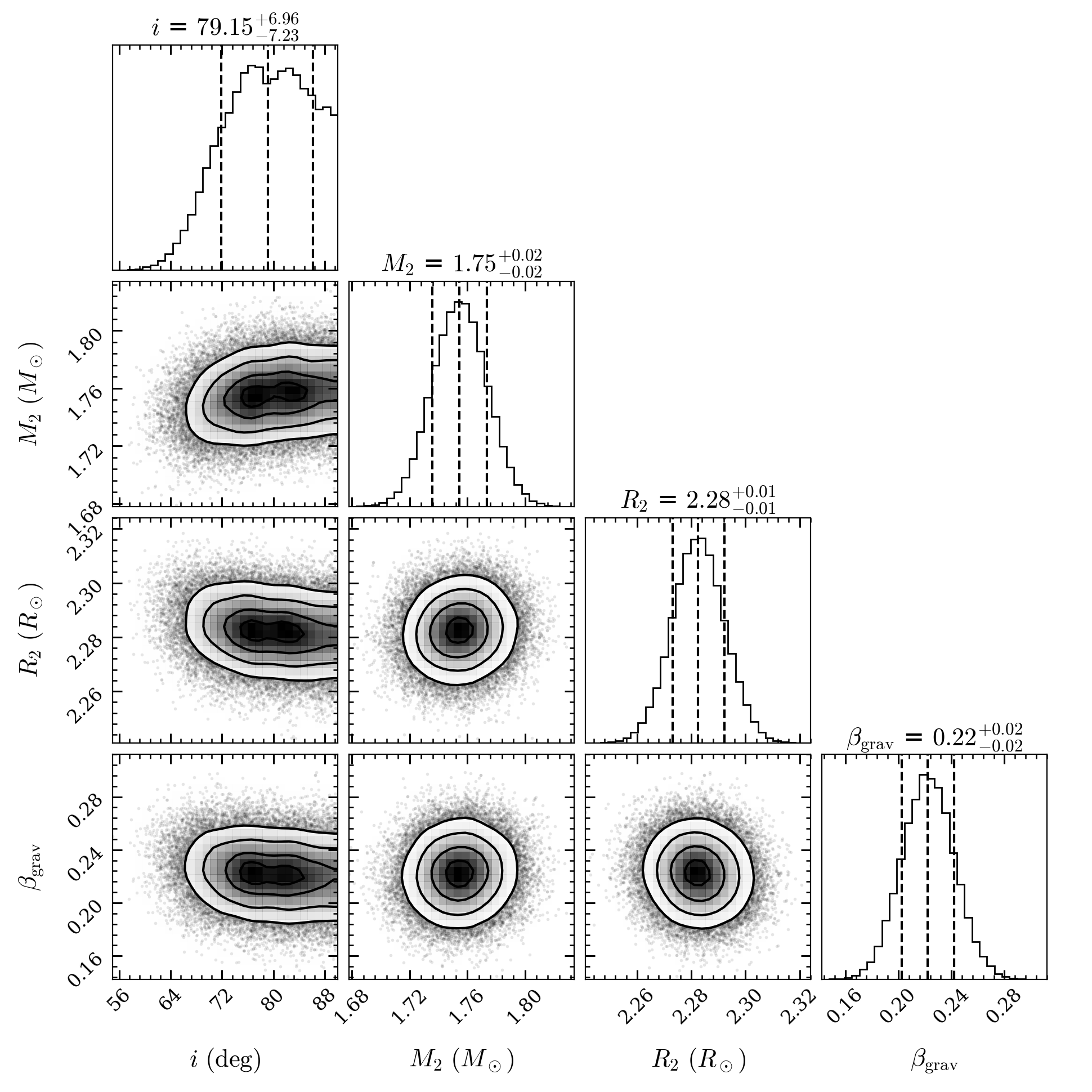}
    \caption{Best-fit \texttt{PHOEBE} light curve model compared with the TESS photometry (left), and posterior parameter distributions from the \texttt{PHOEBE} light curve fitting (right).}
    \label{fig:J2023}
\end{figure}

The best-fit \texttt{PHOEBE} light curve model and the corresponding posterior distributions for J2023 are presented in this appendix. Figure~\ref{fig:J2023} (left) shows the comparison between the TESS photometry and the \texttt{PHOEBE} model based on canonical ellipsoidal modulation. The small differences between the observed data and the model fits could be attributed to additional physical effects in the TESS photometry, such as surface inhomogeneities (e.g., starspots), which are not included in the current model. To avoid introducing additional free parameters and potential systematic uncertainties, we adopt the simplest ellipsoidal modulation model for J2023. Despite the imperfect fit, the inferred stellar parameters shown in Figure~\ref{fig:J2023} (right) are consistent with those listed in Table~\ref{tab:parameter} within uncertainties, supporting the robustness of the derived inclination and stellar properties.

\end{document}